\documentclass{jfm}

\makeatletter
\@ifundefined{mathindent}{\newdimen\mathindent}{}
\makeatother

\AtBeginDocument{%
  \mathindent=0pt
  \setlength{\displayindent}{0pt}
  \setlength{\displaywidth}{\linewidth}
}

\usepackage{graphicx}
\usepackage{newtxtext}
\usepackage{newtxmath}
\usepackage{natbib}
\usepackage{hyperref}
\hypersetup{
    colorlinks = true,
    urlcolor   = blue,
    citecolor  = black,
}

\newcommand{\RomanNumeralCaps}[1]
\linenumbers
\usepackage{amsmath} 
\usepackage{wrapfig} 
\usepackage{subcaption} 
\newcommand{\bvec}[1]{\mathbf{#1}}
\newcommand{\n}[1]{\mathrm{#1}} 

\newcommand{\eq}[1]{\label{eq:#1}}
\newcommand{\refeq}[1]{$\eqref{eq:#1}$}
\newcommand{\fig}[1]{\label{fig:#1}}
\newcommand{\reffig}[1]{$\ref{fig:#1}$}
\newcommand{\sect}[1]{\label{sec:#1}} 
\newcommand{\refsect}[1]{$\ref{sec:#1}$}

\newcommand{\R}{\mathbb{R}}

\newcommand{\cO}{\textit{O}}

\graphicspath{ {./images_folder/} }

\title{An inertial slender-body theory}

\author{Anmol Joshi\aff{1},
  Anubhab Roy\aff{2},
  Arjun Sharma\aff{3}
 \and Donald L. Koch\aff{1}
 \corresp{\email{dlk15@cornell.edu}}}

\affiliation{\aff{1}Robert Frederick Smith School of Chemical and Biomolecular Engineering, Cornell University, Ithaca, NY 14853, USA 
\aff{2}Department of Applied Mechanics, Indian Institute of Technology Madras, Chennai, Tamil Nadu 600036, India 
\aff{3}Center for Computing Research, Sandia National Laboratories, Albuquerque, NM 87185, USA}

\begin{document}
\maketitle

\begin{abstract}
We present a fully inertial slender-body theory (SBT) that incorporates the effect of fluid inertia on the scale of the length (the ``outer" region) as well as the characteristic diameter (the ``inner" region) of a steadily translating slender particle. This is achieved by matching the solution of the quasi-two-dimensional full Navier-Stokes equations in the inner region to an outer solution that consists of a superposition of a solution of the linearized Navier-Stokes equations driven by a line of forces and a potential flow solution driven by a line distribution of sources and source dipoles. The drag and lift forces result from the distribution of Oseen force singularities. These Oseenlets also predominantly govern the torque at small Reynolds numbers and large aspect ratios.  However, the potential flow singularities play a crucial role in yielding a torque that grows with increasing Reynolds number at large Reynolds numbers and finite aspect ratios. By comparing the forces and torque on the steadily translating particle with those obtained from a finite difference Navier-Stokes solution, we demonstrate the accuracy of the resulting inertial SBT for $\n{Re}_D$ up to 10, where $\n{Re}_D$ is the Reynolds number based on the smallest dimension, i.e., the characteristic cross-sectional diameter of the slender particle. 
\end{abstract}

\begin{keywords}

\end{keywords}


\section{Introduction}
\sect{intro}

The objective of this work is to develop a slender-body theory (SBT) to predict the dynamics of slender particles in fluid flows, taking into account the effects of finite fluid inertia. Traditional SBT calculations often rely on simplifications that do not fully capture the complex interactions between the particle and the surrounding fluid, especially at Reynolds numbers that are not small. This limitation is crucial in industrial and environmental processes like sedimentation, pollutant transport, and particle dynamics in manufacturing, where inertial effects significantly influence process efficiency and outcomes. Besides, numerical simulations for slender particles are computationally challenging due to the need to resolve thin particle boundaries and long-range flow disturbances, providing further motivation for a theoretical framework. By developing and validating a more comprehensive SBT that incorporates fluid inertia effects, we aim to provide a robust framework for better understanding and predicting these dynamics, thereby enhancing the accuracy of theoretical solutions used in both scientific research and practical applications.  
\\[6pt] 
The interactions of spherical particles with imposed flows and with one another have been extensively studied through experiments, theoretical analyses, and numerical simulations. Much of this research has focused on hydrodynamic interactions, particularly in the Stokes flow regime and under the influence of weak fluid inertia, to understand their impact on the average settling velocity of suspensions (\cite{batchelor1972sedimentation,nguyen2005sedimentation,subramanian2008evolution}). While these studies have significantly advanced our knowledge of spherical particle dynamics, real-world applications often involve non-spherical, high-aspect-ratio particles, whose interactions with fluid flows are far more complex.  
\\[6pt] 
Elongated particles introduce additional degrees of freedom due to their orientation, leading to unique fluid-particle interactions that spherical particle models cannot fully capture. Understanding these dynamics is essential for optimizing various industrial and environmental processes. For example, in fluidized beds, precise control over particle motion enhances efficiency, while in textile and paper manufacturing, the alignment of fibers in high-speed flows directly affects product quality (\cite{holm2005fluid,carlsson2007fiber}). Similarly, in the environment, the transport and distribution of microplastics, algae, and plankton in rivers and oceans are governed by their interactions with fluid flows (\cite{khatmullina2017settling}). Mixed-phase cloud systems, including cirrus clouds, contain sedimenting ice crystals whose orientation distributions and terminal velocities play a crucial role in thermal energy, cloud albedo and radiation forcing in climate models (\cite{heymsfield2000cirrus,stillwell2019radiative}). Inertia strongly influences the orientation distributions of the ice crystals, often modeled as slender particles, in the ambient turbulent flow (\cite{roy2023orientation}). Addressing these complexities is key to improving wastewater treatment, refining industrial processes, better modeling ice crystal microphysics and understanding the behavior of natural and synthetic fibers in aquatic ecosystems. 
\\[6pt] 
Numerical simulations have provided valuable insights into the influence of fluid inertia on the drag, lift, and torque experienced by non-spherical particles translating in quiescent fluids. However, these studies have largely been limited to particles with aspect ratios of at most 20 to 30,  and in some cases to inclination angles below $30^o$ (\cite{vakil2009drag, jiang2021inertial, fintzi2023inertial, PhysRevFluids.6.044308}). In this context, the finite-difference Navier-Stokes solver developed by \cite{sharma2023finite} formulated in prolate spheroidal coordinates, is particularly well-suited for simulating fluid flow past prolate spheroids with moderately large aspect ratios—up to approximately 100. In this study, we employ this numerical method using body-fitted coordinate system to compare force and torque predictions with those obtained from the inertial slender-body theory (SBT). 
\\[6pt] 
A convenient framework for understanding the dynamics of non-spherical particles with high aspect ratio has been built upon the method of matched asymptotic expansions. Commonly known as slender-body theory, this approach has led to several useful insights on the behavior of these particles in the low Reynolds number limit. Some examples include the anisotropy in the resistance to motion (\cite{batchelor1970slender}), the instability of the homogeneous state of sedimenting particle suspensions (\cite{koch1989instability}) and the resulting enhancement in the mean settling rate (\cite{butler2002dynamic}). This understanding is based on the translational and rotational velocities of the particles in local linear velocity fields; the forces, torques, and stresslets they exert on the fluid, and the far-field velocity disturbances they induce. 
\\[6pt]
Slender-body theory has also been extended to account for weak inertial effects of the surrounding fluid at the scale corresponding to the particle length, i.e. its largest dimension. Developed by \cite{khayat1989inertia}, this is based on the asymptotic matching of a Stokes flow solution on the scale of the characteristic particle diameter $D$ (the ``inner" region) to an ``outer" solution (on the scale of the particle length $L$) of the linearized Navier-Stokes equation for a steadily translating particle. However, a critical limitation of the weakly inertial theory of \cite{khayat1989inertia} is that it involves an expansion in $\n{Re}_L/\n{ln}(2 \kappa)$ assuming the inertial contributions to the force per unit length to be small in comparison with the viscous solution. Here, $\n{Re}_{L} = \rho U L/\mu$ is the Reynolds number based on the particle length and $\kappa = L/D$ is the aspect ratio of the particle. Also, $\rho$ and $\mu$ denote the fluid density and viscosity respectively. Comparisons with experiments (\cite{roy2019inertial,lopez2017inertial}) show that the weakly inertial theory accurately describes the force and torque on translating fibers of moderately large aspect ratio for $\n{Re}_D < 0.1$ and $\n{Re}_L < 2$. More recently, \cite{khair2018higher} extended the weakly inertial theory of \cite{khayat1989inertia} to obtain the next higher-order correction, $O(\n{Re}_L/(\n{ln}(2\kappa))^3)$, to the drag on an axisymmetric particle held parallel to a uniform stream. But this analysis is also limited to the asymptotic regime $\n{Re}_L \ll \n{ln}(2\kappa)$, and not applicable for oblique fiber orientations relative to the uniform stream.  
\\[6pt] 
However, fluid inertia, characterized by its nonlinearities and time-dependent nature, often exerts a significant influence on particle motion in low viscosity fluids like water and air, aspects that the weakly inertial theory fails to capture. For example, the Reynolds number $\n{Re}_D = \rho U D/\mu$, based on the diameter $D$ and velocity $U$ of a slender fiber with density difference $1000 \ \n{kg/m}^3$ settling in air or water exceeds $\n{Re}_D = 0.1$ for $D$ greater than $20 \ \n{\mu m}$ and $30 \ \n{\mu m}$, respectively. Above this threshold, nonlinear inertial effects are important even on the length scale of the smallest dimension $D$ of the particle. For particles with large aspect ratio $\kappa = L/D$, the Reynolds number based on the particle length $\n{Re}_{L} = \rho U L/\mu$ is then larger than one indicating that inertia dominates. Particles experiencing strong inertial effects are important in a variety of industrial applications including fluidization and pneumatic conveying of fibers, as well as in fibrous filters (\cite{henthorn2005measurement,papathanasiou2001computational}). 
\\[6pt]
In this work, we introduce an inertial slender-body theory (SBT) that considers the interplay between fluid inertia and viscous stresses in both the inner region (at the scale of the particle diameter) and the outer region (at the scale of the particle length) for a steadily translating slender particle. We compare our predictions for the force and torque on the particle with experimental data, the weakly inertial theory of \cite{khayat1989inertia}, and results obtained from the finite difference solver of \cite{sharma2023finite} which solves the full Navier-Stokes equations around finite aspect ratio spheroids. 
\\[6pt]
In the next section, we show the development of slender body theory integral equation for the force per unit length along the particle axis, followed by the numerical method to solve this integral equation. Section \refsect{oseen_flow_SBT_results} presents a comparison of the force and torque on a steadily translating fiber with experimental data and finite-difference Navier-Stokes solutions. We also validate the leading-order flow disturbance around the fiber in the matching region against the full Navier-Stokes solution and highlight the limitations of inertial SBT with an Oseen flow outer solution for $\n{Re}_D > 1$. To overcome these shortcomings, in section \refsect{pot_flow}, we superpose the outer Oseen solution with a potential flow disturbance around the fiber and derive the source and source dipole distribution along the particle axis that drives this potential flow. This modification enables accurate prediction of the inertial torque on the fiber when $\n{Re}_D$ exceeds 1. Finally, we present our conclusions in section \refsect{conclusions}. 

\section{Finite $\bvec{Re_D}$ slender body theory for uniform fluid flow relative to an oblique fiber based on an Oseen flow outer solution}
\sect{Finite_ReD_SBT}

Slender-body theory (SBT), originally developed by \cite{batchelor1970slender} for Stokes flow, is a method of matched asymptotic expansions valid in the limit of large particle aspect ratio, $\kappa \gg 1$, where $\kappa = L/D$, $L$ and $D$ being the particle length and a characteristic value of its cross-sectional diameter respectively. It consists of obtaining a quasi-two-dimensional solution to the governing equations in the “inner” region (where the length scale is on the order of the cross-sectional particle radius, $a_0 = D/2$) and matching it to the “outer” solution (where the distance from the particle center is on the order of the particle half-length, $l = L/2$). Previously, slender-body theory has been commonly applied to Stokes flow around slender fibers with Reynolds number based on both fiber length and diameter being asymptotically small (\cite{cox1970motion,cox1971motion,keller1976slender}). This has been based on the solutions of Stokes equations in both the inner and outer regions. The first effort toward incorporating inertia in the slender-body theory was made by \cite{khayat1989inertia} who matched the solution of the linearized Navier-Stokes equations (Oseen equations) in the outer region with the Stokes flow inner solution (see figure \reffig{fig1}).
\begin{figure}  
\centering
\includegraphics[width=0.8\textwidth]{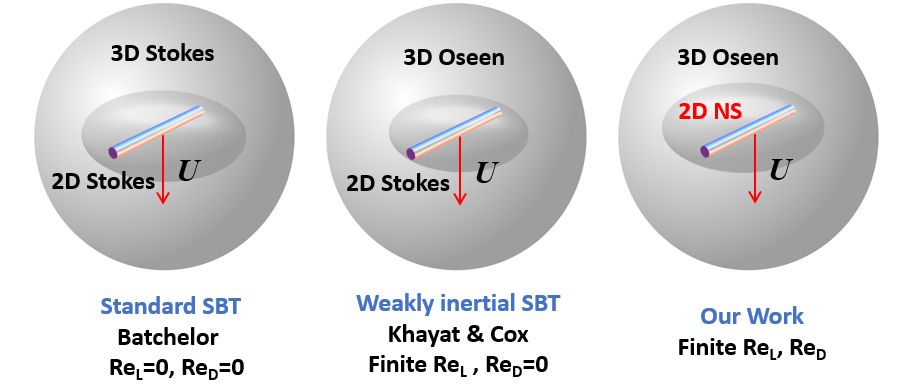}
\caption{ Slender body theories for various Reynolds numbers; (left) \cite{batchelor1970slender}: $\n{Re_D,Re_L = 0}$; (middle) \cite{khayat1989inertia}: $\n{Re_D = 0, Re_L \neq 0}$; (right) our work: $\n{Re_D,Re_L \neq 0}$ }
\fig{fig1}  
\end{figure} 
\\[6pt] 
In this study, we develop a slender-body theory to study the forces and torque experienced by a fully inertial particle translating in a fluid, a problem for which fluid inertia plays an important role on length scales based both on fiber diameter and length. This will occur when even $\n{Re}_D$ is \cO(1). 
\\[6pt]
We start by matching a two-dimensional solution of the full Navier-Stokes equations in the inner region with a three-dimensional solution of the Oseen approximation to the Navier-Stokes equations in the outer region. This leads to an integral equation in which the force per unit length responds to the local relative velocity of the fluid and fiber through an inertial drag coefficient. We go on to show that these Oseenlets capture the drag and lift on the translating fiber accurately, but fail to capture the inertial torque on the fiber when $\n{Re}_D$ exceeds unity. The Oseen flow outer solution is then complemented with a potential flow solution driven by a line distribution of fluid mass sources and source dipoles along the fiber axis. This leads to an inertial torque on the fiber that is accurate up to an $\n{Re}_D$ of 10. 

\subsection{Formulation of the slender body theory integral equation for $\n{Re}_D = \cO(1)$}

The leading order outer solution in the slender body theory for a straight fiber translating with a velocity $\bvec{U}$ in a quiescent fluid satisfies the linearized Navier-Stokes equations (Oseen equations) with a force distribution $\bvec{f}(s)$ exerted by the fiber on the fluid that depends upon a coordinate $s$ measured along the fiber axis. 
\begin{gather}
-\frac{\n{Re}_L}{2} \ \bvec{e}_{\bvec{U}} \cdot \nabla \bvec{u} + \nabla p - \nabla^2\bvec{u} = \int_{-1}^{1} \n{d}s \bvec{f}(s) \delta (\bvec{r}-s\bvec{p}) \eq{oseen_outer_1} \\
\nabla \cdot \bvec{u} = 0
\eq{oseen_outer_2} 
\end{gather} 
In the above equations, the fluid velocity $\bvec{u}$ is evaluated with respect to a reference frame that translates with the fiber. Here, we have non-dimensionalized lengths with the fiber half-length $l = L/2$, the fluid velocity $\bvec{u}$ with the magnitude of the fiber velocity $U$, and the force per unit length with $\mu U$. We define $\bvec{e}_{\bvec{U}}$ as a unit-vector parallel to $\bvec{U}$, $s$ as a coordinate measured along the fiber length, and $\bvec{p}$ as a unit vector parallel to the fiber axis. The solution to equations \refeq{oseen_outer_1}-\refeq{oseen_outer_2} is:
\begin{gather}
\bvec{u}(\bvec{r}) = \int_{-1}^{1} \n{d}s \bvec{f}(s) \cdot \bvec{G}(\bvec{r}-s\bvec{p}) \eq{oseen_outer_vel} 
\end{gather}
where,
\begin{gather}
\bvec{G}(\bvec{r}) = \frac{1}{4\pi r} \n{e}^{-\beta} \bvec{I} + \frac{1}{8 \pi} \left[ \frac{\n{Re}_L}{4}\left(\frac{\beta \n{e}^{-\beta} + \n{e}^{-\beta} -1}{\beta^2}\right)\left(\frac{\bvec{r}}{r}-\bvec{e}_U\right)\left(\frac{\bvec{r}}{r}-\bvec{e}_U\right) + \left(\frac{\n{e}^{-\beta}-1}{\beta}\right)\left(\frac{\bvec{I}}{r}-\frac{\bvec{r}\bvec{r}}{r^3}\right) \right] \eq{oseenlet1} \\
\beta = \frac{\n{Re}_L}{4}(r-\bvec{r} \cdot \bvec{e}_U) \eq{oseenlet2} \\
\n{Re}_L = \frac{UL}{\nu} \eq{ReL} 
\end{gather} 
Here, $\bvec{G}(\bvec{r})$ is the Green’s function associated with the Oseen equations (\cite{pozrikidis2011introduction}). The Oseen solution for a translating point force presents a uniformly-valid first approximation for the flow-field. It reduces to a Stokeslet field for $r << l_O$, where $l_O = \nu/U$ is the inertial screening length, or the Oseen length. On the other hand, for distances considerably larger than $l_O$, it exhibits characteristics of a source flow, accompanied by a wake trailing behind the translating point force. 
\\[6pt] 
The use of linearized Navier-Stokes equations is justified for the outer region because the velocity disturbance due to the particle diminishes significantly at distances from the particle axis much larger than $D$. In Stokes flow, the velocity in the outer region scales with $1/\n{ln}(\kappa)$, and has a logarithmic dependence on the radial distance $h$ from the fiber axis when $a_0 \ll h \ll l$. At finite Reynolds numbers, the velocity decays beyond the Oseen length in the matching region, following a $1/h^{1/2}$ trend in the wake and $1/h$ outside the wake in the source flow region. Going further away, the flow field becomes three-dimensional and varies as $1/r$ in the wake and $1/r^2$ in the source flow region. The validity of using the Oseen approximation for flows far from a particle, even at higher Reynolds numbers, has been supported by studies such as \cite{daniel2009clusters}, \cite{subramanian2008evolution}. These works demonstrated that the experimentally observed spreading rate of a cluster of spherical particles could be predicted by their Oseen source flow interactions even for Reynolds numbers as large as $\n{Re}_D = 300$, provided that the non-linear drag law was used to determine the relationship between the particle force and the relative velocity of the fluid and particle. 
\\[6pt] 
The matching of this outer solution with the inner solution is facilitated by two observations. First, as noted by \cite{keller1996asymptotics}, the two-dimensional solution to the full Navier-Stokes equations approaches the two-dimensional Oseen solution due to a point force in the matching region, $l>>h >> l_O$.
\begin{gather}
\bvec{u}_{Oseen}^{2D}(\bvec{h},s) = \frac{1}{4 \pi \mu} ( \n{K}_0(ch) e^{c(\bvec{h}\cdot \bvec{e}_1)}(\bvec{I} + \bvec{pp}) \cdot \bvec{f}(s) - ( \n{K}_1(ch) e^{c(\bvec{h}\cdot \bvec{e}_1)} - \frac{1}{ch} ) (\bvec{e}_h \cdot \bvec{e}_1) (\bvec{I} - \bvec{pp}) \cdot \bvec{f}(s) \nonumber \\
+ \ 2(\n{K}_1(ch) e^{c(\bvec{h}\cdot \bvec{e}_1)} - \frac{1}{ch} ) (\bvec{e}_h \cdot \bvec{e}_1) \bvec{e}_1 \bvec{e}_1 \cdot \bvec{f}(s) + ( \n{K}_1(ch) e^{c(\bvec{h}\cdot \bvec{e}_1)} - \frac{1}{ch} ) (\bvec{e}_h \cdot \bvec{e}_2) (\bvec{e}_1 \bvec{e}_2 + \bvec{e}_2 \bvec{e}_1) \cdot \bvec{f}(s)) \eq{2D_oseen} 
\end{gather} 
Here, $K_0$ and $K_1$ are the zeroth and the first order modified Bessel functions of the second kind respectively, and $\bvec{e}_1$ denotes the unit vector along $(\bvec{I} - \bvec{pp}) \cdot \bvec{U}$ with $\bvec{e}_2 = \bvec{p} \times \bvec{e}_1$. Moreover, $\bvec{h}$ denotes the position relative to the fiber centre in the plane perpendicular to the fiber axis at coordinate $s$. Finally, $\bvec{e}_h$ is the unit vector along $\bvec{h}$ and $c = |\bvec{U}|/2\nu$.  
\\[6pt] 
Second, the singular part of the two-dimensional Oseen solution due to a point force as $h$ approaches zero coincides with the singular part of the two-dimensional Stokes solution.
\begin{gather}
\lim_{h \rightarrow 0} \bvec{u}_{Oseen}^{2D}(\bvec{h},s)=\bvec{u}_{Stokes}^{2D}(\bvec{h},s) = - \frac{1}{4\pi \mu} \n{ln}(h) (\bvec{I}+\bvec{pp})\cdot \bvec{f}(s) + \bvec{c}(\bvec{h})  \eq{stokes_inner} 
\end{gather} 
where, $\bvec{c}(\bvec{h})$ is an \cO(1) constant. Taken together, these observations indicate that the full two-dimensional Navier-Stokes solution transitions to the same Stokes solution as that approached by the two-dimensional Oseen solution provided that the force per unit length $\bvec{f}$ and the constant $\bvec{c}$ in the two solutions are the same. This mathematical limit occurs as $h$ approaches zero and may require $h$ to be smaller than the local fiber radius $a(s)$. Nonetheless insuring the coincidence of the Navier-Stokes solution to the Oseen and Stokes solutions as $h \rightarrow 0$ by matching the values of $\bvec{f}$ and $\bvec{c}$ ensures   that the two-dimensional Navier-Stokes solution smoothly transitions to the Oseen 2-D solution in the matching region $a_0 \ll h \ll l$ when $\n{Re}_D$ is finite.  This matching Oseen 2-D solution then approaches the Oseen solution used in the three-dimensional analysis when $h$ is of the order of fiber half-length (see figure \reffig{matching_explained}).  The effectiveness of this matching will be demonstrated in a later section, where we compare the flow field predicted by our inertial SBT in the matching region with the finite difference results obtained using the method developed by \cite{sharma2023finite}. 
\begin{figure}  
\centering
\includegraphics[width=0.8\textwidth]{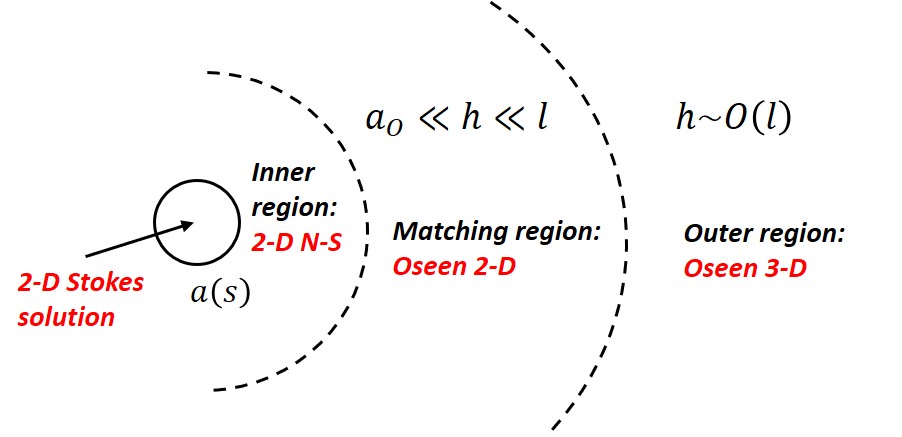}
\caption{ Flow characteristics around a local fiber cross-section at various distances for finite $\n{Re}_D$ } 
\fig{matching_explained} 
\end{figure} 
\\[6pt]
Using the observations noted above, the singular terms in the inner and the outer solutions are matched automatically provided the same $\bvec{f}(s)$ is obtained in both of them.  The matching of the non-singular terms yields an integral equation for $\bvec{f}(s)$. 
\begin{gather}
4\pi (\eta_{\perp}(s)(\delta_{ij}-p_i p_j)+\eta_{\parallel}(s) p_i p_j) (U_j-u_j^I) = f_i\left(\n{ln}(2\kappa) + \n{ln}\left(\frac{(1-s^2)^{1/2}}{\tilde{a}(s)}\right)\right) \nonumber \\
+ \ \frac{1}{2}\int_{-1}^{1}\frac{f_i(s')-f_i(s)}{|s-s'|} \n{d}s' + \frac{1}{2}(\delta_{ik}- 2p_i p_k)f_k \eq{SBT_integral_eqn} 
\end{gather}
The above equation is written in index notation and $\tilde{a}(s)$ is the radius of the fiber cross-section at $s$ scaled with the maximum fiber radius $a_0 = D/2$.  The non-singular or the inertial part of the fluid velocity disturbance produced by the fiber obtained by integrating over the force distribution along the fiber axis is denoted by $\bvec{u}^I$ and is given by 
\begin{gather}
\bvec{u}^I (s) = \int_{-1}^{1} \n{d}s' \bvec{f}(s') \cdot \bvec{G}^I((s-s')\bvec{p}) \eq{u_inertial_definition} 
\end{gather}
where
\begin{gather}
\bvec{G}^I(\bvec{r}) = \bvec{G}(\bvec{r}) - \bvec{G}^{stokes}(\bvec{r}) \eq{G_inertial}
\end{gather}
and
\begin{gather}
\bvec{G}^{Stokes}(\bvec{r}) = \frac{1}{8 \pi r}(\bvec{I} + \frac{\bvec{r} \bvec{r}}{r^2}) \eq{G_stokes}
\end{gather}
is the Stokes flow Green's function. The local matching coefficients $\eta_{\perp}(s)$ and $\eta_{\parallel}(s)$ for flow perpendicular and parallel to the fiber axis capture the effect of using a full Navier-Stokes solution in the inner region and they assure that the local $\bvec{f}(s)$ from the outer and inner solutions coincide. They are functions of the local Reynolds number
\begin{gather}
Re_{D\perp}(s) = \frac{|(\bvec{I}-\bvec{p}\bvec{p})\cdot \bvec{U} |D \ \tilde{a}(s)}{\nu} \eq{ReD_perp} 
\end{gather}
based on the local cross-sectional diameter and the fluid-fiber net relative velocity perpendicular to the fiber at axial position $s$. The expressions for the matching coefficients $\eta_{\perp}(s)$ and $\eta_{\parallel}(s)$ are derived in the next subsection. 

\subsection{Derivation of the expressions for $\eta_{\perp}$ and $\eta_{\parallel}$}

We now derive expressions for the matching coefficients $\eta_{\perp}$ and $\eta_{\parallel}$ appearing in the slender-body-theory integral equation \refeq{SBT_integral_eqn}. These coefficients are introduced to account for inertial effects at the scale of the local fiber diameter and are chosen so that the force per unit length predicted by slender-body theory at a given fiber cross-section coincides with that obtained from an appropriate two-dimensional description of the local flow at finite $\n{Re}_D$ as $\kappa \rightarrow \infty$. 
\\[6pt] 
Our strategy is as follows. We begin by demonstrating that, in the limit of large $\n{Re}_L$, the force per unit length at a local fiber cross-section in response to the local non-singular velocity, as obtained from the SBT integral equation \refeq{SBT_integral_eqn} coincides with the corresponding two-dimensional solution at that cross-section. Establishing this result is important because it shows that, for sufficiently large $\n{Re}_L$, equation \refeq{SBT_integral_eqn} naturally yields a transition to a locally two-dimensional drag law, in which the force depends only on the flow in the immediate vicinity of each cross-section. In particular, when $\n{Re}_L$ is large and $\n{Re}_D$ is small, we will show that the local force components perpendicular and parallel to the fiber axis coincide with the classical two-dimensional results of \cite{lamb1924hydrodynamics} and \cite{tomotika1953forces}, respectively. These results define the small-$\n{Re}_D$ drag coefficients $C_{\perp s}$ and $C_{\parallel s}$. This correspondence provides the foundation for introducing matching coefficients that allow the local force at finite $\n{Re}_D$ to be made consistent with that obtained from a quasi-two-dimensional Navier–Stokes solution. Accordingly, we then introduce $\eta_{\perp}$ and $\eta_{\parallel}$ so that, at finite $\n{Re}_D$, the local force predicted by the inertial slender-body theory matches that obtained from quasi-two-dimensional Navier--Stokes solutions, which define the finite-$\n{Re}_D$ drag coefficients $C_{\perp f}$ and $C_{\parallel f}$. Finally, by eliminating intermediate quantities, we obtain $\eta_{\perp}$ and $\eta_{\parallel}$ explicitly in terms of $C_{\perp s}$, $C_{\perp f}$, $C_{\parallel s}$, and $C_{\parallel f}$ and the fiber aspect ratio $\kappa$. 
\\[6pt] 
We start by considering the regime of small $\n{Re}_D$ and large $\n{Re}_L$. When the fiber aspect ratio $\kappa$ is large, we can expand \refeq{SBT_integral_eqn} for a fiber translating steadily with velocity $U_i$ in a quiescent fluid in the small parameter $\epsilon = 1/\ln(2 \kappa)$, and retain the leading two terms of this expansion. This gives, 
\begin{gather}
f_i(s) + \frac{\varepsilon f_k(s)}{2}(\delta_{ik}-2 p_i p_k) =4\pi \varepsilon \left(\eta_{\perp}(s)(\delta_{ij}- p_i p_j) + \eta_{\parallel}(s)p_i p_j \right) \left(U_j- u_j^I (s) \right) \eq{SBT_approx_eqn} 
\end{gather} 
In equation \refeq{SBT_approx_eqn}, $u_j^I$ is the non-singular or the inertial component of the fluid velocity disturbance, which can be written in a form similar to equation \refeq{u_inertial_definition}. It is to be noted that the presence of $u_j^I (s)$ in equation \refeq{SBT_approx_eqn} indicates matching of the Oseen flow outer solution (see equation \refeq{oseen_outer_vel}) to the Stokes flow inner solution (equation \refeq{stokes_inner}). Although $u_j^I(s)$ could, in principle, be evaluated at any axial position $s$, for the purpose of the present derivation it is evaluated at the midpoint along the fiber axis ($s=0$) without loss of generality, since $\n{Re}_L$ is large and the local force per unit length on the fiber depends only on the flow at the cross-section considered.  
\\[6pt]
We now proceed to evaluate the force per unit length $f_i$ in equation \refeq{SBT_approx_eqn}. Because the inertial fluid velocity is related to a convolution integral of the force per unit length and the Green's function, it is convenient to do this analysis in Fourier space. Defining the Fourier transform for a scalar field in three-dimensional space as:
\begin{gather}
\hat{g}(\bvec{k}) = \int_{\R^3} g(\bvec{r})\n{e}^{- 2 \pi i \bvec{k} \cdot \bvec{r} } \n{d} \bvec{r} \eq{FT_definition} 
\end{gather}
we can write, for $s=0$, 
\begin{gather}
u_j^I (s) = \int_{\R^3} \hat{u}_j^I(\bvec{k}) \n{d} \bvec{k} = \int_{\R^3} \hat{G}_{jm}^I (\bvec{k}) \hat{F}_m(\bvec{k}) \n{d} \bvec{k} \eq{u_inertial_local} 
\end{gather} 
Here, ${\hat{G}}_{jm}^I$ denotes the Fourier transform of $G_{jm}^I$ (see equation \refeq{G_inertial}). The Fourier transforms of the inertial velocity and the body force exerted by the fiber on the fluid are denoted as ${\hat{u}}_j^I\left(\mathbf{k}\right)$ and ${\hat{F}}_m\left(\mathbf{k}\right)$ respectively. Furthermore, ${\hat{F}}_m\left(\mathbf{k}\right)$ can be written as, 
\begin{gather}
\hat{F}_m (\bvec{k}) = \int_{\R^3} \n{d} \bvec{r} \n{e}^{-2 \pi i \bvec{k} \cdot \bvec{r}} \int_{-1}^{1} f_m(s') \delta (\bvec{r}-s'\bvec{p}) \n{d} s' \eq{F_hat_definition} 
\end{gather} 
When $\n{Re}_L \gg 1$, the major contribution to the integral in equation \refeq{F_hat_definition} comes from the region in the wave number space (non-dimensionalized by the reciprocal of fiber half-length, $1/l$) with $1 \ll k \ll \n{Re}_L$.  This implies that the local force per unit length is only dependent on the forces exerted at nearly the same axial position ($s=0$) and is not coupled with $f_i(s)$ at other values of $s$ on the fiber axis. This indicates that we can approximate the integral in equation \refeq{F_hat_definition} as,  
\begin{gather}
\hat{F}_m (\bvec{k}) = f_m (s) \int_{-1}^{1} \n{e}^{-2 \pi i \bvec{k} \cdot s'\bvec{p}} \n{d} s' = 2f_m (s) j_0(2\pi k_p) \eq{F_hat_approximation} 
\end{gather} 
where  $j_0 (2 \pi k_p) = \sin (2 \pi k_p)/(2 \pi k_p)$ is the spherical Bessel function of the first kind. 
Thus, 
\begin{gather}
u_j^I\left(s\right)= f_m(s) J_{jm} 
\end{gather} \eq{u_inertial_approx}
where
\begin{gather}
J_{ij}=2\int{\hat{G}}_{ij}^I\left(\mathbf{k}\right)j_0\left(2\pi k_p\right)d\mathbf{k}=a_1\delta_{ij}+a_2p_ip_j+a_3e_{Ui}e_{Uj}+a_4{(p}_ie_{Uj}+e_{Ui}p_j)  \eq{J_tensor} 
\end{gather} 
where, $e_{Ui}$ denotes the $i$-th component of $\bvec{e}_U$. The constants $a_1$ through $a_4$ appearing in equation \refeq{J_tensor} are dependent on $\n{Re}_L$ and the inclination $\theta$ between the fiber axis and velocity direction. The expressions for these constants are provided in Appendix \ref{appA}.  
\\[6pt] 
Finally, using the above results and the fact that the matching coefficients $\eta_{\perp} = 1$ and $\eta_{\parallel} = 0.5$ when $\n{Re}_D$ is small (\cite{batchelor1970slender,khayat1989inertia}), we can write equation \refeq{SBT_approx_eqn} as
\begin{gather}
f_i+\frac{\varepsilon f_k}{2}\left(\delta_{ik}-2p_ip_k\right)=
4\pi \varepsilon \left(\delta_{ij}-\frac{1}{2} p_i p_j \right) \left(U_j- J_{jm}f_m \right) \eq{SBT_approx_eqn_2} 
\end{gather} 
We now evaluate the components of the force per unit length $f_i$ perpendicular and parallel to the fiber axis at the considered cross-section by using equations \refeq{J_tensor} and \refeq{SBT_approx_eqn_2}. Multiplying both sides of equation \refeq{SBT_approx_eqn_2} with $(\delta_{il}-p_i p_l)$ and $p_i p_l$ separately, we obtain,  
\begin{gather}
\left(\delta_{il}-p_ip_l\right)f_l\left(1+\frac{\varepsilon}{2}+4\pi \varepsilon(a_1+a_3)\right) = 4 \pi \varepsilon (\delta_{il} - p_i p_l) U_l 
\eq{F_perp} \\ 
p_i p_l f_l\left(1-\frac{\varepsilon}{2}+2\pi \varepsilon(a_1+a_2+a_3+a_4)\right) = 2 \pi \varepsilon p_i p_l U_l 
\eq{F_par} 
\end{gather} 
The constants $a_1$ through $a_4$ in equations \refeq{F_perp} and \refeq{F_par} obtained in Appendix \ref{appA} by evaluating the integral $J_{ij}$ in equation \refeq{J_tensor} and given in equations \refeq{LS_1}-\refeq{LS_4} and \refeq{M1_large_ReL}-\refeq{M4_large_ReL}. We find that, for $\n{Re}_L \gg 1$, 
\begin{gather}
a_1 + a_3 = - \frac{1}{4\pi}\gamma - \frac{1}{4\pi} \n{ln} \Big( \frac{\n{Re}_L \n{sin} \theta}{2} \Big) + \frac{1}{4 \pi} \n{ln}2 \eq{a1_plus_a3}  
 \\
a_1 + a_2 + a_3 + a_4 = - \frac{1}{2\pi}\gamma - \frac{1}{2\pi} \n{ln} \Big( \frac{\n{Re}_L \n{sin} \theta}{2} \Big) + \frac{1}{4 \pi} + \frac{1}{2 \pi} \n{ln}2 \eq{sum_all_a} 
\end{gather} 
Substituting equation \refeq{a1_plus_a3} into \refeq{F_perp} and \refeq{sum_all_a} into equation \refeq{F_par}, we obtain the following expressions for perpendicular and parallel components of the local force per unit length $f_{{\perp}s}$ and $f_{{\parallel}s}$, 
\begin{gather}
\frac{f_{{\perp}s}}{U_{\perp}} = \frac{4 \pi \varepsilon}{(1 + \frac{\varepsilon}{2}+ 4 \pi \varepsilon (a_1 + a_3))} = \frac{4 \pi}{\frac{1}{2}-\gamma+\n{ln}(\frac{8}{\n{Re}_{D\perp}})} = C_{\perp s} \eq{C_perp_small_lamb} \\ 
\frac{f_{{\parallel}s}}{U_{\parallel}} = \frac{2 \pi \varepsilon}{(1 - \frac{\varepsilon}{2}+ 2 \pi \varepsilon (a_1 + a_2 + a_3 + a_4))} = \frac{2 \pi}{-\gamma + \n{ln}(\frac{8}{\n{Re}_{D\perp}})} = C_{\parallel s} \eq{C_par_small_tomotika} 
\end{gather} 
In the above equations, $C_{\perp s}$ and $C_{\parallel s}$ denote the local drag coefficients at the considered fiber cross-section, where the symbol $s$ in the subscripts indicates that these quantities are evaluated in the limit of small $\n{Re}_D$. Moreover, as already mentioned, $\n{Re}_{D \perp} = |(\bvec{I - pp}) \cdot \bvec{U}| D /\nu$ denotes the local Reynolds number based on the component of the imposed flow (relative to the fiber motion) perpendicular the fiber axis, evaluated at $s=0$ in the context of this discussion. It can be seen that equations \refeq{C_perp_small_lamb} and \refeq{C_par_small_tomotika} give  local drag coefficients which are the same as those obtained from the solutions of \cite{lamb1924hydrodynamics} and \cite{tomotika1953forces} for oblique Oseen flow past an infinite cylinder in which the fluid velocity disturbance is only a function of the position perpendicular to the cylinder axis. These classical results were obtained using a similar asymptotic matching of the two-dimensional Oseen flow and Stokes flow equations in the outer $(h \sim l_O)$ and inner $(h << l_O)$ regions respectively. The result from \cite{lamb1924hydrodynamics} was originally derived in the context of a cylinder translating perpendicular to its axis, but linear superposition along with the observation that the velocity parallel to the cylinder axis does not convect the momentum disturbance implies that it also yields $C_{\perp s}$ for an oblique imposed velocity. Finally, $\gamma$ in equations \refeq{C_perp_small_lamb} and \refeq{C_par_small_tomotika} is the Euler's constant whose numerical value is 0.5772. The small-$\n{Re}_D$ local drag coefficients $C_{\perp s}$ and $C_{\parallel s}$ will later be used to determine the matching coefficients. 
\\[6pt] 
We have now established that the slender-body theory yields a local force per unit length at a given  fiber cross-section which depends only on the local fiber-fluid relative velocity when $\n{Re}_L$ is large and $\n{Re}_D$ is small. The matching coefficients $\eta_{\perp}$ and $\eta_{\parallel}$ will now be chosen to assure that the force per unit length when $\n{Re}_D$ is \cO(1) and $\n{Re}_L \gg 1$ continues to be equal to the drag due to the local relative velocity. Upon multiplying both sides of equation \refeq{SBT_approx_eqn} with $(\delta_{il}-p_ip_l)$ and $p_ip_l$ separately, we obtain separate equations for the local force per unit length perpendicular and parallel to the fiber axis when $\n{Re}_L$ is large and $\n{Re}_D$ is finite. When $\n{Re}_D$ is finite, the resulting force per unit length from the inertial slender-body theory is 
\begin{gather}
f_{{\perp}f} = \frac{4 \pi \varepsilon \eta_{\perp}} {1 + \varepsilon H_{01} + \frac{\varepsilon}{2} + 4\pi\varepsilon(a_1+a_3)\eta_{\perp} } U_{\perp} \eq{F_perp_finite} \\ 
f_{\parallel f} = \frac{4 \pi \varepsilon \eta_{\parallel}}{1+\varepsilon H_{01}-\frac{\varepsilon}{2} + 2\pi \varepsilon (a_1 + a_2 + a_3 + a_4) \eta_{\parallel}} U_{\parallel} \eq{F_par_finite}
\end{gather} 
where, $H_{01} = \n{ln}(2)-1$. The force per unit length from equations \refeq{F_perp_finite} and \refeq{F_par_finite} should agree with that obtained  from a quasi-two-dimensional Navier-Stokes equations at the considered cross-section. These quasi-2D equations, obtained by neglecting the variation of pressure and velocity parallel to the fiber axis, are given as, 
\begin{gather} 
\nabla \cdot \bvec{u}_{\perp} = 0 \eq{2D_full_NS_1} \\
\n{Re}_{D\perp} \bvec{u}_{\perp} \cdot \nabla_{\perp} \bvec{u}_{\perp} = -\nabla_{\perp} p + \nabla_{\perp}^2\bvec{u}_{\perp} \eq{2D_full_NS_2} \\
\n{Re}_{D\perp} \bvec{u}_{\perp} \cdot \nabla_{\perp} u_{\parallel} = \nabla_{\perp}^2 u_{\parallel} \eq{2D_full_NS_3} 
\end{gather} 
where, $u_{\parallel}$ and $\bvec{u}_{\perp}$ denote the fiber-induced velocity disturbance parallel and perpendicular to the fiber axis. We observe that equations \refeq{2D_full_NS_1}-\refeq{2D_full_NS_2} can be solved for $\bvec{u}_{\perp}$ independent of \refeq{2D_full_NS_3}. Here, we use the solution of equations \refeq{2D_full_NS_1}-\refeq{2D_full_NS_2} for flow past a transversely oriented infinite cylinder by \cite{espinosa2012particle} to find the drag coefficient that relates the force on the cross-section $f_{{\perp}f}$ to the local fiber-fluid relative velocity perpendicular to the fiber axis when $\n{Re}_D$ is \cO(1).  
\begin{gather}
\frac{f_{{\perp}f}}{U_{\perp}} = C_{\perp f} = 4 \pi q \eq{C_perp_finite_espinosa} 
\end{gather}
where,
\begin{gather}
    q= 
\begin{cases}
    \delta - 0.8669\delta^3,& \text{if } \n{Re}_{D \perp} \leq 0.01\\
    0.148 + (2.15\times10^{-2})m + (3.05\times10^{-3})m^2 + (2.13\times10^{-4})m^4,              & 0.01 < \n{Re}_{D \perp} \leq 10 
\end{cases}
\eq{q_espinosa} 
\end{gather} 
In equation \refeq{q_espinosa}, $\delta = \frac{1}{\frac{1}{2} - \gamma - \n{ln}(\frac{\n{Re}_{D \perp}}{8})}$ and, $m = \n{ln}(\frac{\n{Re}_{D \perp}}{0.01})$. The expression for \( q \), as presented in the first line of Equation \refeq{q_espinosa}, was originally derived by \cite{kaplun1957low} and was later extended by \cite{keller1996asymptotics} to larger values of \( \delta \) and, consequently, higher \( \text{Re}_D \) using a hybrid approach. However, here we use the empirical expression proposed by \cite{espinosa2012particle}, based on a numerical solution of the two-dimensional Navier-Stokes equation, in the second line of equation \refeq{q_espinosa}. 
\begin{figure}
\centering
\begin{subfigure}{0.48\textwidth}
  \includegraphics[width=\textwidth]{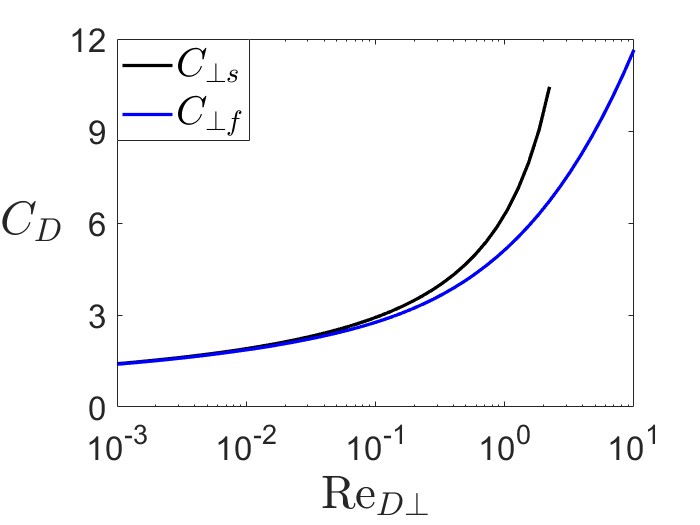}
  \caption{}
  \label{fig:sub1}
\end{subfigure}
\begin{subfigure}{0.48\textwidth}
  \includegraphics[width=\textwidth]{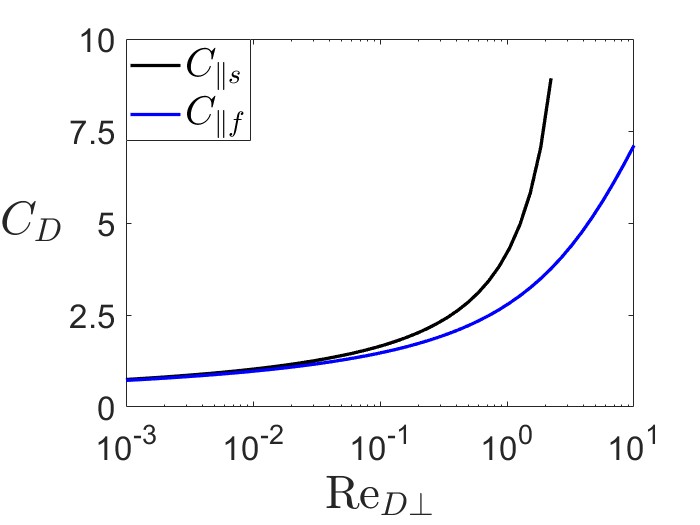}
  \caption{}
  \label{fig:sub2}
\end{subfigure}
\caption{(a) Variation of $C_{\perp s}$ and $C_{\perp f}$ with $\n{Re}_{D \perp}$. $C_{\perp s}$ exhibits a vertical asymptote around $\n{Re}_D$ = 7.4. (b) Variation of $C_{\parallel s}$ and $C_{\parallel f}$ with $\n{Re}_{D \perp}$. $C_{\parallel s}$ exhibits a vertical asymptote around $\n{Re}_D$ = 4.49. } 
\fig{drag_coeffs_vs_ReD} 
\vspace{-7pt} 
\end{figure} 
\\[6pt] 
Equation \refeq{2D_full_NS_3} indicates that the parallel fluid momentum disturbance is convected like a passive scalar by the velocity field perpendicular to the fiber axis. Therefore, equations \refeq{2D_full_NS_1}-\refeq{2D_full_NS_3} are analogous to the equations resulting from a two-dimensional problem of heat transfer from a cylinder at finite Reynolds number in a fluid with a Prandtl number of 1, with $u_\parallel$ being analogous to a temperature field. The study of forced heat convection from a circular cylinder at low Reynolds numbers has been extensively explored alongside its counterpart in momentum transport (\cite{cole1954heat, hieber1968low}). Notably, the Nusselt number—a dimensionless measure of heat transfer—exhibits a mathematical analogy with the longitudinal drag coefficient, $C_{\parallel f}$. Solving the analogous problem using a finite element method in COMSOL gives us the drag coefficient $C_{\parallel f}$ that describes the relationship between local fiber forcing and local fiber-fluid relative velocity, parallel to the fiber axis. 
\begin{gather}
\frac{f_{\parallel f}}{U_{\parallel}} = C_{\parallel f} = \frac{2 \pi (d + b + \gamma)}{(d^2 + bd + c)} \eq{C_par_finite_anubhab} 
\end{gather}
where, $d = \n{ln}(\frac{8}{\n{Re}_{D\perp}} + a)$ and $a = 0.8042$, $b = 2.4248$, $c= 1.7022$ are numerical constants obtained by least squares fitting of the simulation data over the range $\n{Re}_{D\perp} \in [0,10]$. Building on our discussion of the analogy between forced convection and longitudinal momentum transport, the chosen fit form is inspired by the known asymptotic expressions for the Nusselt number, as derived by \cite{cole1954heat} and \cite{hieber1968low}. 
\\[6pt] 
Comparing equation \refeq{F_perp_finite} with equation \refeq{C_perp_finite_espinosa}, and equation \refeq{F_par_finite} with equation \refeq{C_par_finite_anubhab}, the drag coefficients from the slender-body theory are related to the matching coefficients as 
\begin{gather} 
C_{\perp f} = \frac{4 \pi \varepsilon \eta_{\perp}}{1 + \frac{\varepsilon}{2}+4 \pi \varepsilon(a_1+a_3)\eta_{\perp}} \eq{C_perp_finite_SBT} \\ 
C_{\parallel f} = \frac{4 \pi \varepsilon \eta_{\parallel}}{1-\frac{\varepsilon}{2}+ 2\pi \varepsilon (a_1 + a_2 + a_3 + a_4) \eta_{\parallel}} \eq{C_par_finite_SBT} 
\end{gather} 
With expressions now available for the finite-$\n{Re}_D$ drag coefficients $C_{\perp f}$ and $C_{\parallel f}$, and earlier results in this section providing the small-$\n{Re}_D$ coefficients $C_{\perp s}$ and $C_{\parallel s}$, we can eliminate the intermediate constants $a_1$ through $a_4$ and obtain explicit formulas for the matching coefficients in terms of these drag coefficients and the fiber aspect ratio $\kappa$. The dependence on the fiber aspect ratio can be absorbed into the drag coefficients for forces that a translating fiber experiences perpendicular and parallel to its axis in Stokes flow (\cite{batchelor1970slender}), 
\begin{figure}
\centering
\begin{subfigure}{0.48\textwidth}
  \includegraphics[width=\textwidth]{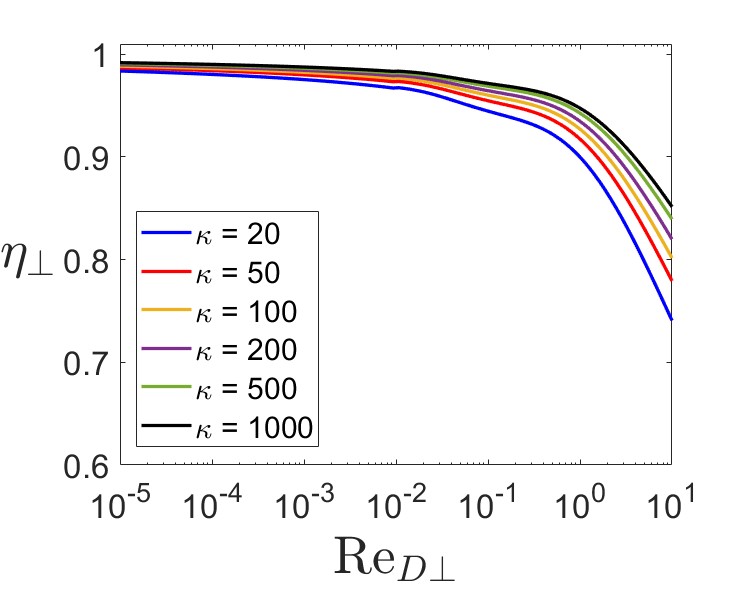}
  \caption{}
  \label{fig:sub1}
\end{subfigure}
\begin{subfigure}{0.48\textwidth}
  \includegraphics[width=\textwidth]{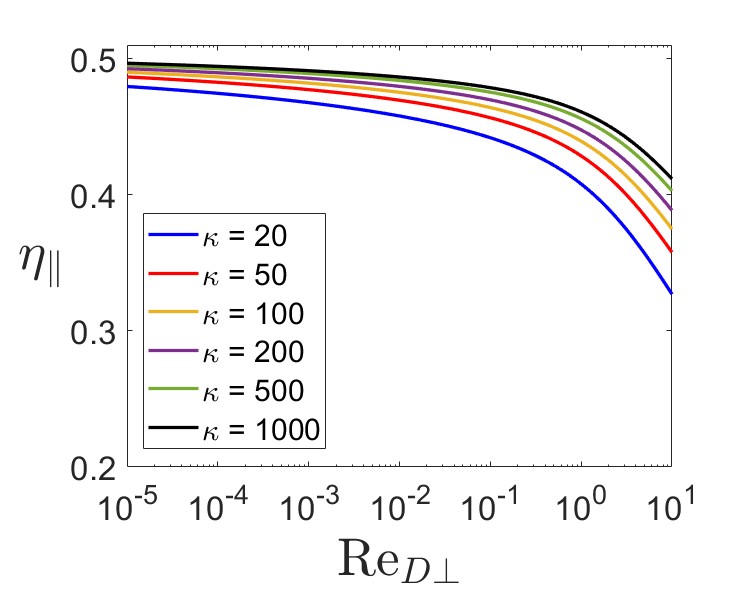}
  \caption{}
  \label{fig:sub2}
\end{subfigure}
\caption{Variation of the matching coefficients $\eta_{\perp}$ and $\eta_{\parallel}$ with $\n{Re}_{D \perp}$. } 
\fig{matching_coeffs_vs_ReD} 
\end{figure} 
\begin{gather} 
C_{\perp z} = \frac{f_{\perp z}}{U_{\perp}} = \frac{4 \pi \varepsilon}{(1 + \frac{\varepsilon}{2})} \eq{C_perp_batchelor} \\ 
C_{\parallel z} = \frac{f_{\parallel z}}{U_{\parallel}} = \frac{2 \pi \varepsilon}{(1 - \frac{\varepsilon}{2}) } \eq{C_par_batchelor}
\end{gather} 
Eliminating $(a_1+a_3)$ from equations \refeq{C_perp_small_lamb}, \refeq{C_perp_finite_SBT} and \refeq{C_perp_batchelor} leads us to the expression for the matching coefficient $\eta_\perp$. Similarly, eliminating $(a_1+a_2 +a_3+a_4)$ from equations \refeq{C_par_small_tomotika}-\refeq{C_par_finite_SBT} and \refeq{C_par_batchelor} leads to the expression for $\eta_{\parallel}$. They are given as, 
\begin{gather} 
\eta_{\perp} = \frac{1}{(1-\frac{C_{\perp z}}{C_{\perp s}}+\frac{C_{\perp z}}{C_{\perp f}})} \eq{eta_perp_eqn} \\ 
\eta_{\parallel} = \frac{0.5}{(1-\frac{C_{\parallel z}}{C_{\parallel s}}+\frac{C_{\parallel z}}{C_{\parallel f}})} \eq{eta_par_eqn} 
\end{gather} 
It is to be noted that the drag coefficients $C_{\perp s}$, $C_{\perp f}$, $C_{\parallel s}$ and $C_{\parallel f}$ in the expressions for the matching coefficients $\eta_{\perp}$ and $\eta_{\parallel}$ given in equations \refeq{eta_perp_eqn} and \refeq{eta_par_eqn} are functions of the local Reynolds number $\n{Re}_{D\perp} (s) =|(\bvec{I}-\bvec{p}\bvec{p})\cdot \bvec{U}|\tilde{a}(s)D/\nu$ based on the radius and the imposed flow relative to fiber motion perpendicular to the fiber axis at axial position $s$. Figure \reffig{matching_coeffs_vs_ReD} shows the variation of $\eta_{\perp}$ and $\eta_{\parallel}$ with $\n{Re}_{D\perp}$ for several fiber aspect ratios between 20 and 1000. It can be seen that $\eta_{\perp}$ and $\eta_{\parallel}$ approach their Stokes flow values of 1 and 0.5 respectively as $\n{Re}_{D\perp}$ approaches 0. 

\subsection{Finite difference method for solving the inertial SBT equation for the force per unit length} 
The integral equation for the local force per unit length that the fiber exerts on the fluid (equation \refeq{SBT_integral_eqn}) is solved numerically using a finite difference method. Using the definition \refeq{u_inertial_definition} for the inertial component of the velocity disturbance induced by the fiber, we can write \refeq{SBT_integral_eqn} in the following form, 
\begin{gather}
4\pi (\eta_{\perp}(s)(\delta_{ij}-p_i p_j)+\eta_{\parallel}(s) p_i p_j) U_j 
= f_i\left(\n{ln}(2\kappa) + \n{ln}\left(\frac{(1-s^2)^{1/2}}{\tilde{a}(s)}\right)\right)  \nonumber \\  
+ \ \frac{1}{2}\int_{-1}^{1}\frac{f_i(s')-f_i(s)}{|s-s'|} \n{d}s'
+ \frac{1}{2}(\delta_{ik}- 2p_i p_k)f_k \nonumber \\ 
+ 4\pi (\eta_{\perp}(s)(\delta_{ij}-p_i p_j)+\eta_{\parallel}(s) p_i p_j) \int_{-1}^{1} G_{jk}^{I}f_k(s') \n{d}s' \eq{SBT_matrix_eqn} 
\end{gather} 
The integrals on the right-hand side of the above equation are approximated using a middle Riemannian sum (mid-point rule), i.e.,
\begin{gather}
\int_{-1}^{1}{\frac{f_i\left(s^\prime\right)-f_i\left(s\right)}{\left|s-s^\prime\right|}ds^\prime=\sum_{n\neq m}\frac{f_i\left(x_n\right)-f_i\left(x_m\right)}{\left|x_n-x_m\right|}\Delta x_n} \\
\int_{-1}^{1}{G_{jk}(s-s^\prime)f_k\left(s^\prime\right)ds^\prime}=\sum_{n\neq m}{G_{jk}\left(x_m-x_n\right)f_k\left(x_n\right)\Delta x_n} 
\end{gather} 
We discretize the fiber axis uniformly with $N$ grid points. The finite difference approximation of the integrals on the right-hand side transforms equation \refeq{SBT_matrix_eqn} into a linear system of the form $\bvec{AX} = \bvec{B}$, where $\bvec{A}$ is a square matrix of size $(3N \times 3N)$, and $\bvec{X}$ and $\bvec{B}$ are column vectors of size $3N$. These dimensions reflect that we are solving for the three components of the local force density at each of the $N$ chosen grid points on the fiber. The resulting linear system is solved using an LU decomposition method. The choice of the number of grid points $N$ is determined by the value of $\n{Re}_L$, ensuring sufficient resolution to capture the rapid force variations within the Oseen length $l_O$. 

\section{Results and validation of the Oseen-flow inertial slender-body theory}
\sect{oseen_flow_SBT_results}

Having established the integral equation for our inertial slender-body theory (SBT) and a technique for its numerical solution, we now present results for the variation of force per unit length along the fiber, as well as the integrated quantities such as the net force and torque. We begin by briefly examining the variation of the local force distribution along the fiber axis and its dependence on $\n{Re}_D$, while also comparing these results for spheroidal and cylindrical fibers. Subsequently, we compare the transverse drag on a cylindrical fiber at small and $\cO(1)$ values of $\n{Re}_D$ to experimental data and the quasi-2D full Navier-Stokes solution. Finally, we compute the total force and torque on an obliquely translating spheroidal fiber and compare them with numerical simulation results obtained using the finite difference code of \cite{sharma2023finite} to assess the accuracy of our inertial SBT formulation. 

\begin{bottomfigure}
\centering
\includegraphics[width=0.35\textwidth]{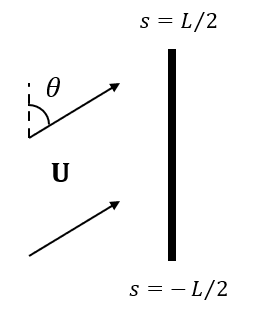}
\caption{ Schematic of uniform flow past a cylindrical fiber  }
\fig{flow_schematic} 
\end{bottomfigure}

\subsection{Variation of the fiber force distribution along the axis of the slender body} 

Figure \reffig{force_distribution} shows the variation of the normalized force per unit length along the fiber axis for three different values of $\n{Re}_D$. We consider the case of a uniform flow past a fiber with $\kappa = 50$ held at rest. 
\begin{figure}  
\centering
\begin{subfigure}{0.31\textwidth}
  \includegraphics[width=\textwidth]{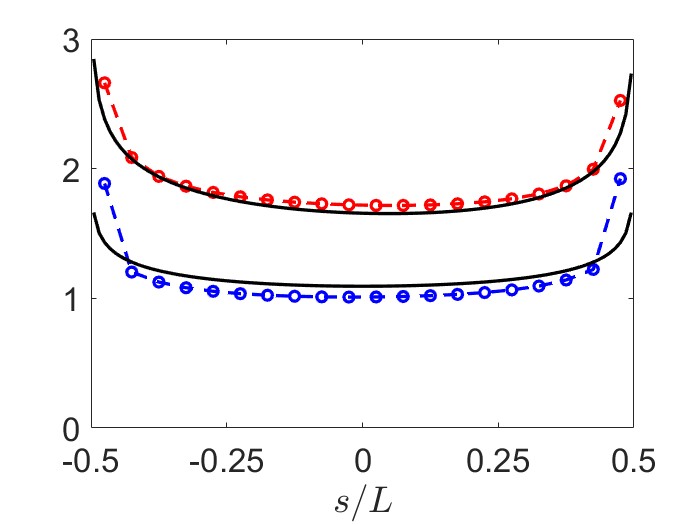}
  \caption{}
  \label{fig:sub1}
\end{subfigure}
\begin{subfigure}{0.31\textwidth}
  \includegraphics[width=\textwidth]{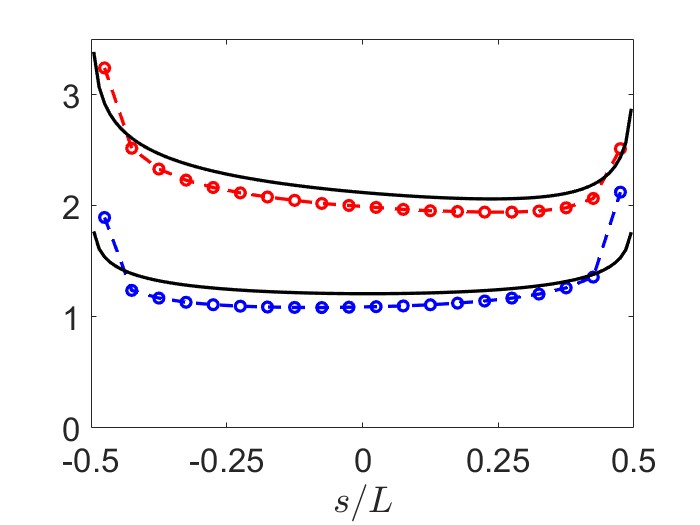}
  \caption{}
  \label{fig:sub2}
\end{subfigure}
\begin{subfigure}{0.31\textwidth}
  \includegraphics[width=\textwidth]{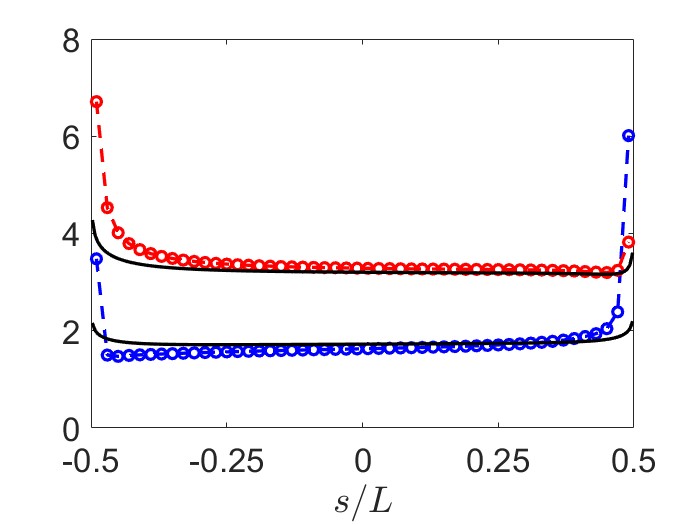}
  \caption{}
  \label{fig:sub3}
\end{subfigure}
\begin{subfigure}{0.31\textwidth}
  \includegraphics[width=\textwidth]{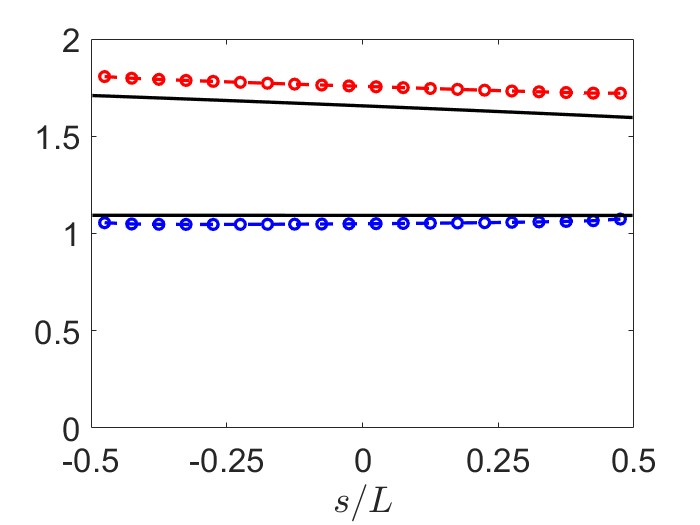}
  \caption{}
  \label{fig:sub4}
\end{subfigure}
\begin{subfigure}{0.31\textwidth}
  \includegraphics[width=\textwidth]{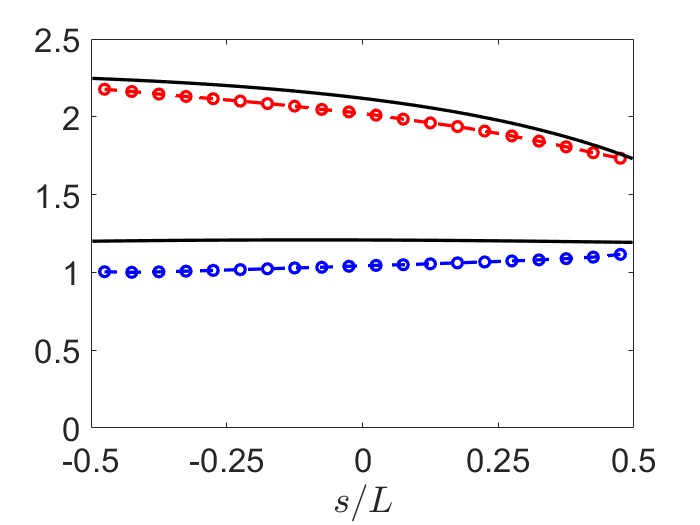}
  \caption{}
  \label{fig:sub5}
\end{subfigure}
\begin{subfigure}{0.31\textwidth}
  \includegraphics[width=\textwidth]{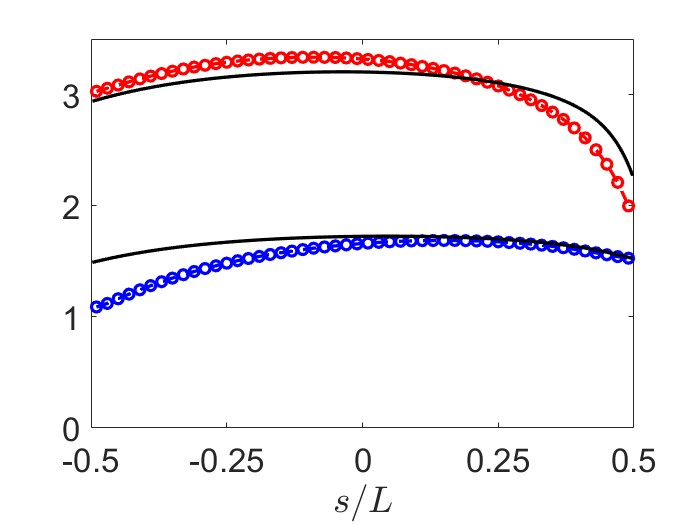} 
  \caption{}
  \label{fig:sub6}
\end{subfigure}
\caption{Variation of the normalized force per unit length ($f(s)/\mu U$), along directions parallel (using blue curves) and perpendicular (using red curves) to the fiber axis, for a stationary $\kappa = 50$ fiber in a uniform flow inclined at $\theta = 45^o$ to the fiber axis (see figure \reffig{flow_schematic}). Also shown for comparison are the force per unit length predictions of \cite{khayat1989inertia} using black curves. Plots (a), (b) and (c) are for a cylindrical fiber with $\n{Re}_D =  0.01, 0.1 \ \n{and} \ 1 $ respectively. The corresponding plots (d), (e) and (f) are for a spheroidal fiber. }
\fig{force_distribution}
\end{figure}
The flow is inclined at an angle of $\theta = 45^o$ to the fiber axis (see figure \reffig{flow_schematic}). Separate sets of plots are presented for cylindrical and spheroidal fibers. A key observation from these plots is the divergence of the force distribution at the endpoints ($s=-L/2$ and $s=L/2$) for cylindrical fibers, but not for spheroidal fibers. This divergence stems from the abrupt termination of cylindrical fibers, causing a sudden discontinuity in the cross-sectional radius and leading to singularities in the force distribution. In contrast, spheroidal fibers have smoothly tapered ends, which allow for a more gradual transition in geometry, mitigating these singularities and resulting in a finite force distribution at the endpoints. This logarithmic divergence at the endpoints for cylindrical fibers is also noted in the weakly inertial theory of \cite{khayat1989inertia} (see their equation 6.2) where it appears in the second term of the expansion in $\n{Re}_L/\n{ln}(\kappa)$ for the force distribution along the fiber axis. However, the divergence obtained from our inertial SBT is stronger than the prediction from the weakly inertial theory at $\n{Re}_D=1$, which can be attributed to the higher-order effects in $\n{Re}_L$ than the weakly inertial theory captures. Another observation is the growing asymmetry in the distribution of the force per unit length perpendicular to the fiber axis as $\n{Re}_D$ increases. Specifically, the force per unit length on the leading half of the fiber exceeds that on the trailing half. This results in an inertial torque on the fiber which tends to rotate the fiber broadside to the direction of uniform flow. This will be discussed in more detail in the next subsection. 

\subsection{Net force and torque on the fiber from inertial SBT } 

Having obtained the fiber force distribution by numerically solving equation \refeq{SBT_matrix_eqn}, we compute the net force and torque on the fiber resulting from this distribution.  
\begin{gather}
\bvec{F}_{\mathrm{oseen}} = \int_{-l}^{l} \bvec{f}(s) \mathrm{d}s = \sum_{n} \bvec{f}(x_n) \Delta x_n \eq{F_oseen} \\
\bvec{T}_{\mathrm{oseen}} = \int_{-l}^{l} (s\bvec{p} \times \bvec{f}(s)) \mathrm{d} s = \sum_{n} (x_n \bvec{p} \times \bvec{f}(x_n)) \Delta x_n \eq{T_oseen} 
\end{gather}
We examine the variation of orientation-dependent drag and lift forces on a steadily translating fiber with the aspect ratio ($\kappa$) and $\n{Re}_D$, and compare them with experiments and numerical simulations.

\subsubsection{Transverse drag on a settling cylindrical fiber}

The variation of the drag force, normalized by \(\mu U l\), on a cylindrical fiber settling perpendicular to its axis in a quiescent fluid is shown in Figure \reffig{transverse_drag_comparison} for \(\text{Re}_D\) up to 10. Separate curves are plotted for fiber aspect ratios of 20 and 100. The drag force predicted by the inertial slender-body theory (SBT) equation \refeq{SBT_integral_eqn} aligns well with the weakly inertial theory of \cite{khayat1989inertia} for \(\text{Re}_D < 0.2\). However, at higher \(\text{Re}_D\), the fully inertial theory predicts a larger force per unit length that becomes independent of fiber length. This behavior is consistent with the experimental findings of \cite{jayaweera1965behaviour}, who measured drag coefficients of falling slender cylinders with \(\kappa > 100\). In contrast, the weakly inertial theory of Khayat and Cox underestimates the transverse drag for \(\text{Re}_D > 0.2\).  Additionally, the fully inertial SBT agrees well with the two-dimensional Navier-Stokes simulations of \cite{espinosa2012particle}. Notably, for high \(\text{Re}_D\), the drag force per unit length is nearly identical for fibers with aspect ratios of 20 and 100. This is because, at large \(\text{Re}_L\), the dominant contribution to drag arises from interactions within the Oseen length (\(l_O \ll l\)), making the force per unit length nearly independent of fiber length.  Two key aspects of the new inertial slender-body theory contribute to these results. First, the theory explicitly accounts for inertial effects at the scale of the fiber diameter by incorporating a two-dimensional solution to the Navier-Stokes equations and the resulting matching coefficients \(\eta_\parallel\) and \(\eta_\perp\). Second, unlike the approach of Khayat and Cox, which relied on an expansion valid only for small \(\text{Re}_L / \ln(\kappa)\), the present method numerically solves the integral equation, making it applicable across all \(\text{Re}_L\) values. 

\begin{figure}
\centering
\includegraphics[width=0.75\textwidth]{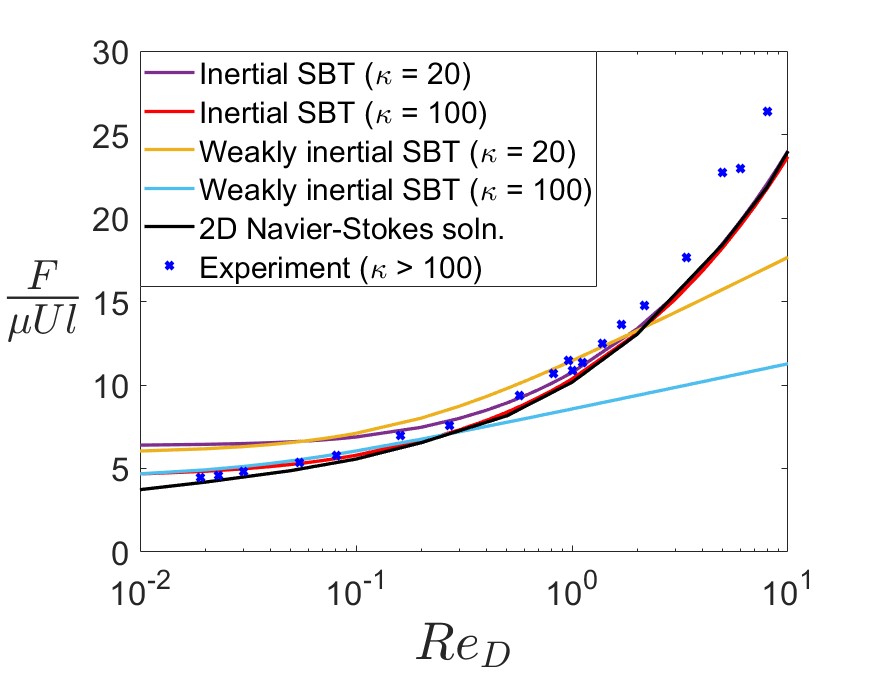}
\caption{ Comparison of transverse drag on a translating fiber obtained from the inertial SBT with the weakly inertial theory of \cite{khayat1989inertia}, 2-D numerical solution of \cite{espinosa2012particle}, and experimental results of \cite{jayaweera1965behaviour}}   
\fig{transverse_drag_comparison} 
\end{figure} 

\subsubsection{Force and torque on a spheroidal fiber translating oblique to its axis}

In this section, we present results for the force and torque on a spheroidal fiber translating in a quiescent fluid for various angles between the fiber axis and its velocity direction. We also compare them to the complementary numerical simulation results using the finite difference code of \cite{sharma2023finite} written in prolate spheroidal coordinates for a translating spheroidal fiber. 
\\[6pt]   
We consider a spheroidal fiber with aspect ratio $\kappa=50$ oriented along the 1-direction and translating with a velocity $\bvec{U} = (U_1,U_2,0)$ in the 1-2 plane in a quiescent fluid. In this case, the fiber experiences a drag force in the 1-2 plane opposite to its velocity and a lift force in the 1-2 plane perpendicular to its velocity direction. It also experiences an inertial torque in the 3-direction. Figure \reffig{drag_lift_asp50} shows the variation of the drag($D_r$) and lift ($L_f$) (both normalized by $\mu UL$) with Reynolds number for various inclination angles ($\theta$) between the fiber axis and its velocity direction. 
\begin{bottomfigure}
\centering
\begin{subfigure}{0.48\textwidth}
  \includegraphics[width=\textwidth]{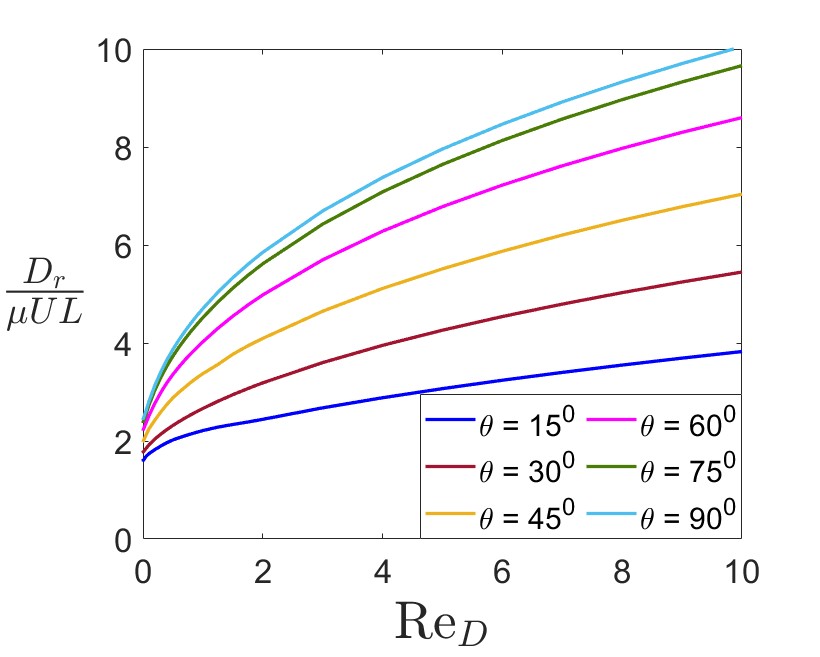}
  \caption{}
  \label{fig:sub1}
\end{subfigure}
\begin{subfigure}{0.48\textwidth}
  \includegraphics[width=\textwidth]{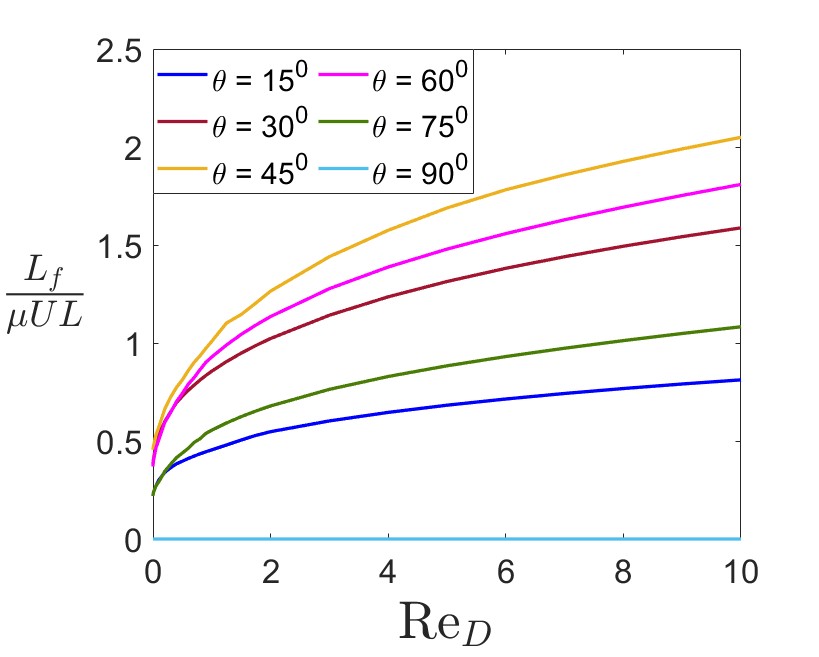}
  \caption{}
  \label{fig:sub2}
\end{subfigure}
\caption{Variation of the (a) normalised drag and (b) normalised lift on a translating fiber ($\kappa = 50$) with Reynolds number and angle of inclination between fiber axis and velocity direction} 
\fig{drag_lift_asp50} 
\end{bottomfigure}
For a spheroidal fiber, we define $\n{Re}_D$ based on the largest cross-sectional diameter (at the midpoint along the fiber axis). We find that the drag and lift forces on the fiber increase monotonically with $\n{Re}_D$. However, the increase becomes more gradual as the Reynolds number is increased. This can be attributed to the fact that at large $\n{Re}_D$ values, the inertial screening length, $l_O = \nu/|\mathbf{U}|$ becomes much smaller than the length of the fiber ($L$). This means that the outer Oseen solution then becomes locally two-dimensional at each fiber cross-section for which the drag and lift forces are logarithmic functions of Reynolds number $\n{Re}_D$. Hence, the nature of the curves for drag and lift changes from linear to logarithmic as $\n{Re}_D$ is increased. Moreover, we also note that at a particular $\n{Re}_D$ value, the drag force is maximum when the fiber is oriented perpendicular to its velocity direction, whereas the lift force is maximum at an inclination angle of $45^o$.  
\begin{figure}
\centering
\begin{subfigure}{0.53\textwidth}
  \includegraphics[width=\textwidth]{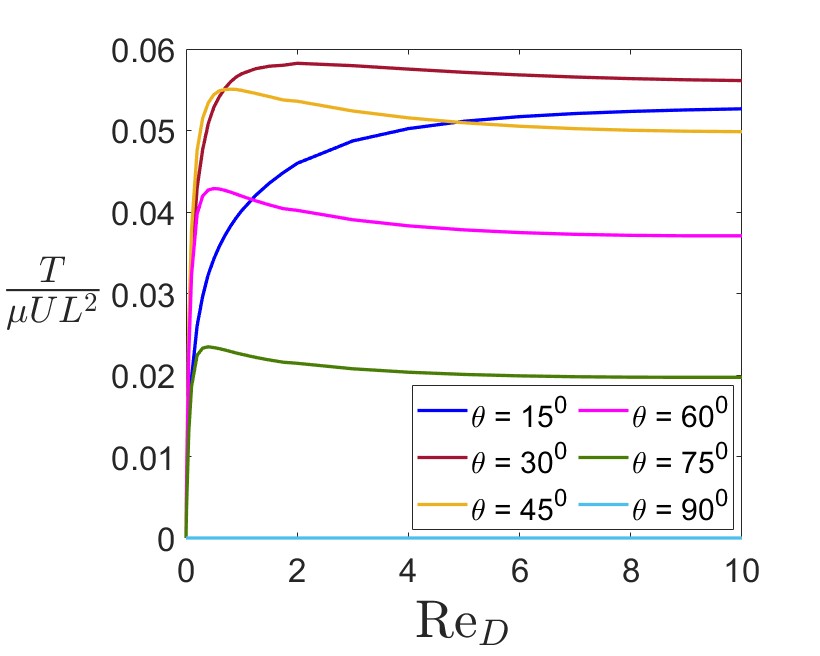}
  \caption{}
  \fig{torque_asp50_sub1}
\end{subfigure}
\begin{subfigure}{0.4\textwidth}
  \includegraphics[width=\textwidth]{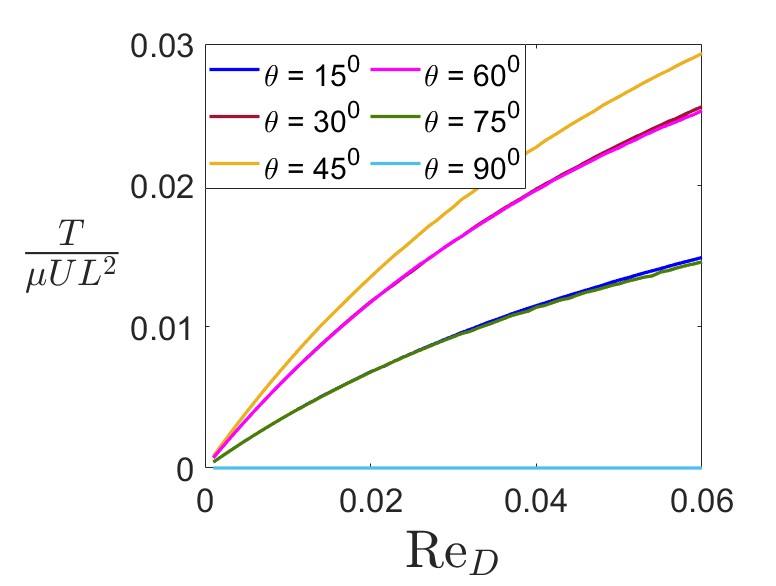}
  \caption{}
  \fig{torque_asp50_sub2}
\end{subfigure} 
\caption{ Variation of the torque (from inertial SBT with Oseen flow outer solution) on a translating fiber ($\kappa = 50$) with Reynolds number and angle of inclination between fiber axis and velocity direction. (a) shows the full range of $\n{Re}_D$ and (b) shows the initial behavior at small $\n{Re}_D$ values. } 
\fig{torque_asp50} 
\end{figure} 
\\[6pt] 
Figure \reffig{torque_asp50_sub1} shows the variation of the torque ($T$) (normalized by $\mu UL^2$) on the fiber with $\n{Re}_D$. The torque, which must be zero at $\n{Re}_D=0$ based on the linearity of Stokes flow, increases at first with increasing $\n{Re}_D$ but then very gradually starts to decrease at moderate $\n{Re}_D$. The initial increase at small $\n{Re}_D$ is shown in figure \reffig{torque_asp50_sub2}, which illustrates that the inertial torque is symmetric for angles equidistant from $45^o$. The subsequent decrease at finite $\n{Re}_D$ values is again due to the aforementioned transition of the Oseen disturbance field in the outer region into a local two-dimensional field at each cross-section. We also note that at sufficiently low $\n{Re}_D$ values, the torque is maximum at an inclination angle of $45^o$. However, as the Reynolds number increases, as observed in figure \reffig{torque_asp50_sub1}, the torque at an inclination angle of $30^o$ becomes greater than that at an inclination angle of $45^o$. Moreover, for an inclination angle of $15^o$, the torque goes on increasing at least until an $\n{Re}_D$ value of 5. This is because, with decreasing inclination angle between the fiber axis and its velocity direction, an increasing fraction of the fiber stays in its own wake, leading to a significant torque even at larger $\n{Re}_L$ values. This means the inclination angle at which the torque becomes maximum should decrease from $45^o$ and asymptotically approaches $0^o$ (see equation (8.17) of \cite{khayat1989inertia}) as the Reynolds number is increased from 0 to infinity. 
\\[6pt] 
It is important to note that the non-monotonicity in the inertial torque when $\n{Re}_D$ is \cO(1), as noted in the previous paragraph, is observed when the flow field in the outer region is modeled as an Oseen disturbance. However, as we shall see later, the inertial torque instead grows monotonically with increasing $\n{Re}_D$ when the flow around the fiber is significantly influenced by the finite fiber thickness when $\n{Re}_D$ is $O(1)$. 
\begin{figure}
\centering
\begin{subfigure}{0.31\textwidth}
  \includegraphics[width=\textwidth]{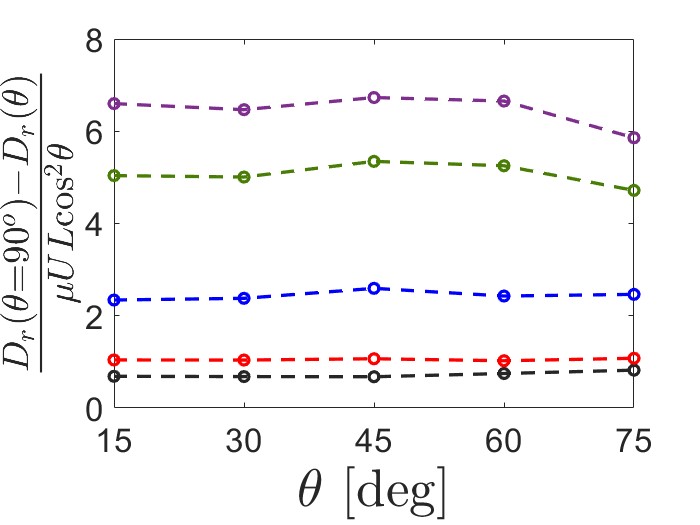}
  \caption{}
  \label{fig:sub1}
\end{subfigure}
\begin{subfigure}{0.31\textwidth}
  \includegraphics[width=\textwidth]{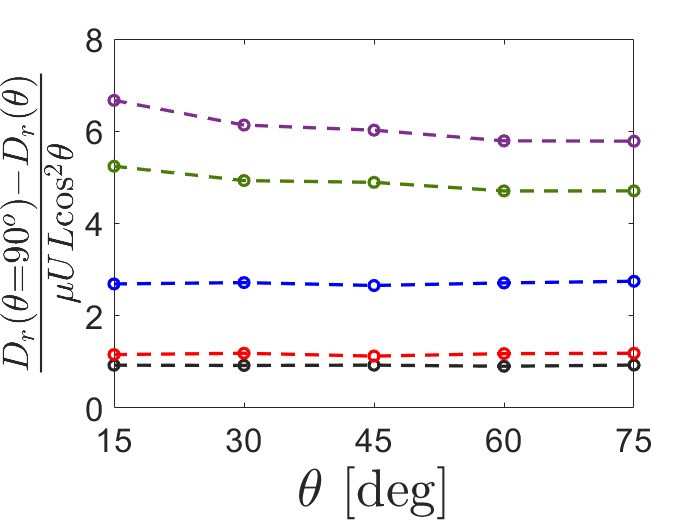}
  \caption{}
  \label{fig:sub2}
\end{subfigure}
\begin{subfigure}{0.31\textwidth}
  \includegraphics[width=\textwidth]{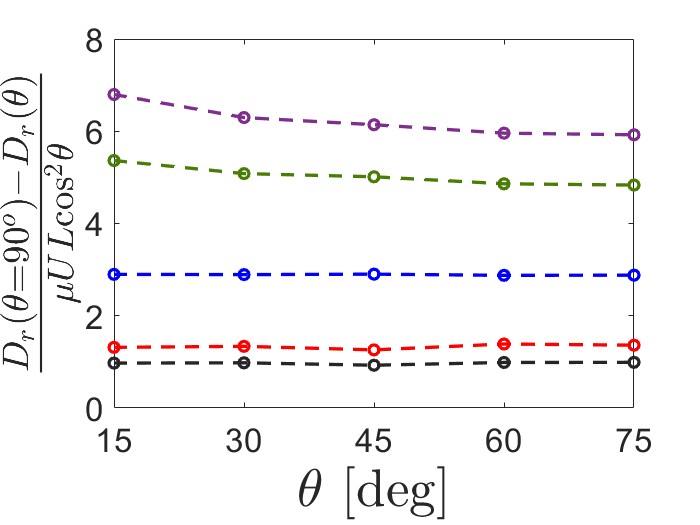}
  \caption{}
  \label{fig:sub3}
\end{subfigure}
\begin{subfigure}{0.31\textwidth}
  \includegraphics[width=\textwidth]{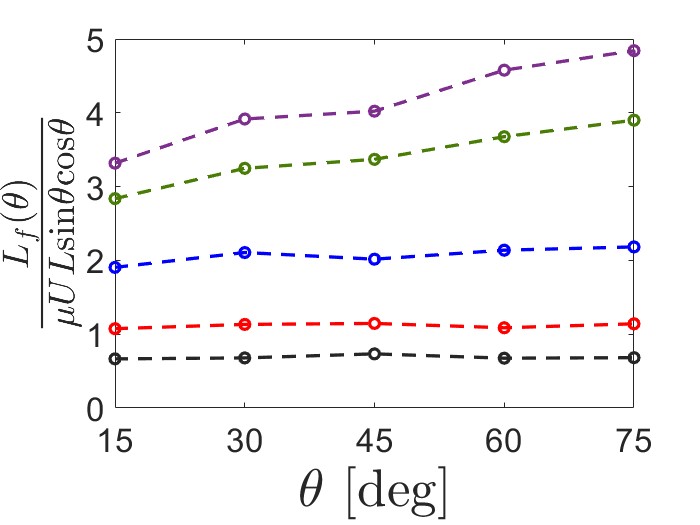}
  \caption{}
  \label{fig:sub4}
\end{subfigure}
\begin{subfigure}{0.31\textwidth}
  \includegraphics[width=\textwidth]{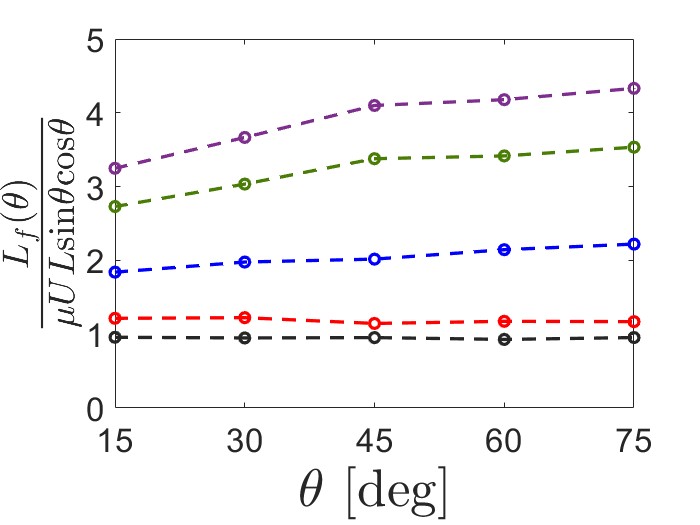}
  \caption{}
  \label{fig:sub5}
\end{subfigure}
\begin{subfigure}{0.31\textwidth}
  \includegraphics[width=\textwidth]{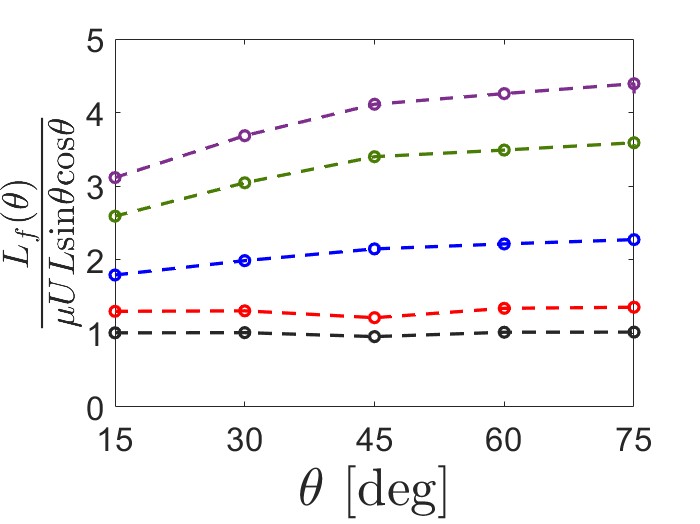} 
  \caption{}
  \label{fig:sub6}
\end{subfigure}
\caption{Variation of the normalised drag ((a), (b) and (c)) and lift ((d),(e) and (f)) forces obtained from the inertial SBT with the inclination angle between the fiber orientation and its velocity direction. Spheroidal fibers with aspect ratios 20 ((a) and (d)), 50 ((b) and (e)), and 100 ((c) and (f)) are considered. Curves colored black, red, blue, green and purple correspond to $\n{Re}_D$ values of 0.01,0.1,1,5 and 10 respectively. }
\fig{drag_lift_vs_theta} 
\end{figure}
\\[6pt]  
The variation of drag ($D_r$) and lift ($L_f$) forces on an obliquely translating slender fiber with the inclination angle ($\theta$) has been extensively investigated in previous studies on slender-body theory. In Stokes flow, the force acting on an axisymmetric translating slender fiber can be expressed by a simple linear superposition in terms of those caused by the longitudinal and transverse motions. These considerations imply that the drag will take the form $A+B\n{sin}^2 \theta$ and the lift will be $B \,\n{sin}\theta\, \n{cos}\theta$, where the constants $A$ and $B$ depend on the particle aspect ratio. While linear superposition does not hold at finite Reynolds numbers, it has been proposed (\cite{roy2023orientation,lopez2017inertial}) that a similar angular dependence of the drag and lift with $\n{Re}_L$-dependent $A$ and $B$ provides a good approximation to experimental measurements with inertial fibers when $\n{Re}_D$ is small. This means that the constants $A$ and $B$ should be independent of the inclination ($\theta$) between fiber orientation and velocity when $\n{Re}_L$ is finite and $\n{Re}_D$ is small. To validate this hypothesis at small $\n{Re}_D$ and explore its robustness with increasing $\n{Re}_D$, we plot the normalised drag and lift forces on the fiber as a function of the inclination angle $\theta$ in figure \reffig{drag_lift_vs_theta}. Specifically, we look at the variation of $(D_r(\theta = 90^o) - D_r (\theta))/(\mu U L \n{cos}^2 \theta)$ and $L_f(\theta)/(\mu U L \n{sin}\theta \n{cos}\theta)$ against $\theta$, in order to determine if $B$ is indeed independent of $\theta$ at small $\n{Re}_D$, and observe if this behavior persists at finite $\n{Re}_D$. Fiber aspect ratios of 20, 50 and 100 are considered. We observe that up to an $\n{Re}_D$ value of 1, $B$ is almost independent of $\theta$, signifying the validity of the assumption made in \cite{roy2023orientation}, \cite{lopez2017inertial} for small $\n{Re}_D$. However, at $\n{Re}_D$ values of 5 and 10, $B$ varies considerably with changing $\theta$. 

\subsubsection{Comparison with the force and torque obtained from a full Navier-Stokes solution}

We now compare the force and torque on a steadily translating spheroidal fiber with complementary numerical simulation results using the finite difference Navier-Stokes solver of \cite{sharma2023finite}. This solver uses a prolate spheroidal coordinate system with the shape of the fiber exactly resolved as one of the coordinate surfaces representing the inner boundary of the computational domain. Since the prolate spheroidal grid is naturally clustered near the fiber surface, this helps to resolve the flow close to the particle. The grid then smoothly transitions to an almost spherical outer boundary where the imposed flow is applied. The computational framework has been thoroughly detailed in \cite{sharma2023finite}. 
\\[6pt] 
For the simulations presented in this work, we have modified the in-house numerical code from its original implementation in \cite{sharma2023finite}. Instead of employing the Schur complement method to iteratively solve the coupled system of momentum and incompressibility equations as described in the original paper, we use a splitting method. 
\begin{bottomfigure}
\centering
\begin{subfigure}{0.31\textwidth}
  \includegraphics[width=\textwidth]{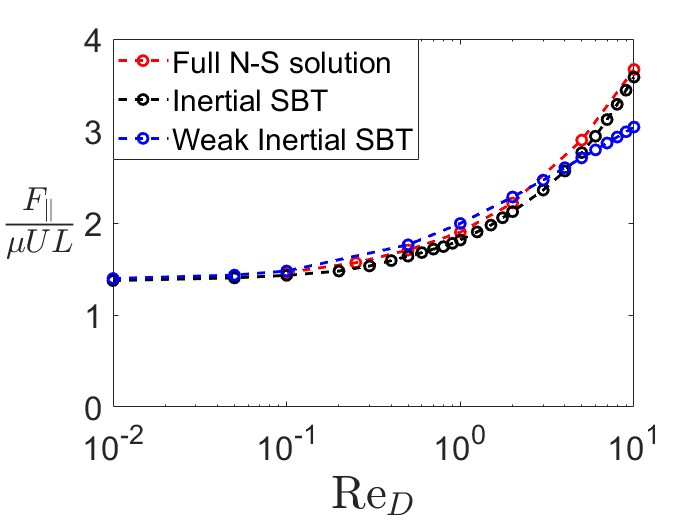}
  \caption{}
  \label{fig:sub1}
\end{subfigure}
\begin{subfigure}{0.31\textwidth}
  \includegraphics[width=\textwidth]{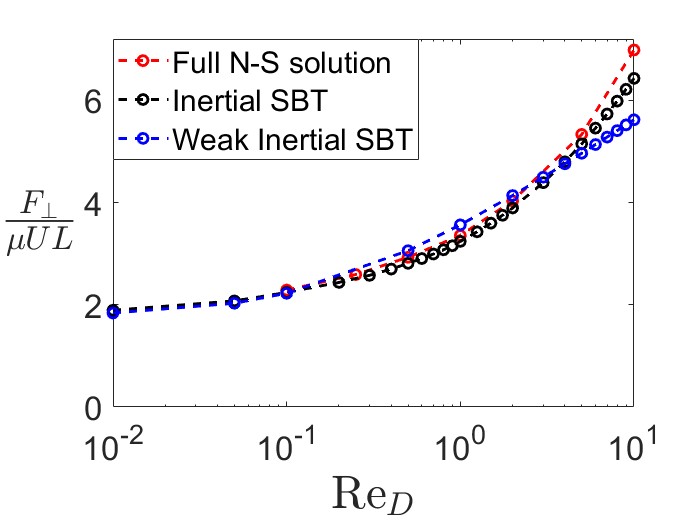}
  \caption{}
  \label{fig:sub2}
\end{subfigure}
\begin{subfigure}{0.31\textwidth}
  \includegraphics[width=\textwidth]{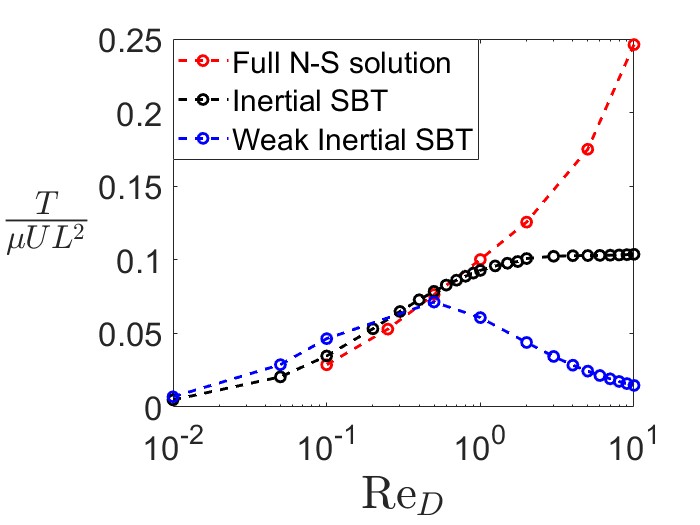}
  \caption{}
  \label{fig:sub3}
\end{subfigure}
\begin{subfigure}{0.31\textwidth}
  \includegraphics[width=\textwidth]{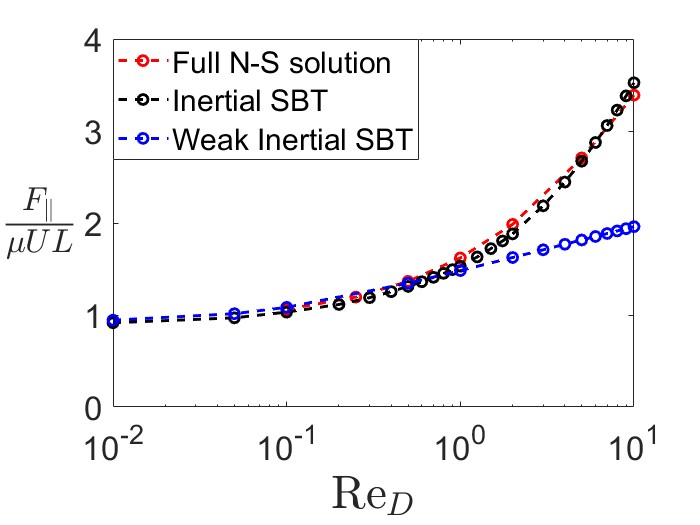}
  \caption{}
  \label{fig:sub4}
\end{subfigure}
\begin{subfigure}{0.31\textwidth}
  \includegraphics[width=\textwidth]{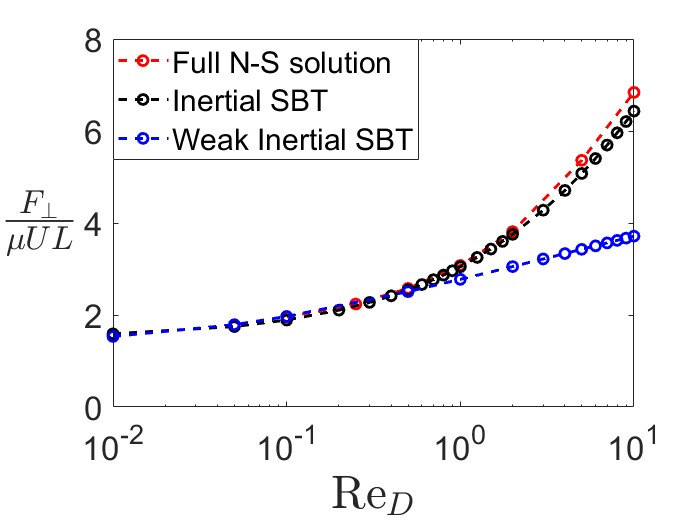}
  \caption{}
  \label{fig:sub5}
\end{subfigure}
\begin{subfigure}{0.31\textwidth}
  \includegraphics[width=\textwidth]{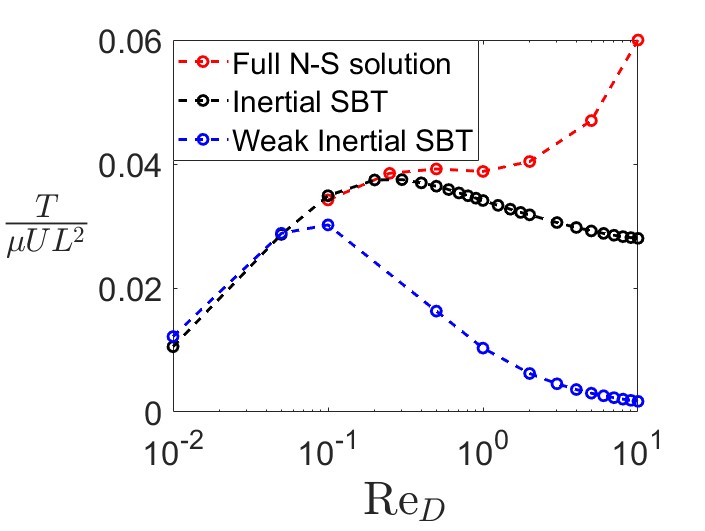}
  \caption{}
  \label{fig:sub6}
\end{subfigure}
\caption{Variation of the normalised forces parallel [(a) and (d)] and perpendicular [(b) and (e)] to the fiber axis and the normalised inertial torque [(c) and (f)] with $\n{Re}_D$. Curves (a),(b),(c) are for $\kappa=20$ and (d),(e),(f) are for $\kappa=100$ }
\fig{force_torque_vs_ReD_comparison} 
\end{bottomfigure}
In this approach, the momentum equations are advanced in time, and the incompressibility condition is enforced through a pressure Poisson equation. This modification is particularly suitable for the current study, as we deal with finite $\n{Re}_D$ (and correspondingly large $\n{Re}_L$) values for high particle aspect ratios ($\kappa$). Consequently, the mass-momentum system of equations (equations (13) and (14) from \cite{sharma2023finite}, excluding the polymer stress term in equation (14)) is solved using the following procedure. 
\\[6pt] 
\textbf{Step 1:}
We start with time-advancing the momentum equation and obtain an intermediate (non-divergence-free) velocity field. This step ignores the incompressibility constraint. 
\begin{gather}
\rho \left( \frac{\bvec{u}^* - \bvec{u}^n}{\Delta t} + \frac{3 \n{ADV}^n - \n{ADV}^{n-1}}{2} \right) = - \nabla p^n + \mu \nabla^2 \bvec{u}^*   
\end{gather}
\textbf{Step 2:} 
Next, we derive the pressure correction equation by enforcing incompressibility. The resulting Poisson equation is solved for the updated pressure $p^{n+1}$. 
\begin{gather}
\nabla^2 p^{n+1} = \frac{\rho}{\Delta t} \nabla \cdot \bvec{u}^*  
\end{gather}
\textbf{Step 3:} 
Finally, we update the velocity field to enforce incompressibility using the newly computed pressure. 
\begin{gather}
\bvec{u}^{n+1} = \bvec{u}^n - \frac{\Delta t}{\rho} \nabla p^{n+1}  
\end{gather}  
To establish the convergence of the simulation results obtained from the solver, we performed tests for the grid resolution and the influence of the outer boundary of the computational domain. This is detailed in Appendix \ref{appB}. We validated the above mentioned pressure Poisson procedure by matching the results with the original procedure from \cite{sharma2023finite} at smaller aspect ratios where it remains computationally feasible. The close match between the large aspect ratio numerical results and our SBT, shown below, serves as another source of numerical validation.  
\\[6pt] 
We now begin the comparison by showing the $\n{Re}_D$-dependence of the forces parallel and perpendicular to the fiber axis and the inertial torque in figure \reffig{force_torque_vs_ReD_comparison}. Curves are plotted for fibers with aspect ratios 20 and 100 translating at an angle of $45^o$ to their orientation. Also plotted are the corresponding quantities obtained from the weakly inertial slender body theory of \cite{khayat1989inertia}. 
\begin{figure}
\centering
\begin{subfigure}{0.31\textwidth}
  \includegraphics[width=\textwidth]{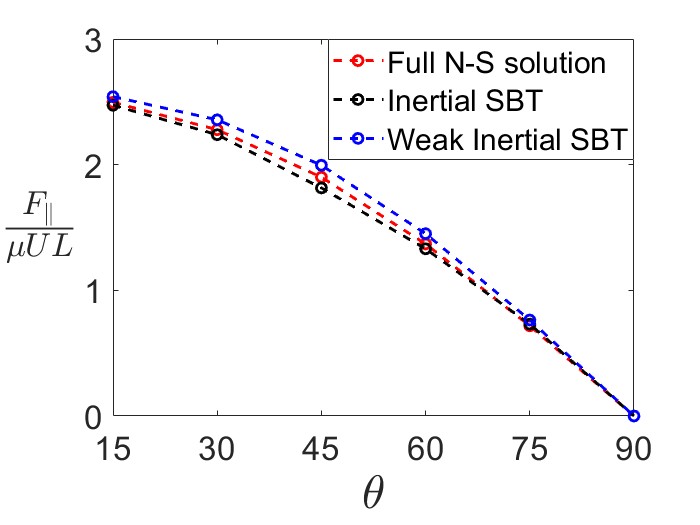}
  \caption{}
  \label{fig:sub1}
\end{subfigure}
\begin{subfigure}{0.31\textwidth}
  \includegraphics[width=\textwidth]{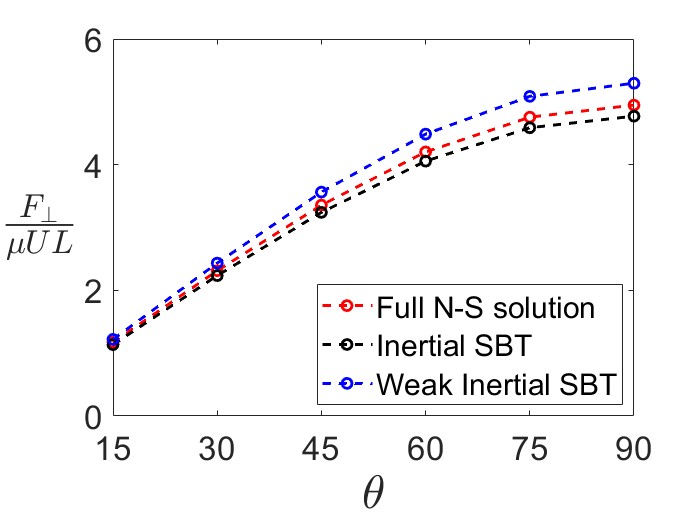}
  \caption{}
  \label{fig:sub2}
\end{subfigure}
\begin{subfigure}{0.31\textwidth}
  \includegraphics[width=\textwidth]{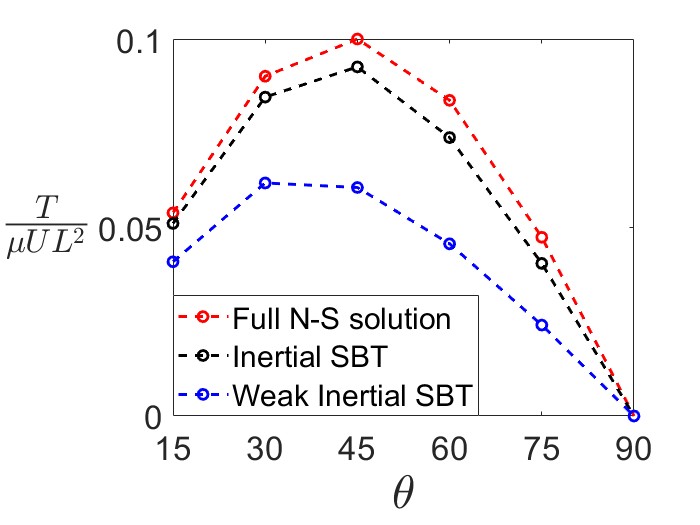}
  \caption{}
  \label{fig:sub3}
\end{subfigure}
\begin{subfigure}{0.31\textwidth}
  \includegraphics[width=\textwidth]{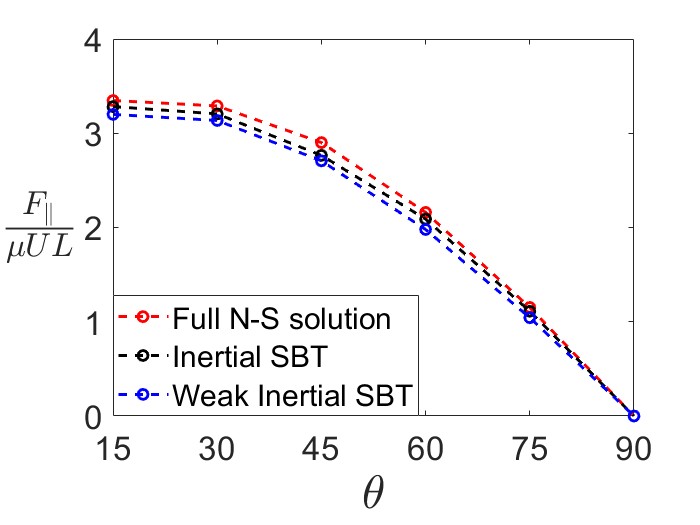}
  \caption{}
  \label{fig:sub4}
\end{subfigure}
\begin{subfigure}{0.31\textwidth}
  \includegraphics[width=\textwidth]{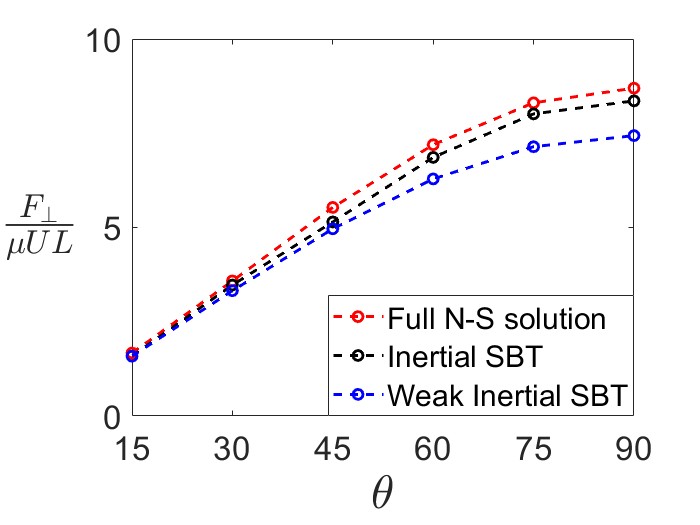}
  \caption{}
  \label{fig:sub5}
\end{subfigure}
\begin{subfigure}{0.31\textwidth}
  \includegraphics[width=\textwidth]{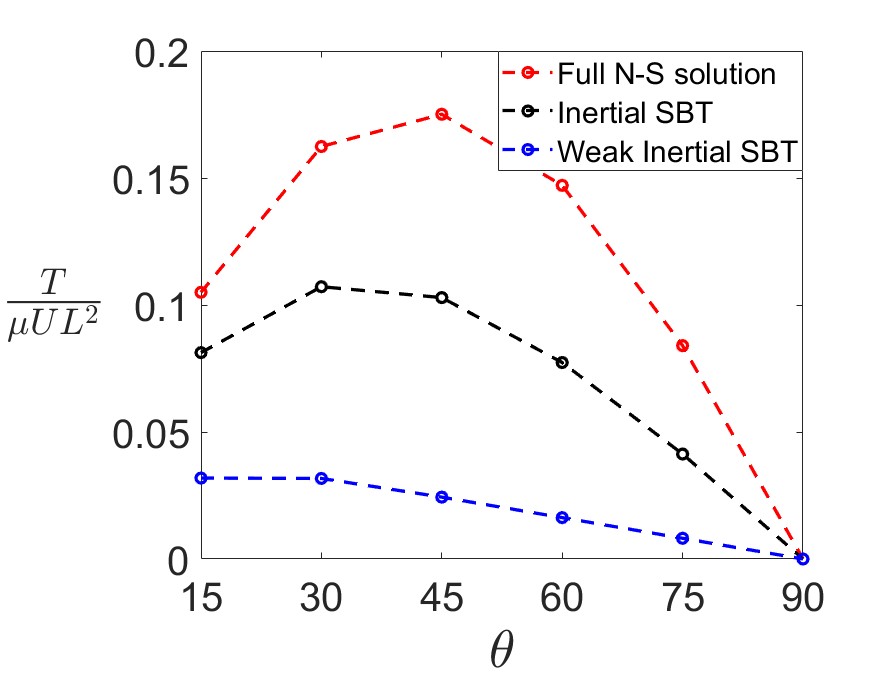}
  \caption{}
  \label{fig:sub6}
\end{subfigure}
\caption{Variation of normalised forces parallel [(a) and (d)] and perpendicular [(b) and (e)] to the fiber axis and the normalised inertial torque [(c) and (f)] with the inclination angle ($\theta$) between the fiber axis and velocity for $Re_D=1$ [(a),(b),(c)] and $Re_D=5$ [(d),(e),(f)]. The fiber aspect ratio ($\kappa$) is 20. } 
\fig{force_torque_vs_theta_comparison_asp20}
\end{figure} 
We observe that the forces parallel and perpendicular to the fiber axis obtained from our inertial slender body theory agree well with the corresponding quantities obtained from the full Navier-Stokes solution for all the $\n{Re}_D$ values considered. On the other hand, the weakly inertial theory underestimates the force on the fiber for both $\kappa = 20$ and $\kappa = 100$ when $\n{Re}_D$ exceeds 1. As far as the inertial torque on the fiber is concerned, we find that our inertial SBT significantly improves upon the weakly inertial SBT of Khayat and Cox when $\n{Re}_D$ exceeds around 0.1. However, beyond $\n{Re}_D = 1$, the predictions for the inertial torque from our inertial SBT with an Oseen outer solution deviate significantly from the full Navier-Stokes solution. 
\begin{figure}
\centering
\begin{subfigure}{0.31\textwidth}
  \includegraphics[width=\textwidth]{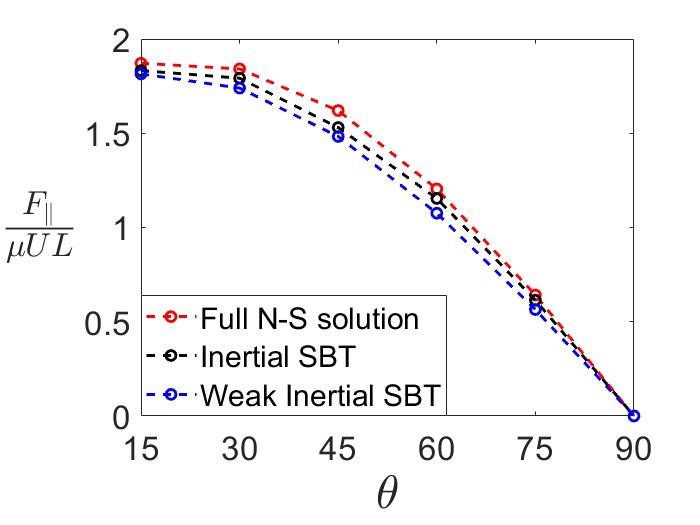}
  \caption{}
  \fig{f_par_vs_theta_asp100_ReD_1}
\end{subfigure}
\begin{subfigure}{0.31\textwidth}
  \includegraphics[width=\textwidth]{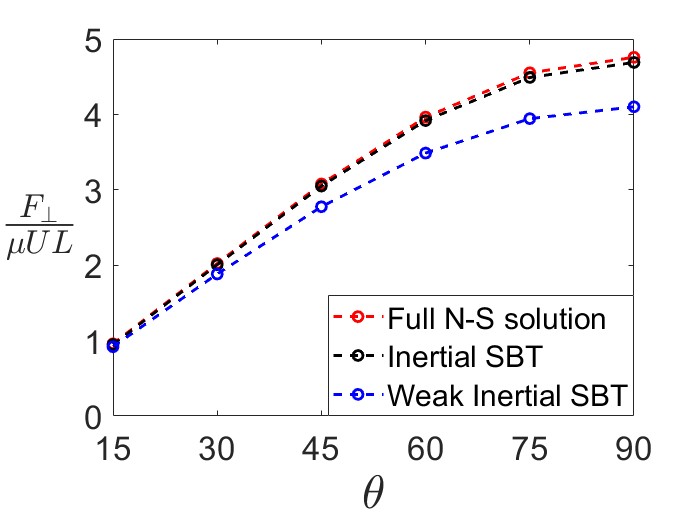}
  \caption{}
  \label{fig:sub8}
\end{subfigure}
\begin{subfigure}{0.31\textwidth}
  \includegraphics[width=\textwidth]{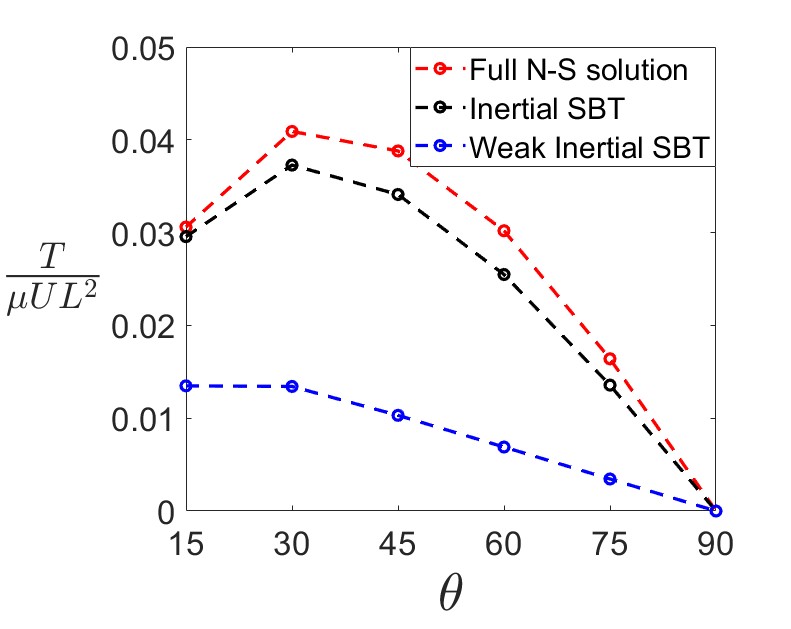}
  \caption{}
  \label{fig:sub9}
\end{subfigure}
\begin{subfigure}{0.31\textwidth}
  \includegraphics[width=\textwidth]{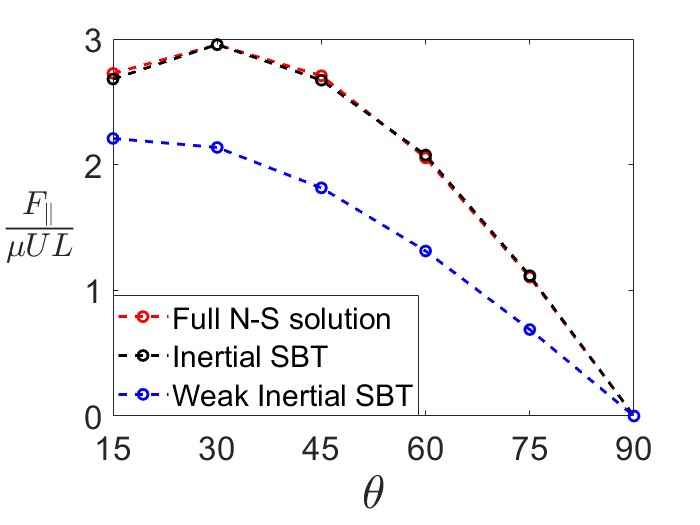}
  \caption{}
  \fig{f_par_vs_theta_asp100_ReD_5}
\end{subfigure}
\begin{subfigure}{0.31\textwidth}
  \includegraphics[width=\textwidth]{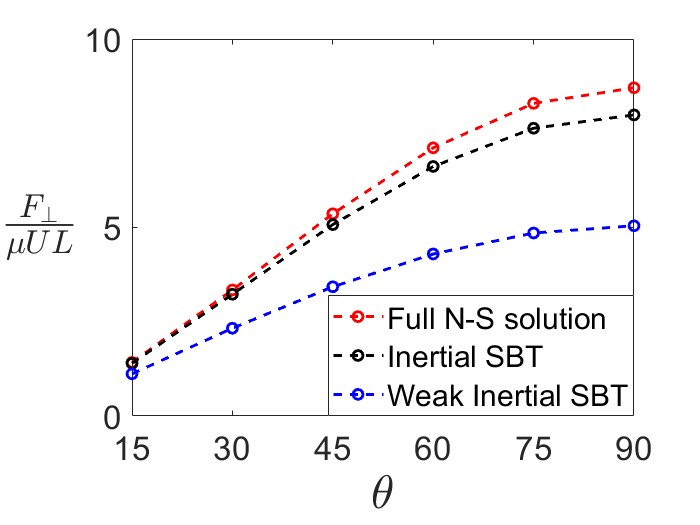}
  \caption{}
  \label{fig:sub11}
\end{subfigure}
\begin{subfigure}{0.31\textwidth}
  \includegraphics[width=\textwidth]{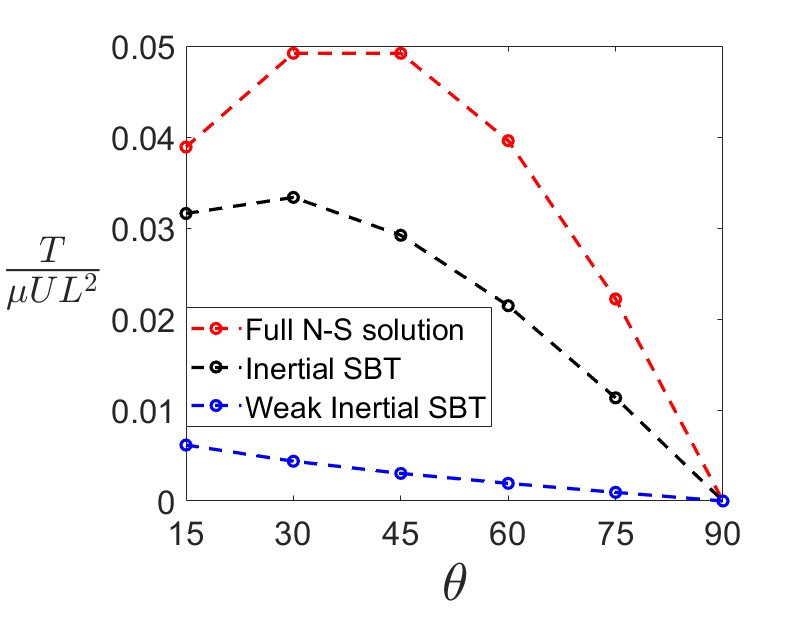}
  \caption{}
  \label{fig:sub12}
\end{subfigure}
\caption{Variation of normalised forces parallel [(a) and (d)] and perpendicular [(b) and (e)] to the fiber axis and the normalised inertial torque [(c) and (f)] with the inclination angle ($\theta$) between the fiber axis and velocity for $Re_D=1$ [(a),(b),(c)] and $Re_D=5$ [(d),(e),(f)]. The fiber aspect ratio ($\kappa$) is 100. } 
\fig{force_torque_vs_theta_comparison_asp100}
\end{figure} 
\\[6pt] 
We also illustrate how the force and torque acting on the fiber vary as a function of the inclination angle between the fiber's velocity and its orientation. Results are presented for Reynolds numbers \(\n{Re}_D = 1\) and \(\n{Re}_D = 5\), considering fiber aspect ratios (\(\kappa\)) of 20 (Figure \reffig{force_torque_vs_theta_comparison_asp20}) and 100 (Figure \reffig{force_torque_vs_theta_comparison_asp100}) and compared with the predictions of the weakly inertial theory and the full Navier-Stokes solution. A key observation from these results is that our inertial slender body theory demonstrates significantly better agreement with the full Navier-Stokes solution for the inertial torque when compared to the predictions made by the weakly inertial theory of \cite{khayat1989inertia}. This improvement highlights the capability of our approach in capturing the essential inertial effects more accurately. 
\\[6pt] 
Moreover, from figures \reffig{force_torque_vs_theta_comparison_asp20} and \reffig{force_torque_vs_theta_comparison_asp100}, we also note that our theory accurately predicts the inertial torque across all inclination angles at \(\n{Re}_D = 1\), but its accuracy declines at \(\n{Re}_D = 5\). This trend is consistent with our earlier findings, which indicate that the applicability of our inertial SBT diminishes for \(\n{Re}_D > 1\). This limitation can be attributed to the increasing role of fluid inertial effects that are not fully captured by our current formulation. To address this issue, we extend the formulation of our inertial slender body theory in Section 4 to enhance the accuracy of torque predictions at higher Reynolds numbers.  
\\[6pt] 
Another notable feature seen in figure \reffig{f_par_vs_theta_asp100_ReD_5} for $\kappa = 100$ and $\n{Re}_D = 5$ is that, unlike the weakly inertial theory, both our inertial SBT and the full Navier–Stokes solution predict a non-monotonic variation of the force parallel to the fiber axis, with a maximum at an intermediate inclination between the fiber axis and the flow direction. This behavior is absent at $\n{Re}_D = 1$ (figure \reffig{f_par_vs_theta_asp100_ReD_1}). When $\n{Re}_L (= \kappa \n{Re}_D)$ is large, this non-monotonic trend can be explained by computing the net force on the fiber resulting from a quasi-2D full Navier-Stokes solution at each fiber cross-section. This is because at large $\n{Re}_L$, as already alluded to, the inertial screening length is much smaller than the fiber length, and the local flow field at each fiber cross-section effectively becomes two-dimensional. We therefore compute the force on the fiber in this manner, and in figure \reffig{f_par_2D_soln_large_ReL}, show the variation of its component parallel to the fiber axis ($F_{\parallel}$) with the inclination angle for both small and \cO(1) values of $\n{Re}_D$.
\\[6pt] 
\begin{figure}
\centering
\includegraphics[width=0.55\textwidth]{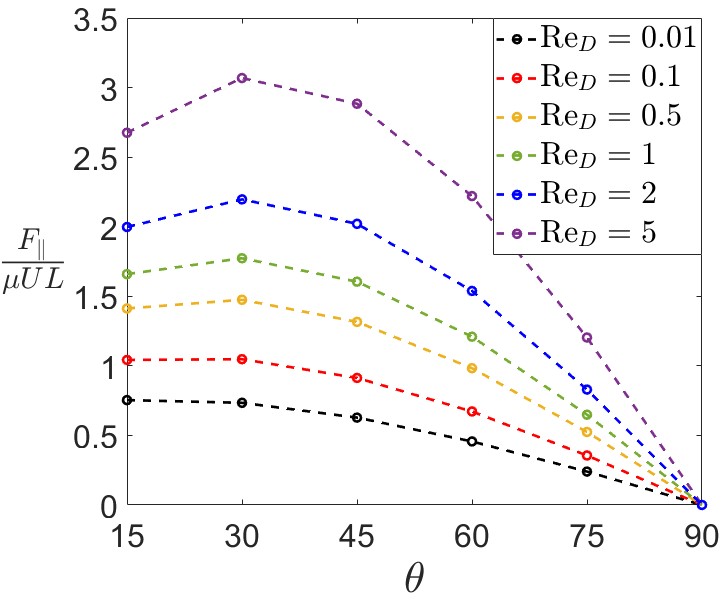}
\caption{ Variation of the normalised force parallel to the fiber axis obtained from the full two-dimensional Navier-Stokes solution at each fiber cross-section. $\n{Re}_L$ is fixed at 500. }   
\fig{f_par_2D_soln_large_ReL} 
\end{figure} 
To interpret the trends in figure \reffig{f_par_2D_soln_large_ReL}, we draw on the analogy (mentioned in section 2.2) between a local two-dimensional fluid flow and heat transfer around an infinite cylinder: the velocity component parallel to the fiber axis plays the role of the temperature field, while the perpendicular velocity acts as the cross-stream motion that drives the convection of heat transfer in the plane normal to the axis. When the Peclet number $Pe$ (analogous to $\n{Re}_D$) is small, the rate of heat transfer (analogous to the local force parallel to the fiber axis) is dominated by conduction, and therefore governed solely by the temperature difference (analogous to the relative fiber–fluid motion along the axis) between the cylinder and its surroundings. This explains the monotonic decrease of $F_{\parallel}$ with inclination $\theta$ at small $\n{Re}_D$. However, when $Pe = $ \cO(1), the rate of heat transfer, quantified by the Nusselt number, depends not only on the temperature difference but also on the strength of convection driven by the perpendicular flow. This means that the heat transfer rate, being analogous to the parallel component of the force, reaches a maximum at an intermediate inclination, where both the parallel component of relative motion and the perpendicular motion (that enhances convection and thus the Nusselt number) are significant. 

\subsection{Comparison of the flow fields in the matching region between inertial SBT and the finite difference Navier-Stokes solution} 

In this section, we examine the validity of our matching procedure by comparing the flow fields in the matching region obtained from our inertial SBT and the finite-difference Navier-Stokes solution using the method of \cite{sharma2023finite}. To illustrate, we consider a case where a spheroidal fiber with aspect ratio 100 is held at rest at an inclination perpendicular to a uniform imposed flow. We are inspecting the flow field around the fiber obtained from our inertial SBT and the Navier-Stokes solution in the midplane ($s=0$) perpendicular to the fiber axis (see figure \reffig{flowfield_schematic}). 
\begin{figure}
\centering
\begin{subfigure}{0.4\textwidth}
  \includegraphics[width=\textwidth]{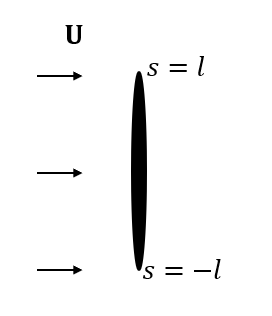}
  \caption{}
  \label{fig:sub1}
\end{subfigure}
\begin{subfigure}{0.4\textwidth}
  \includegraphics[width=\textwidth]{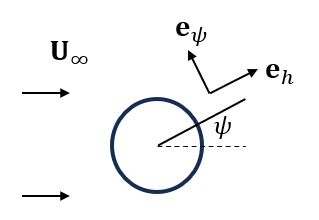} 
  \caption{}
  \label{fig:sub2}
\end{subfigure}
\caption{Illustration of uniform flow past a spheroidal fiber held at rest. (a) Depicts the flow around the entire fiber, with endpoints at $s = \pm l$. (b) Provides a close-up view of the circular cross-section of the fiber at $s = 0$, showing how the flow locally resembles uniform flow past a circle. The unit vectors $\bvec{e}_{\psi}$ and $\bvec{e}_h$ are shown, with $\psi$ denoting the angle relative to the direction of the uniform flow. } 
\fig{flowfield_schematic}
\end{figure} 
\begin{bottomfigure}
\centering
\begin{subfigure}{0.45\textwidth}
  \includegraphics[width=\textwidth]{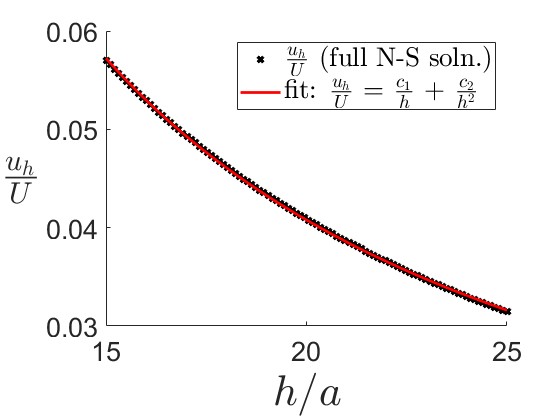}
  \caption{} 
\end{subfigure}
\begin{subfigure}{0.45\textwidth}
  \includegraphics[width=\textwidth]{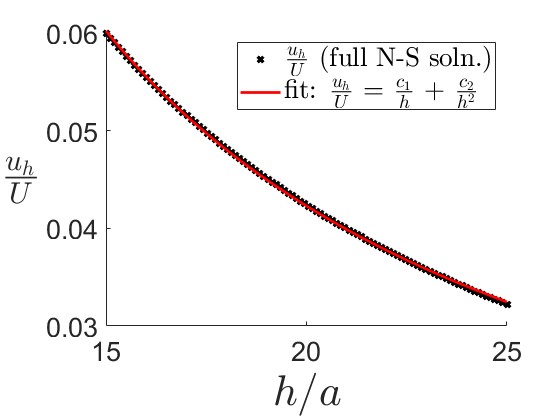}
  \caption{}
\end{subfigure}
\caption{Variation of the radial component of the velocity disturbance with the distance from the fiber axis obtained from the full Navier-Stokes solution for uniform flow past a $\kappa = 100$ fiber together with the fitting function (equation \refeq{u_radial_fit}). Plots (a) and (b) correspond to the directions along $\psi = 90^o$ and $\psi = 180^o$ respectively (see figure \reffig{flowfield_schematic}). The Reynolds number $\n{Re}_D$ is 5.}  
\fig{code_fit}
\end{bottomfigure}
In the ``matching" region $a_0 \ll h \ll l$, the two-dimensional Oseen flow field is expected to vary like $1/h$ (except in a narrow wake downstream of the fiber), where $h$ is the distance away from the fiber axis. To this end, we decompose the radial ($h$) component of the velocity disturbance obtained from the Navier-Stokes solution in the matching region, $a_0 \ll h \ll l$, into a dominant $1/h$ Oseen flow disturbance and a weaker $1/h^2$ disturbance, 
\begin{gather}
u_h(h,\alpha) = \frac{c_1(\psi)}{h} + \frac{c_2(\psi)}{h^2} \eq{u_radial_fit} 
\end{gather} 
where, the azimuthal angle $\psi$ represents the inclination from the direction of the imposed flow. The azimuthal angle ($\psi$) dependent coefficients $c_1$ and $c_2$ are obtained by performing a least squares fitting to the velocity disturbance data from the Navier-Stokes solution in the matching region. 
\begin{figure}
\centering
\begin{subfigure}{0.45\textwidth}
  \includegraphics[width=\textwidth]{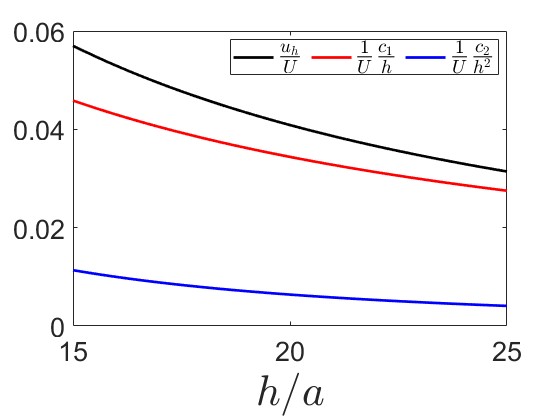}
  \caption{}
\end{subfigure}
\begin{subfigure}{0.45\textwidth}
  \includegraphics[width=\textwidth]{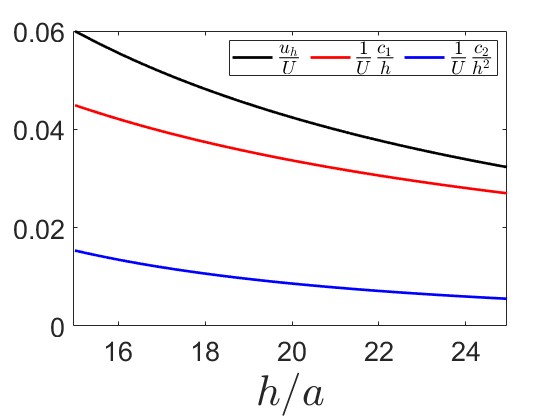}
  \caption{}
\end{subfigure}
\caption{The constituents $c_1/h$ and $c_2/h^2$ of the fit (equation \refeq{u_radial_fit}) shown in figure \reffig{code_fit}. The labels (a) and (b) have the same meaning as in figure \reffig{code_fit}.} 
\fig{fit_constituents}
\end{figure} 
Figure \reffig{code_fit} shows such a fit of the form given in equation \refeq{u_radial_fit} to the radial component of the velocity disturbance in the plane perpendicular to the fiber axis obtained from the full Navier-Stokes solution. 
\begin{bottomfigure}
\centering
\begin{subfigure}{0.49\textwidth}
  \includegraphics[width=\textwidth]{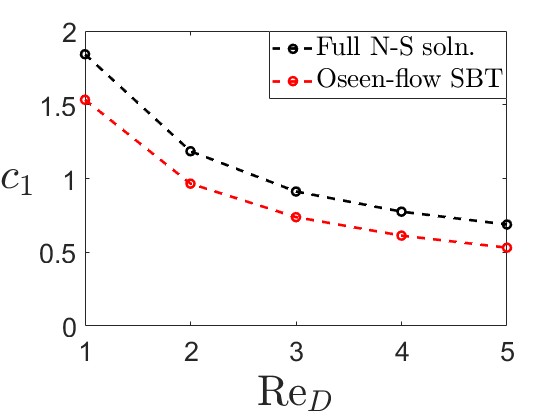}
  \caption{}
  \label{fig:sub1}
\end{subfigure}
\begin{subfigure}{0.49\textwidth}
  \includegraphics[width=\textwidth]{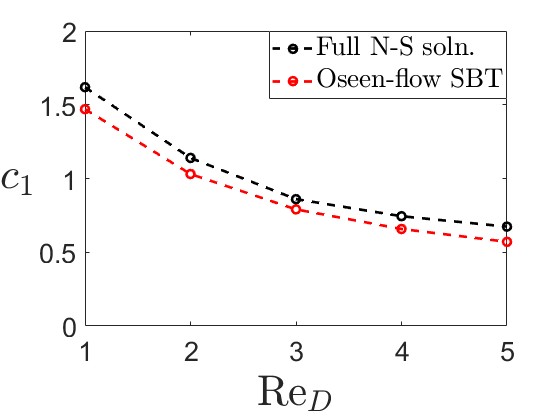} 
  \caption{}
  \label{fig:sub2}
\end{subfigure}  
\caption{Variation with $\n{Re}_D$ of the coefficient of $1/h$ component in the radial velocity disturbance in a plane perpendicular to the fiber axis obtained from inertial SBT (with Oseen flow outer solution) and the Navier-Stokes solution for a $\kappa = 50$ fiber along the directions (a) $\psi = 90^o$ and (b) $\psi = 180^o$ (see figure \reffig{flowfield_schematic}).} 
\fig{c1_comparison}
\end{bottomfigure} 
This is done for one particular $\n{Re}_D$ value of 5 along directions with $\psi = 90^o$ and $\psi = 180^o$ noting the fact that the fits for other values of $\n{Re}_D$ and $\psi$ are equally good. We also show the individual constituents $\frac{c_1(\psi)}{h}$ and $\frac{c_2(\psi)}{h^2}$ in the radial component of the velocity disturbance in figure \reffig{fit_constituents}, indicating that the $1/h^2$ disturbance makes up a significant portion of the overall velocity disturbance around the fiber at this value of $\n{Re}_D$. This additional disturbance stems from the effects of finite fiber diameter and the fluid incompressibility condition, something that our inertial SBT with Oseen flow field due to a line distribution of force singularities does not account for. 
\\[6pt] 
However, it is observed that the $1/h$ component of the velocity disturbance in the matching region is well captured by our inertial SBT. This is illustrated in Figure \reffig{c1_comparison}, which shows that the coefficient of the $1/h$ term in the radial component of the fluid velocity disturbance obtained from our inertial SBT closely matches the corresponding coefficient from the full Navier-Stokes solution in the matching region ($a_0 \ll h \ll l$) around the fiber.  
\\[6pt]  
The predominance of the leading-order Oseen flow disturbance at small $\n{Re}_D$ can be clearly observed in the velocity disturbance field around the fiber, as depicted in figure \reffig{flowfield_comparison_SBT_AC} in a plane perpendicular to the fiber axis at $s = 0$ (see figure \reffig{flowfield_schematic}). 
\begin{figure}
\centering
\begin{subfigure}{0.49\textwidth}
  \includegraphics[width=\textwidth]{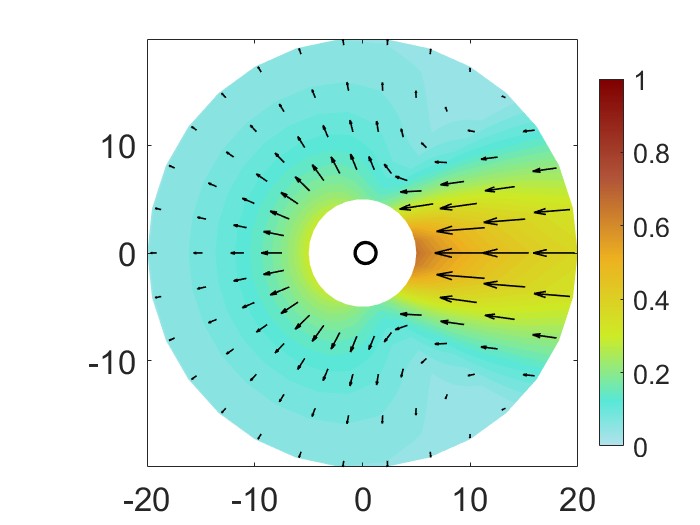}
  \caption{}
  \label{fig:sub1}
\end{subfigure}
\begin{subfigure}{0.49\textwidth}
  \includegraphics[width=\textwidth]{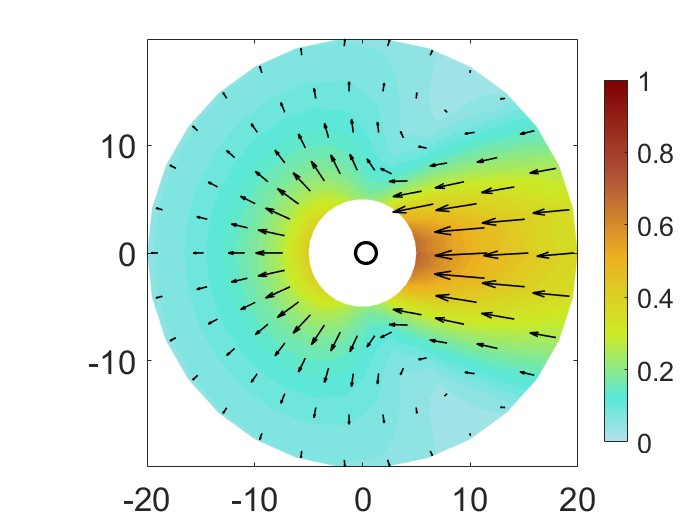} 
  \caption{}
  \label{fig:sub2}
\end{subfigure} 
\begin{subfigure}{0.49\textwidth}
  \includegraphics[width=\textwidth]{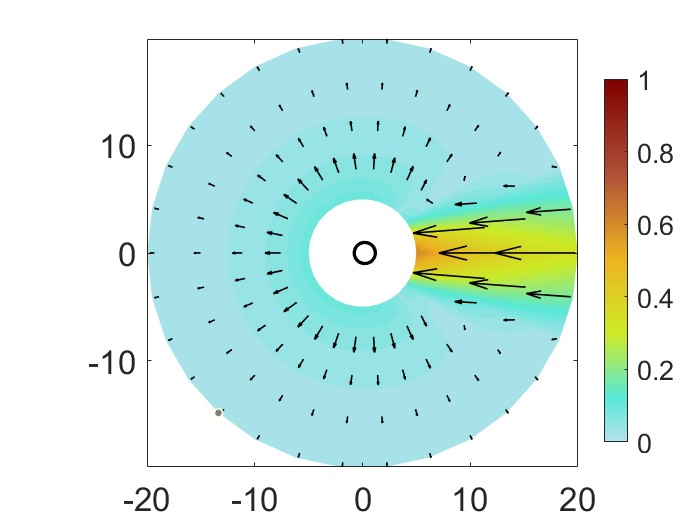}
  \caption{}
  \label{fig:sub1}
\end{subfigure}
\begin{subfigure}{0.49\textwidth}
  \includegraphics[width=\textwidth]{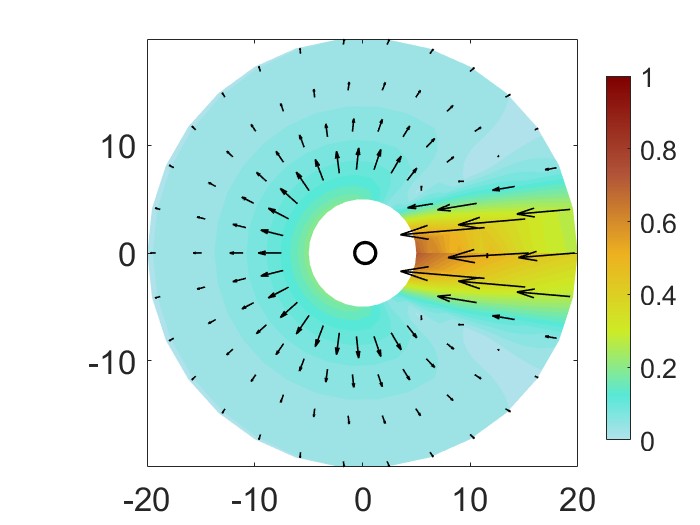} 
  \caption{}
  \label{fig:sub2}
\end{subfigure} 
\caption{Vector plots of the velocity disturbance (scaled with $U_{\infty}$) and contour plots of its magnitude in a plane perpendicular to the fiber axis at $s= 0$ (see figure \reffig{flowfield_schematic}). $\n{Re}_D$ values of 1 ((a) and (b)) and 5 ((c) and (d)) are considered. The fiber aspect ratio is 100. Only the region $h \in [5a_0,20a_0]$ (matching region) is shown. The tickmarks along x and y axes indicate the coordinates (scaled with the maximum fiber radius $a_0$) of a point on the plane with (0,0) being the point of intersection of the fiber major axis with the plane. The black circle at the center of each plot denotes the fiber cross-section. Plots (a) and (c) are obtained from our inertial SBT, while (b) and (d) are obtained using the solver developed by \cite{sharma2023finite}. } 
\fig{flowfield_comparison_SBT_AC}
\end{figure} 
The velocity disturbance field in the matching region, derived from our inertial SBT, closely resembles that obtained from the full Navier-Stokes solution for $\n{Re}_D = 1$. This strong agreement highlights the validity of our inertial SBT in capturing the dominant flow features at lower $\n{Re}_D$, where the Oseen flow approximation is accurate. 
\\[6pt] 
While the qualitative features of a wake and radial source flow are retained at a larger $\n{Re}_D$ of 5, the quantitative correspondence between the inertial SBT results and the Navier-Stokes solution diminishes. This deviation arises because the inertial SBT, with its Oseen flow outer solution, does not account for additional weaker flow disturbances that become increasingly significant as $\n{Re}_D$ increases. These higher-order disturbances, neglected in the current SBT formulation, contribute to the discrepancy observed at higher $\n{Re}_D$. More plots of the flow field for a few $\n{Re}_D, \kappa$ and $\theta$ values from the Navier-Stokes solution of \cite{sharma2023finite} are included in the supplementary materials. 

\section{Inclusion of potential flow disturbances due to the fiber} 
\sect{pot_flow}

In section \refsect{oseen_flow_SBT_results}, we noted that the inertial SBT with an Oseen flow outer solution fails to accurately predict the inertial torque on an obliquely translating fiber beyond $\n{Re}_D = 1$. This discrepancy arises because, the flow field in the outer region arising out of the linear superposition of Oseenlets does not account for the finite thickness of the fiber. The finite fiber diameter introduces additional physical effects, such as the necessity for incompressible flow to incorporate localized sources and source dipoles of fluid volume per unit length to satisfy mass conservation. These contributions, which are absent in the Oseen flow treatment, play a critical role in accurately capturing the torque behavior.
\\[6pt] 
In this context, introducing potential flow disturbances around the fiber provides a straightforward yet meaningful way to address these missing effects. Potential flow inherently accounts for the sources and dipoles generated by the fiber's finite thickness, offering a more comprehensive description of the local flow field around the fiber. While this approach is not rigorous at finite Reynolds numbers and involves simplifying assumptions, such as a linear superposition of the potential flow and the dominant Oseen flow disturbance, it proves effective in capturing the inertial torque trends at higher $\n{Re}_D$. The success of this approach is evidenced by its ability to reproduce the torque behavior obtained from finite difference solutions of the Navier-Stokes equations. 

\subsection{Slender-body theory treatment of potential flow around the fiber and determination of the torque} 

The Oseen flow outer solution that leads to force singularities along the fiber axis is complemented by a weaker potential flow velocity disturbance resulting from the effects of finite fiber diameter. In this section, we lay out the framework based on slender-body theory to compute the potential flow disturbance around a straight fiber held fixed in a uniform flow with velocity $\bvec{U}$.  
\\[6pt] 
One of the earliest developments of a slender-body theory for axially symmetric potential flows was by \cite{tuck1964some}, who focused on selecting an appropriate coordinate system to avoid singularities in slender bodies with blunt ends. A formal slender-body theory for axially symmetric potential flow around an axially symmetric rigid body was later established by \cite{handelsman1967axially}, who modeled the flow using an axial distribution of point sources and solved the corresponding integral equation for the source strengths. \cite{geer1975uniform} extended the analysis of \cite{handelsman1967axially} to include transverse flow relative to the fiber axis, incorporating an axial distribution of source dipoles. However, none of these studies went on to derive the torque acting on a slender fiber held fixed in a uniform flow. 
\\[6pt] 
Our approach for determining the torque on a fiber held fixed in potential flow is based on slender-body theory that involves the derivation of source and source-dipole of fluid volume per unit length along the fiber axis. It follows the same framework used earlier to determine the fiber force distribution using an outer Oseen flow treatment in section \refsect{Finite_ReD_SBT}. 
\\[6pt] 
Since the flow is irrotational, the velocity disturbance $\bvec{u}$ on the fluid induced by the fiber can be expressed as the gradient of a scalar potential $\phi$. 
\begin{gather}
\bvec{u} = \nabla \phi 
\end{gather}
Since the flow is incompressible, the velocity potential $\phi$ satisfies Laplace's equation. 
\begin{gather}
\nabla \cdot \bvec{u} = \nabla^2 \phi = 0 \eq{gov_eqn_for_phi} 
\end{gather} 
We now perform a perturbation expansion for the potential disturbance $\phi$ in $1/\kappa$ in the inner and outer regions and asymptotically match the solutions to determine the coefficients in these expansions. We further determine the fluid disturbance velocity field and hence the pressure on the fiber surface, which is essential for computing the torque acting on the slender fiber. It will be seen that the two leading order terms of the expansion for the potential disturbance produce a torque that agrees with the result of \cite{lamb1924hydrodynamics}.   
\\[6pt] 
In the inner region, we scale the projection of position vector $\bvec{h} = (\bvec{I}-\bvec{pp}) \cdot \bvec{r}$ of a point onto the transverse plane at an axial position $s$ with the local fiber cross-sectional radius $a(s)$. As before, the coordinate $s$ along the fiber axis is scaled with the fiber half-length $l$.  The fiber aspect ratio is $\kappa = l/a_0$. We begin by obtaining the leading-order approximation to the potential disturbance $\phi^i$ in the inner region by solving the two-dimensional Laplace equation at a local fiber cross-section, under the assumption that the variation of the potential disturbance along the fiber axis is negligible. Due to the local source $q(s)$ and source-dipole $\bvec{d}(s)$ of fluid volume per unit length at a cross-section with coordinate $s$, the leading-order potential disturbance in the inner region is given by, 
\begin{gather}
\phi^i(s,\bvec{h}) = \frac{q(s)}{2\pi} \n{ln} h - \frac{\bvec{d}(s) \cdot \bvec{h}}{2 \pi a_0 \tilde{a}(s) h^2} \eq{phi_lo} 
\end{gather}
where $\tilde{a}(s)$ denotes the fiber cross-sectional radius at $s$ scaled with the maximum cross-sectional radius $a_0$. For a prolate spheroidal fiber, 
\begin{gather} 
a(s) = a_0 (1-s^2)^{1/2} \implies \tilde{a}(s) = (1-s^2)^{1/2} \eq{as_variation} 
\end{gather} 
The potential disturbance $\phi^i$ in the inner region satisfies the no-penetration boundary condition at the fiber surface, $\bvec{n}(s,\alpha) \cdot \nabla \phi^i = 0$. Here, 
\begin{gather}
\bvec{n}(s,\alpha) = \frac{-\frac{\tilde{a}'(s)}{\kappa} \bvec{p} + \bvec{e}_h}{\left[1+\frac{1}{\kappa^2}[\tilde{a}'(s)]^2\right]^{1/2}} \eq{local_normal} 
\end{gather} 
is the normal to the fiber surface along an azimuthal coordinate $\alpha$ at the axial position $s$, and  $\tilde{a}'(s) = \n{d}\tilde{a}(s)/\n{d}s$. Let \( U_{\parallel} \) and \( \bvec{U}_{\perp} \) represent the components of the ambient flow $\bvec{U}$ parallel and perpendicular to the fiber axis respectively.  If $k_{\parallel}(s)$ and $\bvec{k}_{\perp}(s)$ denote the imposed fluid flow parallel and perpendicular to the fiber axis respectively at a local cross-section $s$ resulting from the sum of the uniform ambient flow $\bvec{U}$ and the non-singular potential-flow velocity disturbance (to be derived later in this section) induced by the fiber, then the above boundary condition at the fiber surface ($h = 1$), 
\begin{gather}
\left[ -\frac{\tilde{a}'(s)}{\kappa} \bvec{p} + \bvec{e}_h \right] \cdot \left[  \frac{1}{a(s)} \frac{\partial \phi^i}{\partial h} \bvec{e}_h + \bvec{k}_{\perp}(s) + \left( \frac{1}{l} \frac{\partial \phi^i}{\partial s} + k_{\parallel}(s) \right) \bvec{p} \right] = 0 
\end{gather} 
leads to,  
\begin{gather}
\frac{q(s)}{2 \pi a_0 \tilde{a}(s)} + \frac{\bvec{d}(s) \cdot \bvec{e}_h} {2 \pi [a_0 \tilde{a}(s)]^2} + \bvec{k}_{\perp} (s) \cdot \bvec{e}_z = \frac{a_0}{l} \tilde{a}'(s) \left[ \frac{q'(s)}{2 \pi l} \n{ln} [\tilde{a}(s)] - \frac{\bvec{d}'(s) \cdot \bvec{e}_h}{2 \pi l [a_0 \tilde{a}(s)] } + k_{\parallel}(s) \right] \eq{scaled_BC_full} 
\end{gather} 
Since the unit vector $\bvec{e}_h$ at a local fiber cross-section is arbitrary, equation \refeq{scaled_BC_full} can be decomposed into two separate equations for the source and source-dipole strengths, 
\begin{gather}
\frac{q(s)}{2 \pi a(s)} - \frac{a'(s)q'(s)}{2 \pi \kappa^2} \n{ln}[a(s)] = \frac{a_0 a'(s)}{\kappa} k_{\parallel} (s) \eq{scaled_BC_par} \\
\frac{\bvec{d}(s)}{2 \pi [a(s)]^2} + \frac{a'(s) \bvec{d}'(s)}{2 \pi \kappa^2 a(s)} = - a_0^2 \bvec{k}_{\perp}(s)  \eq{scaled_BC_perp}
\end{gather}
We now use the leading order expressions, $k_{\parallel}(s) = U_{\parallel}$ and $\bvec{k}_{\perp}(s) = \bvec{U}_{\perp}$ in equations \refeq{scaled_BC_par} and \refeq{scaled_BC_perp} respectively, and the perturbation expansions for $q(s)$ and $\bvec{d}(s)$ in $1/\kappa$, 
\begin{gather}
q(s) = \frac{q_1(s)}{\kappa} + \frac{q_2(s)}{\kappa^2} + \frac{q_3(s)}{\kappa^3} + ... \eq{qs_expansion} \\
\bvec{d}(s) = \bvec{d}_0(s) + \frac{\bvec{d}_1(s)}{\kappa} + \frac{\bvec{d}_2(s)}{\kappa^2} + ... \eq{ds_expansion} 
\end{gather}
to obtain the following equations, 
\begin{gather}
\frac{1}{2\pi \tilde{a}(s)} \left[ \frac{q_1(s)}{\kappa} + \frac{q_2(s)}{\kappa^2} + ... \right] - \frac{\tilde{a}'(s) \n{ln}[a(s)]}{2 \pi \kappa^2} \left[ \frac{q_1'(s)}{\kappa} + \frac{q_2'(s)}{\kappa^2} + ... \right] = \frac{a_0 \tilde{a}'(s)}{\kappa} U_{\parallel} \eq{BC_qs_expanded} \\
\frac{1}{2 \pi [\tilde{a}(s)]^2} \left[ \bvec{d}_0(s) + \frac{\bvec{d}_1(s)}{\kappa} + \frac{\bvec{d}_2(s)}{\kappa^2} + ... \right] + \frac{\tilde{a}'(s)}{2 \pi \kappa^2 \tilde{a}(s)} \left[ \bvec{d}_0'(s) + \frac{\bvec{d}_1'(s)}{\kappa} + \frac{\bvec{d}_2'(s)}{\kappa^2} + ... \right] = -a_0^2 \bvec{U}_{\perp} \eq{BC_ds_expanded} 
\end{gather} 
Using equations \refeq{BC_qs_expanded} and \refeq{BC_ds_expanded}, we obtain at the leading order, 
\begin{gather}
q(s) = q_1(s)/\kappa = \frac{2 \pi a_0 \tilde{a}(s) \tilde{a}'(s)}{\kappa} U_{\parallel} \eq{lo_qs} \\ 
\bvec{d}(s) = \bvec{d}_0(s) = -2 \pi a_0^2 [\tilde{a}(s)]^2 \bvec{U}_{\perp} \eq{lo_ds}
\end{gather} 
We now perform an expansion in $1/\kappa$ of the potential disturbance in the inner region, 
\begin{gather}
\phi^i = \phi_0^i + \frac{\phi_1^i}{\kappa} + \frac{\phi_2^i}{\kappa^2} + ... \eq{phi_inner_expansion} 
\end{gather}
and substitute the leading order expressions for $q(s)$ and $\bvec{d}(s)$
into equation \refeq{phi_lo} to obtain the first two terms of the expansion, thereby resulting in,  
\begin{gather}
\phi^i(s,\bvec{h}) = \phi_0^i (s,\bvec{h}) + \frac{1}{\kappa} \phi_1^i(s,\bvec{h}) + O(1/\kappa^2) = - \frac{a_0 s}{\kappa} U_{\parallel} \n{ln} h + a_0 \tilde{a}(s) \frac{\bvec{U}_{\perp} \cdot \bvec{h}}{h^2} + O(1/\kappa^2) \eq{phi_first_two_terms} 
\end{gather} 
With the first two terms $\phi_0^i$ and $\phi_1^i/\kappa$ of the expansion in $1/\kappa$ for the potential disturbance $\phi^i$ in the inner region determined, we can now obtain the first approximation to the force and torque acting on the fiber, without computing the outer solution. To this end, we evaluate the velocity disturbance $\bvec{u}^i$ resulting from the potential disturbance $\phi^i$. This gives the components of the velocity disturbance, 
\begin{gather}
u_{\parallel}^i = \frac{1}{l} \frac{\partial \phi^i}{\partial s} = -\frac{U_{\parallel}}{\kappa^2} \n{ln}h + \frac{\tilde{a}'(s) \bvec{U}_{\perp} \cdot \bvec{h}}{\kappa h^2} + o(1) \eq{u_par_f2t} \\ 
u_{\perp h}^i = \frac{1}{a_0 \tilde{a}(s)} \frac{\partial \phi^i}{\partial h} = \frac{\tilde{a}'(s) U_{\parallel}}{\kappa h} - \frac{\bvec{U}_{\perp} \cdot \bvec{e}_h}{h^2} + o(1) \eq{u_perp_z_f2t} \\
u_{\perp \alpha}^i = -\frac{U_{\perp} \n{sin}\alpha}{h^2} + o(1) \eq{u_perp_alpha_f2t}  
\end{gather} 
along the direction parallel to the fiber axis, and the radial and azimuthal directions in the plane perpendicular to the fiber axis at an axial position $s$. In equation \refeq{u_perp_alpha_f2t}, $\n{cos}\alpha = \bvec{U}_{\perp}\cdot \bvec{e}_h / U_{\perp}$. We now evaluate equations \refeq{u_par_f2t}-\refeq{u_perp_alpha_f2t} at the fiber surface ($h=1$) to obtain $\bvec{u}^i(s,h=1,\alpha)$ and use Bernoulli's equation to obtain the pressure $p(s,\alpha)$. 
\begin{gather}
p(s,\alpha) + \frac{1}{2} \rho \vert \bvec{U} + \bvec{u}^i (s,h=1,\alpha) \vert ^2 = \frac{1}{2}\rho U^2 \eq{bernoulli_eqn} 
\end{gather}  
where the pressure at a large distance away from the fiber is assumed to be zero. The first approximation to the force and torque on the fiber are then obtained as, 
\begin{gather} 
\bvec{F}_{\n{pot}} = - \int_{A_s} p(s,\alpha) \bvec{n}(s,\alpha) \n{d}A_s \\
\implies \bvec{F}_{\n{pot}} = - a_0 l \int_{-1}^{1} \int_{0}^{2 \pi} p(s,\alpha) \bvec{n}(s,\alpha) \tilde{a}(s) \left[ 1+\frac{1}{\kappa^2}(\tilde{a}'(s))^2 \right]^{1/2} \n{d}\alpha \n{d}s \eq{F_f2t} \\
\bvec{T}_{\n{pot}} = - \int_{A_s} (\bvec{r}(s,\alpha) \times p(s,\alpha) \bvec{n}(s,\alpha)) \n{d}A_s \nonumber \\
\implies \bvec{T}_{\n{pot}} = -a_0 l \int_{-1}^{1} \int_{0}^{2 \pi} (\bvec{r}(s,\alpha) \times p(s,\alpha) \bvec{n}(s,\alpha)) \tilde{a}(s) \left[ 1+\frac{1}{\kappa^2}(\tilde{a}'(s))^2 \right]^{1/2} \n{d}\alpha \n{d}s \eq{T_f2t} 
\end{gather} 
where, $\bvec{r}(s,\alpha) = s\bvec{p}l + a_0 [\tilde{a}(s)] \bvec{e}_h$. The integrals in equations \refeq{F_f2t} and \refeq{T_f2t} are evaluated analytically. We find that $\bvec{F}_{\n{pot}} = 0$, consistent with d'Alembert's paradox. On the other hand, the torque $\bvec{T}_{\n{pot}}$ is, 
\begin{gather}
\bvec{T}_{\n{pot}} = \left[ 2\rho a_0^2 l \left( 1+ \frac{1}{\kappa^2} \right) \int_{-1}^{1} \int_{0}^{2 \pi} s^2 \n{cos}^2\alpha \n{d}\alpha \n{d}s \right] (\bvec{U}\cdot \bvec{p}) (\bvec{U} \times \bvec{p} ) + o(1) 
\end{gather}
When $\kappa \gg 1$, we obtain, 
\begin{gather}
\frac{T_{\n{pot}}}{\mu U L^2} = \frac{\pi \n{Re}_D}{12\kappa}\n{sin}2\theta + o(1) \eq{torque_f2t_large_kappa} 
\end{gather}
We find that when $\kappa \gg 1$, the torque $\bvec{T}_{\n{pot}}$ from \refeq{torque_f2t_large_kappa} comes out to be approximately the same as that predicted by \cite{lamb1924hydrodynamics} for steady flow past an oblique spheroid held at rest. The full expression for the torque given by \cite{lamb1924hydrodynamics} is, 
\begin{gather}
\bvec{T}_{\n{Lamb}} = \left| \frac{\tau}{2-\tau} - \frac{\sigma}{2-\sigma} \right| \left(\frac{4}{3}\pi \rho a_0^2 l \right) (\bvec{U}\cdot \bvec{p}) (\bvec{U} \times \bvec{p} ) \eq{T_lamb} 
\end{gather} 
where, 
\begin{gather}
\tau = \frac{\kappa^2}{\kappa^2 -1} - \frac{\kappa}{2(\kappa^2-1)^{3/2}} \n{ln} \left[ \frac{\kappa +(\kappa^2-1)^{1/2}}{\kappa - (\kappa^2-1)^{1/2}} \right] \\ 
\sigma = -\frac{2}{\kappa^2-1} + \frac{\kappa}{(\kappa^2-1)^{3/2}} \n{ln} \left[ \frac{\kappa +(\kappa^2-1)^{1/2}}{\kappa - (\kappa^2-1)^{1/2}} \right] 
\end{gather} 
One should note that the potential flow torque on the fiber from \cite{lamb1924hydrodynamics} is not computed using an explicit evaluation of the surface pressure distribution on the fiber. Instead, the torque is obtained by treating the fiber and fluid as a single dynamical system. The total kinetic energy of the system is expressed as a quadratic form in the translational and rotational velocities of the fiber, with additional “inertia coefficients” representing the fluid’s contribution to the kinetic energy. The hydrodynamic couple then follows from the dynamical equations as the time rate of change of the angular impulse, which can be directly computed from the expression for the kinetic energy. 
\\[6pt] 
From the above discussion, it follows that the leading-order inner solution for $\phi^i$ in equation \refeq{phi_first_two_terms} is the major contributor to the torque on the fiber in potential flow. We next derive the next higher-order corrections to this torque, arising from matching with the outer solution, which are $O((\ln \kappa)/\kappa^2)$ and $O(1/\kappa^2)$ smaller than the leading order result in \refeq{torque_f2t_large_kappa}. 
\\[6pt] 
Before we proceed to compute the outer solution, we recall that the first two terms in the inner expansion of the potential disturbance are obtained by solving a two-dimensional Laplace equation at a local fiber cross-section, neglecting the variation of the potential disturbance along the fiber axis. An $O(1/\kappa^2)$ correction to the potential disturbance arises upon relaxing this assumption and accounting for this axial variation. This requires solving the full three-dimensional Laplace equation for the potential disturbance in the inner region. 
\begin{gather}
\frac{1}{[a(s)]^2} \nabla_{2D}^{2} \phi^i + \frac{1}{l^2} \frac{\partial^2 \phi^i}{\partial s^2} = 0 \nonumber \\ 
\implies \frac{1}{[\tilde{a}(s)]^2} \nabla_{2D}^{2} \phi^i + \frac{1}{\kappa^2} \frac{\partial^2 \phi^i}{\partial s^2} = 0 \eq{gov_eqn_phi_nondim} 
\end{gather} 
Upon substituting the expansion \refeq{phi_inner_expansion} for $\phi^i$ into equation \refeq{gov_eqn_phi_nondim}, we obtain the following equation for the $O(1/\kappa^2)$ correction to the leading order potential disturbance. 
\begin{gather}
\frac{1}{h} \frac{\partial}{\partial h}\left(h \frac{\partial \phi_{2}^i}{\partial h} \right) + \frac{1}{h^2}\frac{\partial^2 \phi_{2}^i}{\partial \alpha^2} = [a_0\tilde{a}(s)] \frac{\bvec{U}_{\perp}\cdot \bvec{h}}{h^2} \eq{gov_eqn_phi2_expanded} 
\end{gather} 
We assume a trial solution, $\phi_{2}^i(s,\bvec{h}) =  f(s,h) \bvec{U}_{\perp} \cdot \bvec{h}$ to equation \refeq{gov_eqn_phi2_expanded}, which upon substitution leads us to arrive at the following ODE for $f(s,h)$, 
\begin{gather}
h \frac{\partial^2 f}{\partial h^2} + 3 \frac{\partial f}{\partial h} - \frac{a_0[\tilde{a}(s)]}{h} = 0 
\end{gather} 
The general solution to the above equation is, 
\begin{gather}
f(s,z) = \frac{a_0[\tilde{a}(s)]}{2} \n{ln} h + \frac{c_1(s)}{h^2} + c_2(s) \eq{f_result_pot_flow} 
\end{gather} 
This gives, 
\begin{gather}
\phi_{2}^i(s,\bvec{h}) = \left[ \frac{a_0[\tilde{a}(s)]}{2} \n{ln} h + \frac{c_1(s)}{h^2} + c_2(s) \right] \bvec{U}_{\perp} \cdot \bvec{h} \eq{phi_inner_third_term} 
\end{gather} 
where, $c_1(s)$ and $c_2(s)$ are functions of $s$ to be determined later in this section. Specifically, the second and third terms in equation \refeq{phi_inner_third_term} form a contribution to $\phi_{2}^i(s,\bvec{h})$ which is driven by the next-order corrections to $q(s)$ and $\bvec{d}(s)$. It will be shown that these corrections are themselves of $O(1/\kappa^2)$, and originate from the non-singular part of the fiber-induced velocity disturbance at a local cross-section. The potential disturbance in the inner region correct to $O(1/\kappa^2)$ is therefore given by, 
\begin{gather}
\phi^i = a_0 \tilde{a}(s) \frac{\bvec{U}_{\perp} \cdot \bvec{h}}{h^2} - \frac{a_0 s}{\kappa} U_{\parallel} \n{ln} h + \frac{\phi_2^i(s,\bvec{h})}{\kappa^2} + O(1/\kappa^3) \eq{phi_inner_soln} 
\end{gather} 
where, $\phi_{2}^i(s,\bvec{h})$ is given in equation \refeq{phi_inner_third_term}. 
\\[6pt]
We now proceed to evaluate the outer solution $\phi^o$ to the potential disturbance. The first approximation to the potential disturbance $\phi^o$ in the outer region is that produced by a line distribution of the leading order expressions for the source and source-dipole strengths given in equations \refeq{lo_qs} and \refeq{lo_ds} respectively. This outer potential disturbance is then matched to the corresponding inner solution at $O(1/\kappa)$ and the non-singular part of the velocity disturbance is evaluated at a local fiber cross-section. Finally, equations \refeq{scaled_BC_par}-\refeq{ds_expansion} are used to obtain the next higher order corrections to $q(s)$ and $\bvec{d}(s)$. 
\\[6pt] 
In the outer region, the projection $\bvec{H} = (\bvec{I}-\bvec{pp})\cdot (\bvec{r}-s \bvec{p})$ of the position vector $\bvec{r}-s \bvec{p}$ relative to a point on the fiber axis into the transverse plane is scaled with the fiber half-length $l$. The potential disturbance $\phi^o(s,\bvec{H})$ in the outer region can be written as a sum of the disturbances induced by the line of fluid volume sources $q(s)$ and those by the line of source-dipoles $\bvec{d}(s)$, i.e., $\phi^o = \phi_{q}^o + \phi_{d}^o$. Using equation \refeq{as_variation} for the cross-section of the prolate spheroidal fiber considered, the potential disturbance $\phi_{q}^o (s,\bvec{H})$ resulting from a distribution of point fluid volume sources along the center-line of the fiber with a linear density $q(s)$ given in equation \refeq{lo_qs} is, 
\begin{gather}
\phi_{q}^o (s,\bvec{H}) = - \frac{1}{4 \pi} \int_{-1}^{1} \frac{q(s')}{[(s-s')^2+H^2]^{1/2}} ds' = \frac{a_0 U_{\parallel}}{2\kappa} \int_{-1}^{1} \frac{s'}{[(s-s')^2+H^2]^{1/2}} ds' \eq{integral_for_phi_q} 
\end{gather} 
Analytical evaluation of the integral in equation \refeq{integral_for_phi_q} gives the following expression for $\phi_{q}^o (s,\bvec{H})$, 
\begin{align}
\phi_{q}^o(s,\bvec{H}) &= \frac{U_{\parallel}a_0 s}{2 \kappa} \Bigg[ \n{ln} \left( \frac{1-s}{H} + \sqrt{1+\left(\frac{1-s}{H}\right)^2} \right) + 
\n{ln} \left( \frac{1+s}{H} + \sqrt{1+\left(\frac{1+s}{H}\right)^2} \right) \Bigg] \nonumber \\ 
&\quad 
+ \frac{U_{\parallel}a_0 }{2 \kappa} \left[ \sqrt{(1-s)^2+H^2} - \sqrt{(1+s)^2+H^2} \right]  \eq{phi_par_outer}
\end{align} 
As we approach the fiber axis, we obtain, 
\begin{gather}
\lim_{H \to 0} \phi_{q}^o(s,\bvec{H}) = \frac{U_{\parallel}a_0s}{2\kappa}\n{ln} \left( \frac{1-s^2}{H^2} \right) + \frac{U_{\parallel}a_0 s}{\kappa} (\n{ln}2 -1) \eq{phi_par_outer_inner_limit}
\end{gather} 
Similarly, we use equation \refeq{as_variation} and compute the potential disturbance $\phi_{d}^o(s,\bvec{H})$ in the outer region resulting from a line distribution of volume source-dipoles $\bvec{d}(s)$ given in equation \refeq{lo_ds}.  
\begin{gather}
\phi_{d}^o (s,\bvec{H}) = -\frac{1}{4 \pi l} \int_{-1}^{1} \frac{\bvec{d}(s') \cdot \bvec{H}}{[(s-s')^2+H^2]^{3/2}} ds' = \frac{a_0 \bvec{U}_{\perp} \cdot \bvec{H}}{2\kappa} \int_{-1}^{1} \frac{1-(s')^2}{[(s-s')^2+H^2]^{3/2}} ds' \eq{integral_for_phi_d}
\end{gather} 
We evaluate the integral in equation \refeq{integral_for_phi_d} analytically to obtain, 
\begin{gather}
\phi_{d}^o (s,\bvec{H}) = \frac{a_0 (1-s^2)}{\kappa} \frac{\bvec{U}_{\perp} \cdot \bvec{H}}{2H^2} \left[ \frac{1-s}{\sqrt{(1-s)^2+H^2}} - \frac{1+s}{\sqrt{(1+s)^2+H^2}} \right] \nonumber \\ 
- \frac{a_0 \bvec{U}_{\perp} \cdot \bvec{H}}{2\kappa} \left[ \n{arsinh}\left(\frac{1-s}{H} \right) + \n{arsinh}\left(\frac{1+s}{H} \right) \right] + \frac{a_0 \bvec{U}_{\perp} \cdot \bvec{H}}{2\kappa} \left[ \frac{1+s}{\sqrt{(1-s)^2+H^2}} - \frac{1+s}{\sqrt{(1-s)^2+H^2}} \right] \eq{phi_perp_outer}
\end{gather} 
As we approach the fiber axis, we obtain, 
\begin{gather}
\lim_{H \to 0} \phi_{d}^o (s,\bvec{H}) = \frac{[\tilde{a}(s)]^2 \bvec{U}_{\perp} \cdot \bvec{H}}{\kappa H^2} + \frac{a_0\bvec{U}_{\perp}\cdot\bvec{H}}{2\kappa}\left( \frac{1+s^2}{1-s^2} \right) - \frac{a_0 \bvec{U}_{\perp}\cdot\bvec{H}}{2\kappa}\n{ln} \left( \frac{1-s^2}{H^2} \right) \eq{phi_perp_outer_inner_limit} 
\end{gather} 
To match the inner limit of the outer solution for the potential disturbance $\bvec{\phi}^o = \bvec{\phi}_q^o + \bvec{\phi}_d^o$ we express equations \refeq{phi_par_outer_inner_limit} and \refeq{phi_perp_outer_inner_limit} in terms of the radial vector $\bvec{h}$ in the transverse plane in the inner region to obtain, 
\begin{gather}
\phi^{o \to i}(s,\bvec{h}) = \phi_q^{o \to i}(s,\bvec{h}) + \phi_d^{o \to i}(s,\bvec{h}) \nonumber \\ 
= -\frac{U_{\parallel}a_0s}{\kappa}\n{ln} h + \frac{U_{\parallel}a_0s}{\kappa}(\n{ln}(2\kappa)-1) + a_0 [\tilde{a}(s)] \frac{\bvec{U}_{\perp}\cdot \bvec{h}}{h^2} \nonumber \\
+  \frac{a_0 [\tilde{a}(s)]\bvec{U}_{\perp}\cdot\bvec{h}}{2\kappa^2 }\n{ln}\left(\frac{h}{\kappa}\right) + \frac{a_0 [\tilde{a}(s)]\bvec{U}_{\perp}\cdot\bvec{h}}{2\kappa^2}\left( \frac{1+s^2}{1-s^2} \right) \eq{phi_outer_inner_variables} 
\end{gather} 
Upon comparing the above equation with equation \refeq{phi_inner_soln}, we find that the singular terms at $O(1)$ and $O(1/\kappa)$ are matched automatically. The non-singular part of the velocity disturbance produces the corrections at $O(1/\kappa^2)$ to the imposed flows $k_{\parallel}$ and $\bvec{k}_{\perp}$ correct up to $O(1/\kappa^2)$, 
\begin{gather}
k_{\parallel}(s) = U_{\parallel} + \frac{U_{\parallel}}{\kappa^2}\n{ln}\kappa + \frac{U_{\parallel}}{2\kappa^2}\left( \frac{1}{1-s} + \frac{1}{1+s} \right) \eq{k_par} \\ 
\bvec{k}_{\perp}(s) = \bvec{U}_{\perp} + \frac{\bvec{U}_{\perp}}{\kappa^2} \left(\frac{1}{1-s} + \frac{1}{1+s} \right) - \frac{2\bvec{U}_{\perp}}{\kappa^2} \n{ln} \kappa \eq{k_perp} 
\end{gather} 
We now substitute the expansions \refeq{qs_expansion} and \refeq{ds_expansion} for $q(s)$ and $\bvec{d}(s)$ into equations \refeq{scaled_BC_par} and \refeq{scaled_BC_perp} and use equations \refeq{k_par} and \refeq{k_perp} for $k_{\parallel}$ and $\bvec{k}_{\perp}$ to compute the next higher corrections to $q(s)$ and $\bvec{d}(s)$. This gives the following corrections to $q(s)$, 
\begin{gather}
q_2 (s) = 0 \\[6pt]  
\frac{q_3(s)}{2\pi [\tilde{a}'(s)]} - \frac{\tilde{a}'(s)\n{ln} [\tilde{a}(s)]}{2\pi} q_1'(s) = a_0 \tilde{a}'(s) \left[ \n{ln}\kappa+\frac{1}{1-s^2} \right] U_{\parallel} \nonumber \\ 
\implies q_3(s) = -2\pi a_0 s U_{\parallel} \left[ \n{ln}[\tilde{a}(s)] + \n{ln}\kappa + \frac{1}{1-s^2} \right] \eq{q3} 
\end{gather} 
and the following corrections to $\bvec{d}(s)$, 
\begin{gather}
\bvec{d}_1 (s) = 0 \\[6pt] 
\frac{\bvec{d}_2(s)}{2\pi[\tilde{a}(s)]^2} + \frac{\tilde{a}'(s)}{2\pi [\tilde{a}(s)]} \bvec{d}_0'(s) = -2 a_0^2\left[ \frac{1}{1-s^2} - \n{ln}\kappa \right] \bvec{U}_{\perp} \nonumber \\[6pt]  
\implies \bvec{d}_2(s) = 4\pi [a_0\tilde{a}(s)]^2 (\n{ln}\kappa - 1) \bvec{U}_{\perp} \eq{d2}
\end{gather} 
It is to be noted that the higher-order corrections, equations \refeq{q3} and \refeq{d2}, drive an outer solution that is of higher order than that needed to match the inner solution at $O(1/\kappa)$, so we will not pursue computing the outer solution to the potential disturbance driven by these corrections. Using the above corrections to $q(s)$ and $\bvec{d}(s)$, we can now determine the unknown functions $c_1(s)$ and $c_2(s)$ in equation \refeq{phi_inner_third_term} to evaluate the $O(1/\kappa^2)$ correction to the potential disturbance in the inner region. This is given by,  
\begin{gather} 
\frac{\phi_2^i(s,\bvec{h})}{\kappa^2} = \frac{1}{\kappa^2} \left[ \frac{a_0 \tilde{a}(s) \n{ln} h}{2} + \frac{\bvec{d}_2(s) \cdot \bvec{h}}{2\pi [a_0 \tilde{a}(s)] h^2} \right] \nonumber \\[6pt] 
\implies \frac{\phi_2^i(s,\bvec{h})}{\kappa^2} = \frac{1}{\kappa^2} \left[ \frac{a_0 \tilde{a}(s) \n{ln} h}{2} + \frac{2 a_0\tilde{a}(s) (\n{ln}\kappa - 1)} {h^2} \right] \bvec{U}_{\perp} \cdot \bvec{h} 
\end{gather} 
We can now compute the contribution of this $O(1/\kappa^2)$ potential disturbance correction to the torque on the fiber. For this, we evaluate the velocity disturbance in the inner region resulting from this $O(1/\kappa^2)$ correction and obtain, 
\begin{gather}
u_{2,\parallel}^i = \frac{1}{l} \frac{\partial \phi_{2}^i}{\partial s} = -\frac{1}{\kappa^3} \left[ \frac{s \n{ln}h}{2[\tilde{a}(s)]} - \frac{2 s (\n{ln}\kappa -1)}{[\tilde{a}(s)]h^2} \right] \bvec{U}_{\perp} \cdot \bvec{h} \eq{u_par_2} \\ 
u_{2,\perp h}^i = \frac{1}{a_0 \tilde{a}(s)} \frac{\partial \phi_{2}^i}{\partial h} = \frac{1}{\kappa^2} \left[ \frac{1+\n{ln}h}{2} + \frac{2(1-\n{ln}\kappa)}{h^2} \right] \bvec{U}_{\perp} \cdot \bvec{e}_h \eq{u_perp_z_2} \\
u_{2,\perp \alpha}^i = \frac{1}{a_0 \tilde{a}(s)h} \frac{\partial \phi_{2}^i}{\partial \alpha} = -\frac{1}{\kappa^2} \left[ \frac{2(\n{ln}\kappa-1)}{h^2} + \frac{\n{ln}h}{2} \right] U_{\perp} \n{sin}\alpha \eq{u_perp_alpha_2}  
\end{gather} 
The net fiber-induced velocity disturbance in the inner region is now given as, 
\begin{gather}
u_{\parallel}^i = \frac{1}{l} \frac{\partial \phi_{f2t}^i}{\partial s} = -\frac{U_{\parallel}}{\kappa^2} \n{ln}h + \frac{\tilde{a}'(s) \bvec{U}_{\perp} \cdot \bvec{h}}{\kappa h^2} + u_{2,\parallel}^i + o(1) \eq{u_par_f3t} \\ 
u_{\perp h}^i = \frac{1}{a_0 \tilde{a}(s)} \frac{\partial \phi_{f2t}^i}{\partial h} = \frac{\tilde{a}'(s) U_{\parallel}}{\kappa h} - \frac{\bvec{U}_{\perp} \cdot \bvec{e}_h}{h^2} + u_{2,\perp h}^i + o(1) \eq{u_perp_z_f3t} \\
u_{\perp \alpha}^i = -\frac{U_{\perp} \n{sin}\alpha}{h^2} + u_{2,\perp \alpha}^i + o(1) \eq{u_perp_alpha_f3t}  
\end{gather} 
Finally, we use equations \refeq{u_par_f3t}-\refeq{u_perp_alpha_f3t} to evaluate the velocity disturbance $\bvec{u}^i$ at the fiber surface ($h=1$), and the Bernoulli's equation to evaluate the pressure $p$. 
\begin{gather}
p(s,\alpha) + \frac{1}{2} \rho \vert \bvec{U} + \bvec{u}^i (s,h=1,\alpha) \vert ^2 = \frac{1}{2}\rho U^2 \eq{bernoulli_eqn} 
\end{gather}  
Using equations \refeq{F_f2t} and \refeq{T_f2t} with $p(s,\alpha)$ as the pressure, we obtain the next higher order corrections to the force and torque on the fiber. The magnitudes of the resulting force $\bvec{F}_{\n{pot}}$ and torque $\bvec{T}_{\n{pot}}$, with these corrections accounted for, are then given as, 
\begin{gather}
F_{\n{pot}} = 0 \\ 
\frac{T_{\n{pot}}}{\mu U L^2} = \frac{\pi \n{Re}_D}{12\kappa}\n{sin}2\theta - \frac{\pi \n{Re}_D}{4\kappa^3}\left(\n{ln}\kappa - \frac{5}{4} \right) \n{sin}2\theta + o(1) \eq{torque_f3t_large_kappa} 
\end{gather} 
We find that the expression for the normalised torque in \refeq{torque_f3t_large_kappa} agrees with the Lamb's result \refeq{T_lamb} up to $O((\n{ln}\kappa)/\kappa^3)$. As mentioned earlier in this section, it can now be seen that the computed corrections to the normalised torque is $O(\n{ln}\kappa/\kappa^2)$ smaller than the leading order term in equation \refeq{torque_f2t_large_kappa}. In all subsequent results we show, the expression for the potential-flow torque given in \refeq{torque_f3t_large_kappa} is added to the torque arising from Oseen flow to determine the net torque on the fiber at finite $\n{Re}_D$.  

\subsection{Torque on the translating fiber upon superposing Oseen flow torque and the potential flow torque} 

We now superpose the force and torque on the fiber computed in Oseen flow and potential flow to determine the net force and torque acting on the fiber in a steady relative uniform motion with the fluid at finite $\n{Re}_D$. Using the observation that the force $\bvec{F}_{\n{pot}}$ on the fiber in potential flow is zero, and the result in equation \refeq{torque_f3t_large_kappa} for the torque $\bvec{T}_{\n{pot}}$, we obtain, 
\begin{gather}
\bvec{F} = \bvec{F}_{\n{oseen}} + \bvec{F}_{\n{pot}} = \bvec{F}_{\n{oseen}} \\
\bvec{T} = \bvec{T}_{\n{oseen}} + \bvec{T}_{\n{pot}} 
\end{gather} 
Since the force on the fiber remains the same upon the introduction of potential flow disturbances, we will only show the results for the torque on the fiber as a function of $\n{Re}_D$ and the inclination between the fiber axis and the flow direction in this section.  
\begin{figure}
\centering
\includegraphics[width=0.56\textwidth]{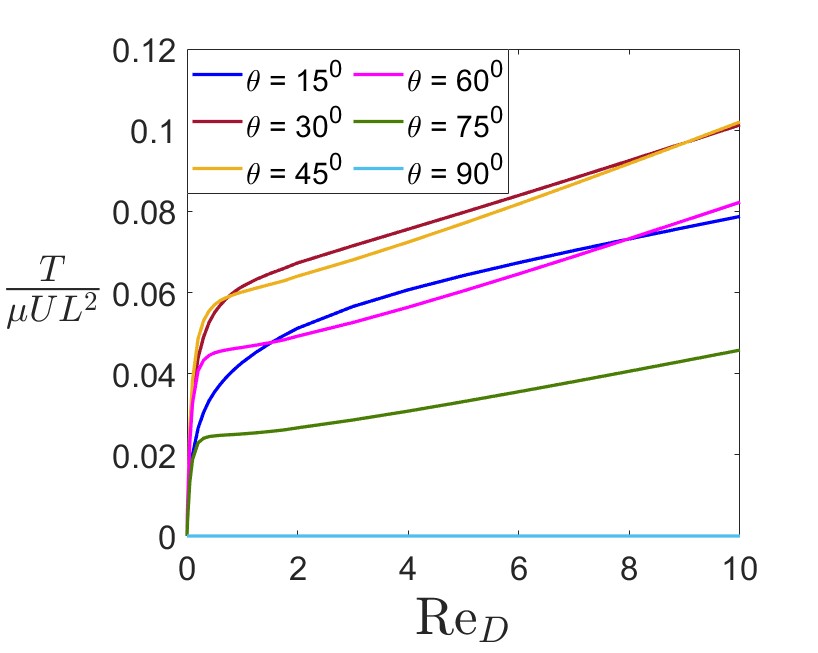} 
\caption{ Variation of the torque (obtained from inertial SBT with Oseen and potential flow disturbances in the outer region) on a translating fiber ($\kappa = 50$) with Reynolds number and angle of inclination between fiber axis and velocity direction}
\fig{T_pot_vs_ReD}
\end{figure} 
\\[6pt] 
The inertial SBT predictions for the variation of the torque on a steadily translating fiber from  with aspect ratio 50 as a function of $\n{Re}_D$ for various inclination angles between the fiber axis and its velocity direction are plotted in figure \reffig{T_pot_vs_ReD}. Unlike the torque from our inertial SBT with just the Oseen flow disturbance in the outer region (figure \reffig{torque_asp50}), we observe that the normalized torque grows with $\n{Re}_D$ for all the inclination angles considered. 
\\[6pt] 
We now proceed to compare the torque obtained from our inertial SBT with the full Navier-Stokes solution obtained using the method of \cite{sharma2023finite} for a steadily translating fiber. 
\begin{figure}
\centering
\begin{subfigure}{0.31\textwidth}
  \includegraphics[width=\textwidth]{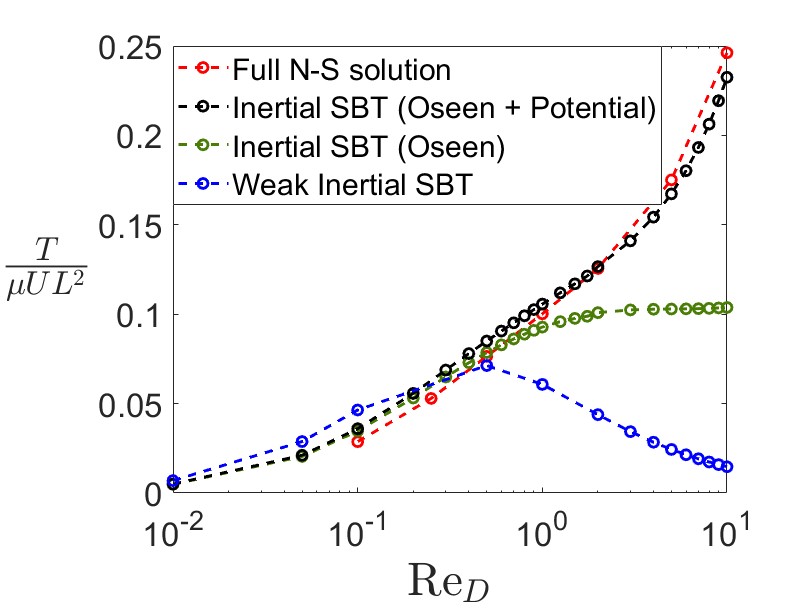}
  \caption{}
  \label{fig:sub1}
\end{subfigure}
\begin{subfigure}{0.31\textwidth}
  \includegraphics[width=\textwidth]{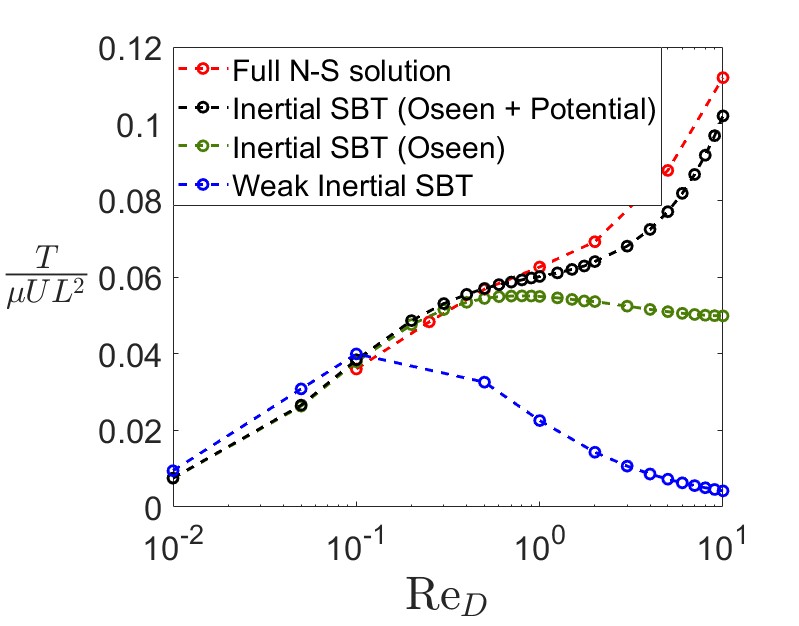} 
  \caption{}
  \label{fig:sub2}
\end{subfigure} 
\begin{subfigure}{0.31\textwidth}
  \includegraphics[width=\textwidth]{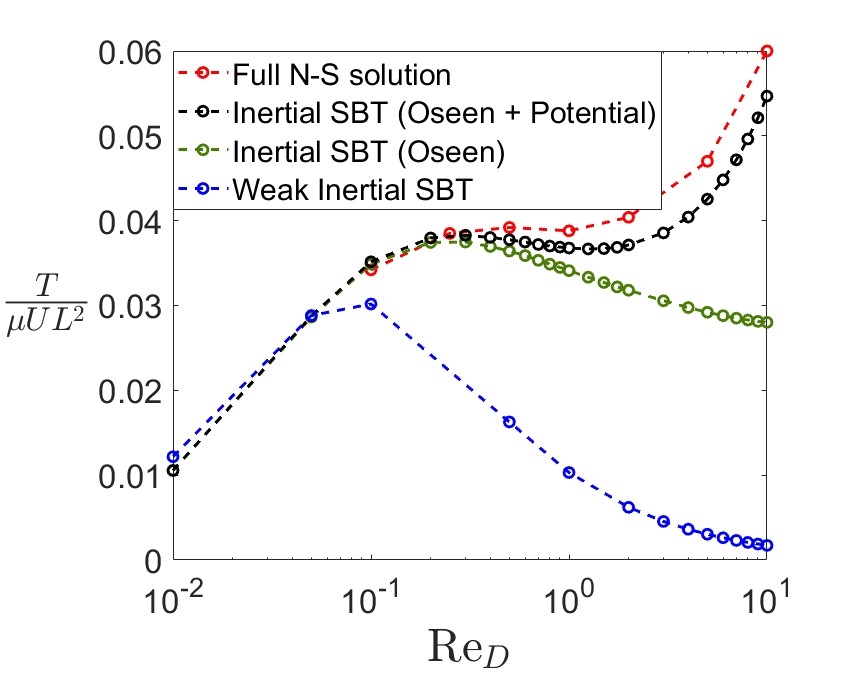} 
  \caption{}
  \label{fig:sub3}
\end{subfigure} 
\caption{Variation with $\n{Re}_D$ of the normalised inertial torque on the fiber translating in a direction inclined at $45^o$ to its orientation. Plots (a), (b) and (c) are for $\kappa = 20, 50$ and $100$ respectively.}  
\fig{T_pot_vs_ReD_comparison} 
\end{figure}  
\begin{figure}
\centering
\begin{subfigure}{0.45\textwidth}
  \includegraphics[width=\textwidth]{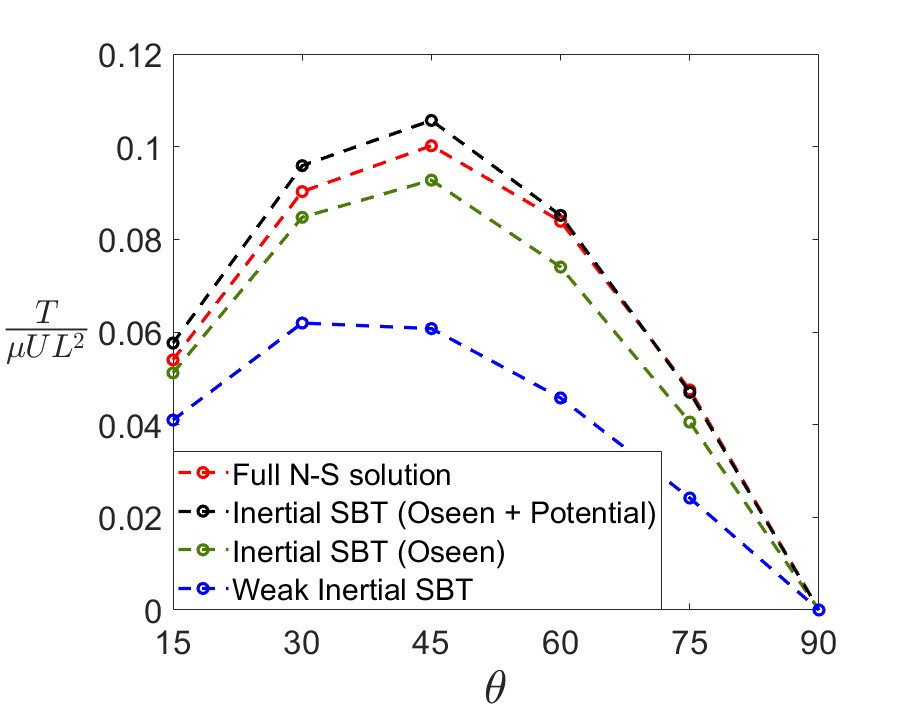}
  \caption{}
  \label{fig:sub1}
\end{subfigure}
\begin{subfigure}{0.45\textwidth}
  \includegraphics[width=\textwidth]{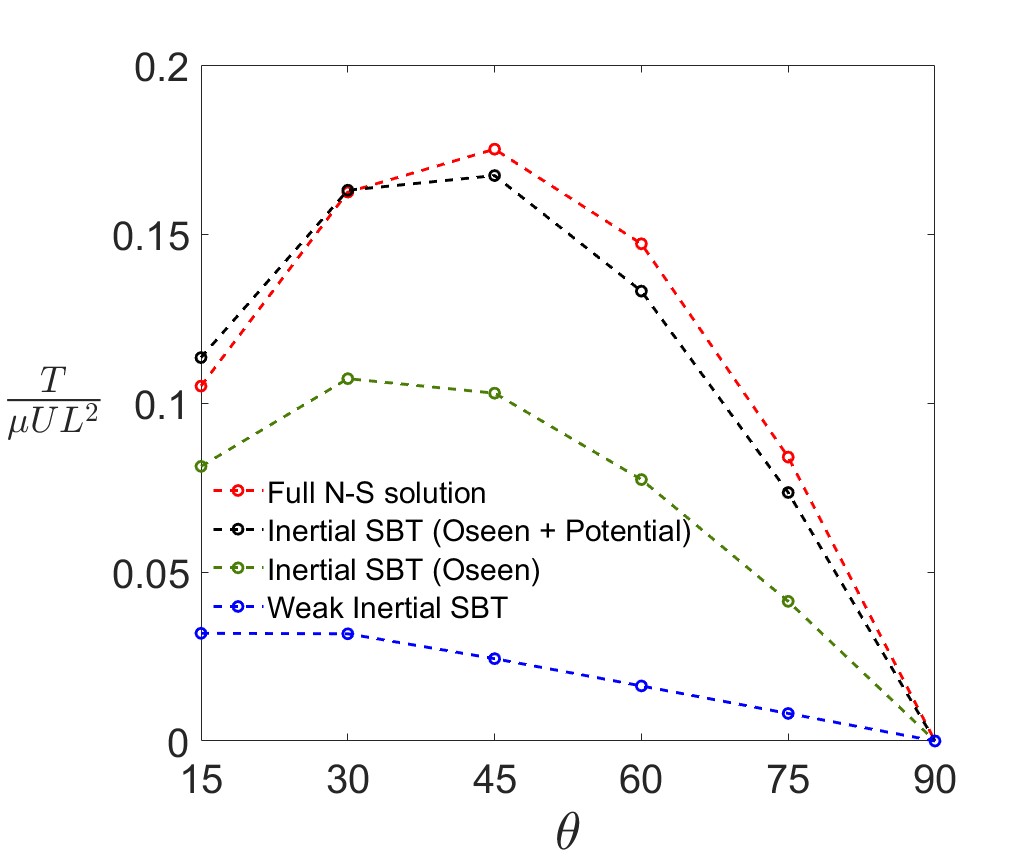} 
  \caption{}
  \label{fig:sub2}
\end{subfigure} 
\begin{subfigure}{0.45\textwidth}
  \includegraphics[width=\textwidth]{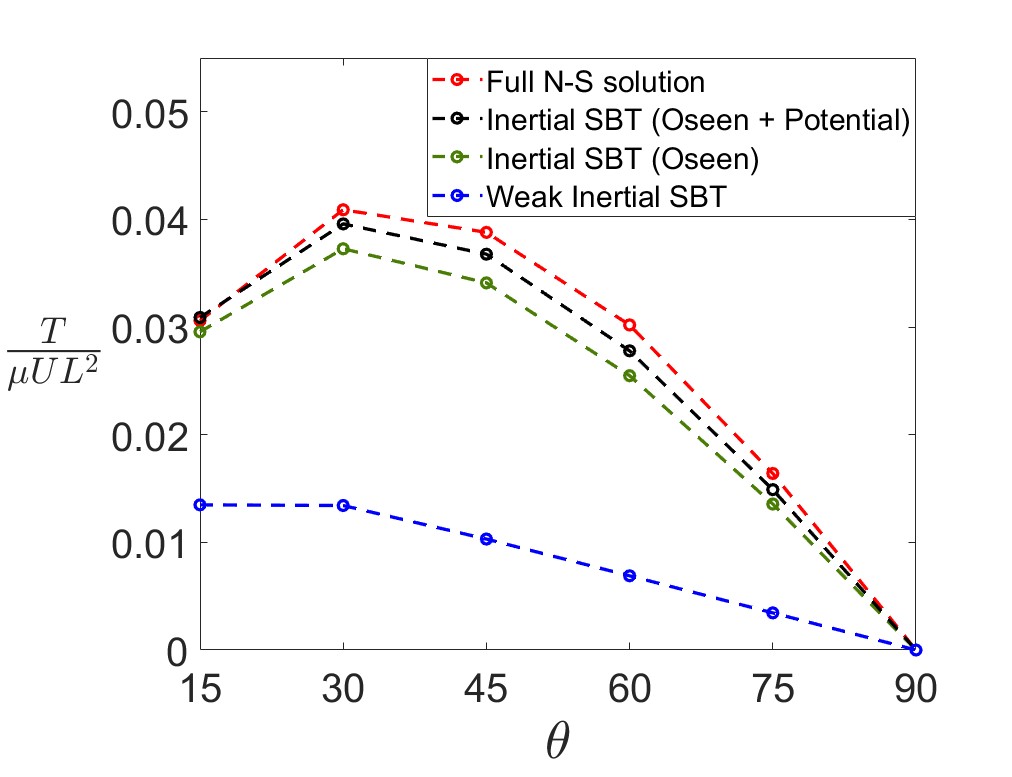}
  \caption{}
  \label{fig:sub1}
\end{subfigure}
\begin{subfigure}{0.45\textwidth}
  \includegraphics[width=\textwidth]{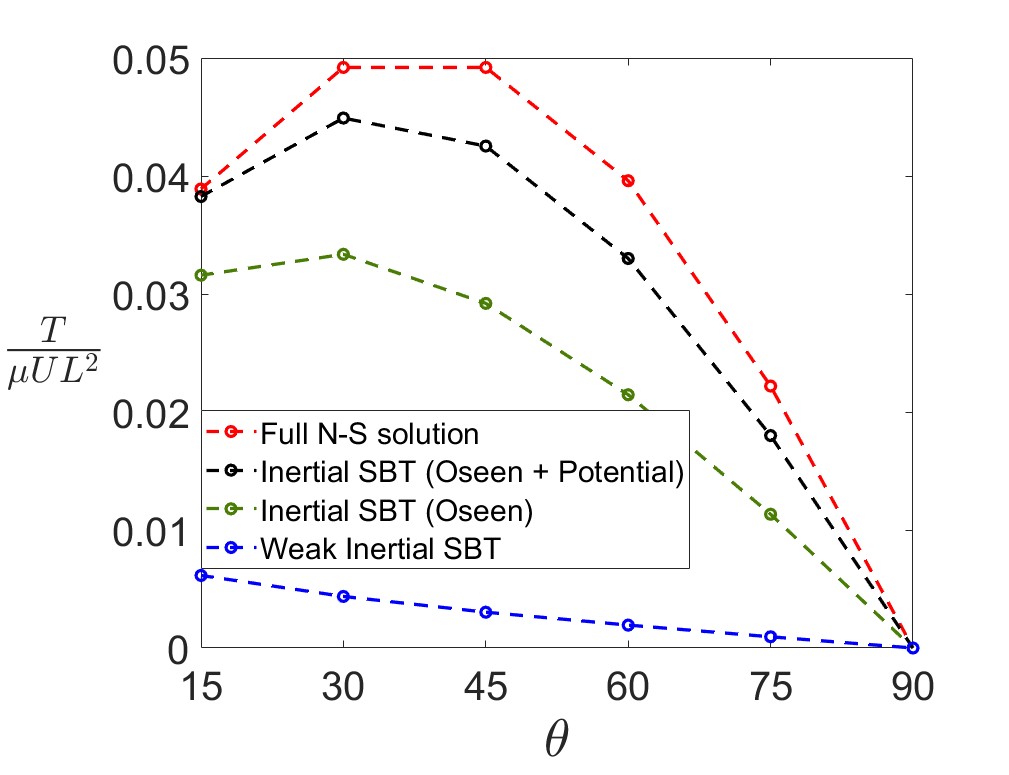} 
  \caption{}
  \label{fig:sub2}
\end{subfigure} 
\caption{Variation of the normalised inertial torque on the fiber with the inclination angle between the fiber velocity and orientation for $Re_D = 1$ [(a) and (c)] and $Re_D = 5$ [(b) and (d)]. Plots (a) and (b) are for $\kappa = 20$, and plots (c) and (d) are for $\kappa = 100$. } 
\fig{T_pot_vs_theta_comparison}
\end{figure} 
Figure \reffig{T_pot_vs_ReD_comparison} shows the variation of the normalized torque as a function of $\n{Re}_D$ for a fiber translating in a direction inclined at $45^o$ to its axis. We observe that the inclusion of potential flow disturbance in the outer region in our inertial SBT leads to a much better agreement of the inertial torque with the full Navier-Stokes solution. The variation of the inertial torque with the inclination angle between the fiber axis and its velocity direction for $\n{Re}_D = 1$ and $\n{Re}_D = 5$ is shown in figure \reffig{T_pot_vs_theta_comparison}. Curves are plotted for fiber aspect ratios of 20 and 100. The effect of the inclusion of potential flow disturbance is particularly evident at $\n{Re}_D = 5$ (for both $\kappa = 20$ and $\kappa = 100$) for  which it leads to a much more accurate prediction of the inertial torque for all the inclination angles considered. 

\section{Conclusions}
\sect{conclusions} 

In this work, we have developed an inertial slender-body theory (SBT) that captures the effects of finite fluid inertia on the scale of the particle length as well as the characteristic particle diameter. This has been achieved by matching the solution of the full quasi-two-dimensional Navier-Stokes equations (assuming negligible convection of the momentum disturbance along the particle major axis) on the scale of the particle diameter to the three-dimensional Oseen solution on the scale of the particle length. We have demonstrated the accuracy of the force and the inertial torque on a steadily translating fiber from our inertial SBT up to $\n{Re}_D$ of 10 by comparisons with experiments and finite-difference numerical simulations. Contrary to the weakly inertial theory of \cite{khayat1989inertia}, our theory predicts an increase in the inertial torque on a steadily translating fiber with $\n{Re}_D$. This is a consequence of the finite fiber diameter and the fluid incompressibility condition, and we model this effect by using a line distribution of fluid volume sources and source dipoles in potential flow.  Superposition of the effects of potential-flow and Oseen velocity disturbances leads to accurate predictions of the variation of the torque with Reynolds number and the angle between the fiber axis and fiber velocity.  While Khayat and Cox's expansion for small $Re_L/\ln(\kappa)$ leads to a force per unit length that depends on the fiber length at all values of $Re_L$, the new inertial slender-body theory exhibits a transition to a drag determined solely by the local fiber cross-section as seen in experiments and simulations. It is crucial to recognize that the assumption of negligible momentum convection along the fiber's major axis no longer holds when the relative velocity between the particle and the fluid is nearly aligned with the particle's axis of symmetry. Consequently, this study does not apply to such scenarios and so we have restricted the angle between the velocity and fiber axis to values greater than 15 degrees.  While the current study considered a steady relative uniform flow between the fiber and fluid, the inertial slender-body theory could be extended to consider the transient flows associated with the translation and rotation of freely suspended fibers settling in quiescent fluids or in imposed linear flows. 

\backsection[Acknowledgements]{A.S. is currently an employee of NTESS. This paper describes objective technical results and analysis. Any subjective views or opinions that might be expressed in the paper do not necessarily represent the views of the U.S. Department of Energy or the United States Government. Sandia National Laboratories is a multimission laboratory managed and operated by National Technology and Engineering Solutions of Sandia, LLC (NTESS), a wholly owned subsidiary of Honeywell International Inc., for the U.S. Department of Energy’s National Nuclear Security Administration under contract DE-NA0003525.}

\backsection[Funding]{This work was supported by the NSF Grants 2206851 and 2535828. }

\backsection[Declaration of interests]{The authors report no conflict of interest.} 

\bibliographystyle{jfm}
\bibliography{jfm} 

\appendix 

\section{Expressions for $a_1$ through $a_4$ in equation \refeq{J_tensor}} \label{appA}

Here, we provide analytical expressions for the $\n{Re}_L$ and $\theta$-dependent constants $a_1$, $a_2$, $a_3$ and $a_4$ mentioned in equation \refeq{J_tensor}. We write down that equation here for convenience. 
\begin{gather}
J_{ij}=2\int{\hat{G}}_{ij}^I\left(\mathbf{k}\right)j_0\left(2\pi k_p\right)d\mathbf{k}=a_1\delta_{ij}+a_2p_ip_j+a_3e_{Ui}e_{Uj}+a_4{(p}_ie_{Uj}+e_{Ui}p_j)  \eq{I_tensor_app}
\end{gather} 
Here, $\hat{\mathbf{G}}_{mn}^I(\mathbf{k})$ denotes the Fourier transform of $\mathbf{G}_{mn}^I(\mathbf{r})$ (see equation \refeq{G_inertial}), and is given as, 
\begin{gather}
\hat{\mathbf{G}}_{mn}^I(\mathbf{k}) = \frac{\pi i \n{Re}_L k_U}{(2\pi k)^2[(2 \pi k)^2 -  \pi i\n{Re}_L k_U]} \left(\bvec{I} - \frac{k_m k_n}{k^2}\right)  
\end{gather} 
where, $k_U = \bvec{k} \cdot \bvec{e}_U$ and $k_p = \bvec{k} \cdot \bvec{p}$ denote the components of the wave-vector $\bvec{k}$ along the directions of fiber-fluid relative velocity and orientation. We now perform double dot products of both sides of the tensor equation \refeq{I_tensor_app} with the second order tensors $\delta_{mn}$, $p_m p_n$, $e_{Um} e_{Un}$ and $p_m U_n$. This reduces equation \refeq{I_tensor_app} to a set of four scalar equations for the coefficients $a_1$ through $a_4$.  
\begin{gather}
3a_1 + a_2 + a_3 + 2\n{cos}\theta a_4 =  4 \int \frac{d\bvec{k}}{(2\pi)^3} \frac{\frac{i}{2}\n{Re}_Lk_U j_o(k_p)}{k^2(k^2-\frac{i}{2}\n{Re}_Lk_U)} = M_1 \eq{LS_1} \\
a_1 + a_2 + \n{cos}^2 \theta a_3 + 2\n{cos}\theta a_4 = 2 \int \frac{d\bvec{k}}{(2\pi)^3}\frac{\frac{i}{2}\n{Re}_L k_U (1-k_p^2/k^2)j_0(k_p)}{k^2(k^2-\frac{i}{2}\n{Re}_L k_U)} = M_2  \eq{LS_2} \\ 
a_1 + \n{cos}^2\theta a_2 + a_3 + 2\n{cos}\theta a_4 = 2 \int \frac{d\bvec{k}}{(2\pi)^3} \frac{\frac{i}{2}\n{Re}_L k_U (1-k_U^2/k^2) j_0(k_p)}{k^2(k^2-\frac{i}{2}\n{Re}_L k_U)} = M_3  \eq{LS_3} \\
\n{cos}\theta (a_1 + a_2 + a_3) + (1+\n{cos}^2\theta) a_4 = 2 \int \frac{d\bvec{k}}{(2\pi)^3} \frac{\frac{i}{2}\n{Re}_L k_U (\n{cos}\theta - k_p k_U/k^2)j_0(k_p)}{k^2(k^2-\frac{i}{2}\n{Re}_L k_U)} = M_4  \eq{LS_4} 
\end{gather} 
The expressions for the scalar integrals $M_1$ through $M_4$ are as follows, 
\begin{align} 
M_1 &= -\frac{1}{2\pi}\left[  E_1\left(\frac{A}{2} \right) + E_1\left(\frac{B}{2} \right) + 2\gamma + \n{ln}\left(\frac{A}{2} \right) + \n{ln}\left(\frac{B}{2} \right) \right]  \eq{M1} \\[10pt]  
M_2 &= -\left( \frac{1}{4\pi} + \frac{\n{sin}^2\theta \n{cos}^2\theta}{8\pi} \right) \left[  E_1\left(\frac{A}{2} \right) + E_1\left(\frac{B}{2} \right) + 2\gamma + \n{ln}\left(\frac{A}{2} \right) + \n{ln}\left(\frac{B}{2} \right) \right] \nonumber \\
&\quad
-\frac{\n{sin}^2\theta}{8\pi} \left[  E_1\left(\frac{B}{2} \right) - E_1\left(\frac{A}{2} \right) + \n{ln}\left(\frac{B}{2} \right) - \n{ln}\left(\frac{A}{2} \right) - 1 \right] - \frac{1}{\pi \n{Re}_L} \eq{M2} \\[10pt] 
M_3 &= -\left( \frac{1}{4\pi} - \frac{\n{sin}^2\theta}{8\pi} \right) \left[  E_1\left(\frac{A}{2} \right) + E_1\left(\frac{B}{2} \right) + 2\gamma + \n{ln}\left(\frac{A}{2} \right) + \n{ln}\left(\frac{B}{2} \right) \right] \nonumber \\
&\quad 
\frac{\n{cos}\theta}{2\pi \n{Re}_L} (e^{-A/2} - e^{-B/2} ) \eq{M3} \\[10pt] 
M_4 &= -\frac{\n{cos}\theta}{4\pi} \left[  E_1\left(\frac{A}{2} \right) + E_1\left(\frac{B}{2} \right) + 2\gamma + \n{ln}\left(\frac{A}{2} \right) + \n{ln}\left(\frac{B}{2} \right) \right] \nonumber \\
&\quad 
+ \frac{1}{4\pi \n{Re}_L}(e^{-B/2} - e^{-A/2}) \eq{M4} 
\end{align} 
In the above equations, $\gamma = 0.5772$ is the Euler's constant, and 
\begin{gather}
E_1(x) = \int_{x}^{\infty} \frac{e^{-\tau}}{\tau} d\tau 
\end{gather}
is the exponential integral. Finally, 
\begin{gather}
A = \frac{1}{2}\n{Re}_L(1-\n{cos}\theta) \\
B = \frac{1}{2}\n{Re}_L(1+\n{cos}\theta) 
\end{gather}
where, $\theta$ denotes the angle between the fiber-fluid relative velocity $\bvec{U}$ and the fiber orientation $\bvec{p}$. The analytical expressions for the constants $a_1$, $a_2$, $a_3$ and $a_4$ can be obtained by solving a system of four linear equations \refeq{LS_1}-\refeq{LS_4} and using the above expressions for $M_1$, $M_2$, $M_3$ and $M_4$. In the limit of $\n{Re}_L \gg 1$, one obtains for $M_1$ through $M_4$, 
\begin{flalign}
& M_1 = -\frac{1}{\pi} \Bigg[ \gamma + \n{ln}\left(\frac{\n{Re}_L \n{sin}\theta}{2} \right) - \n{ln}2 \Bigg] \eq{M1_large_ReL} \\
& M_2 = -\left(\frac{1}{2\pi} + \frac{\n{sin}^2\theta \n{cos}^2\theta}{4\pi} \right) \Bigg[\gamma + \n{ln}\left(\frac{\n{Re}_L \n{sin}\theta}{2} \right) -\n{ln}2 \Bigg] - \frac{\n{sin}^2\theta}{8\pi} \Bigg[ \n{ln} \left( \frac{1-\n{cos}\theta}{1+ \n{cos}\theta} \right) - 1 \Bigg] \eq{M2_large_ReL} \\
& M_3 = -\left(\frac{1}{2\pi} - \frac{\n{sin}^2\theta}{4\pi} \right) \Bigg[\gamma + \n{ln}\left(\frac{\n{Re}_L \n{sin}\theta}{2} \right) -\n{ln}2 \Bigg] \eq{M3_large_ReL} \\ 
& M_4 = -\frac{\n{cos}\theta}{2\pi} \Bigg[\gamma + \n{ln}\left(\frac{\n{Re}_L \n{sin}\theta}{2} \right) -\n{ln}2 \Bigg] \eq{M4_large_ReL}
\end{flalign} 

\section{Influence of the outer boundary and grid spacing in the results of the finite-difference Navier-Stokes solver of \cite{sharma2023finite}} \label{appB} 

In this appendix, we examine the effects of the outer boundary radius ($R_{\infty}$) and the fineness of the computational grid on the forces and torques experienced by a prolate spheroidal particle, as calculated using the finite difference Navier-Stokes solver described in \cite{sharma2023finite}. We have varied the outer radius of the computational domain ($R_{\infty}$) and the number of grid points ($N_1, N_2, N_3$) along the radial, polar, and azimuthal directions of the prolate spheroidal coordinate system for a particle inclined at $45^\circ$ to an imposed uniform flow. The results of these tests are presented for a particle aspect ratio of 100, as this case is expected to be the most computationally challenging due to the need for resolving highly disparate length scales  near the particle surface and in the far field. 
\\[6pt] 
Figure \reffig{FD_code_validity} illustrates the variation of the forces parallel and perpendicular to the fiber axis, as well as the inertial torque, with $R_\infty$ and grid resolution. The labels 1 through 3 for grid resolution represent three different grid spacings, with type 1 being the coarsest and type 3 the finest. 
\begin{figure}
\centering
\begin{subfigure}{0.31\textwidth}
  \includegraphics[width=\textwidth]{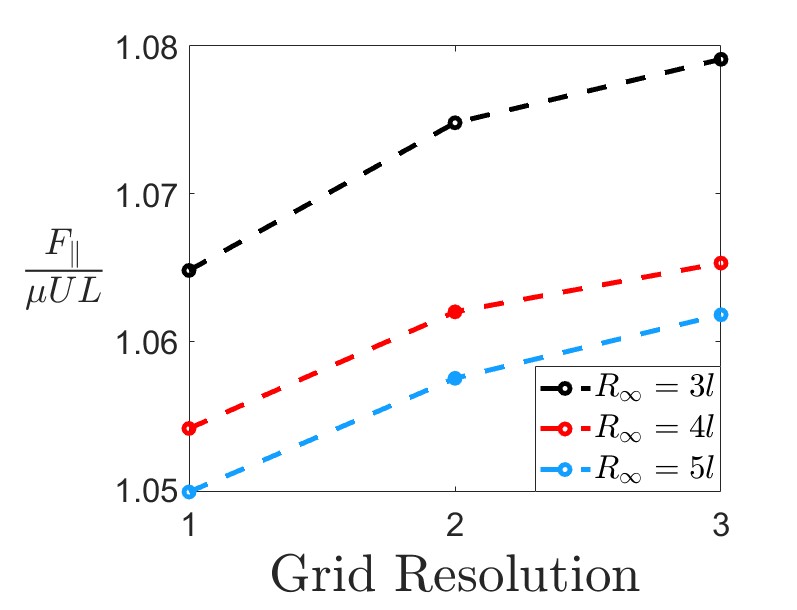}
  \caption{}
  \label{fig:sub1}
\end{subfigure}
\begin{subfigure}{0.31\textwidth}
  \includegraphics[width=\textwidth]{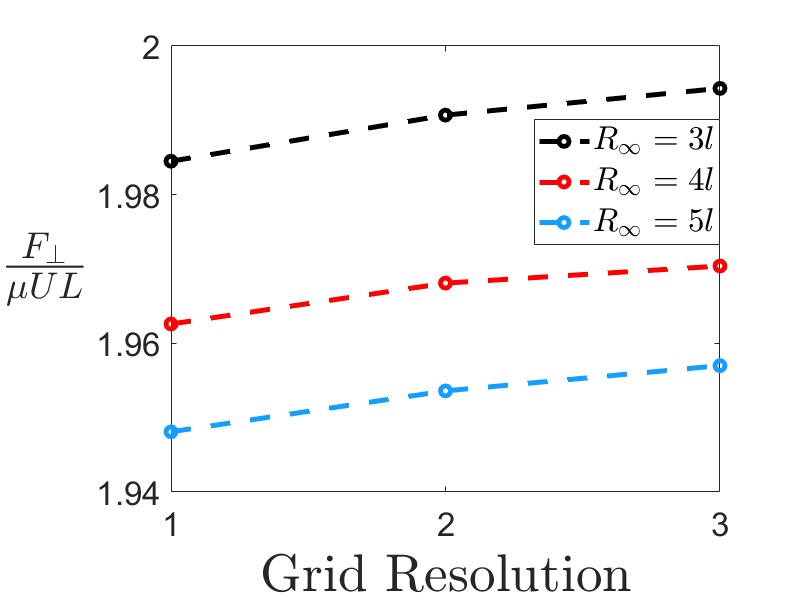}
  \caption{}
  \label{fig:sub2}
\end{subfigure}
\begin{subfigure}{0.31\textwidth}
  \includegraphics[width=\textwidth]{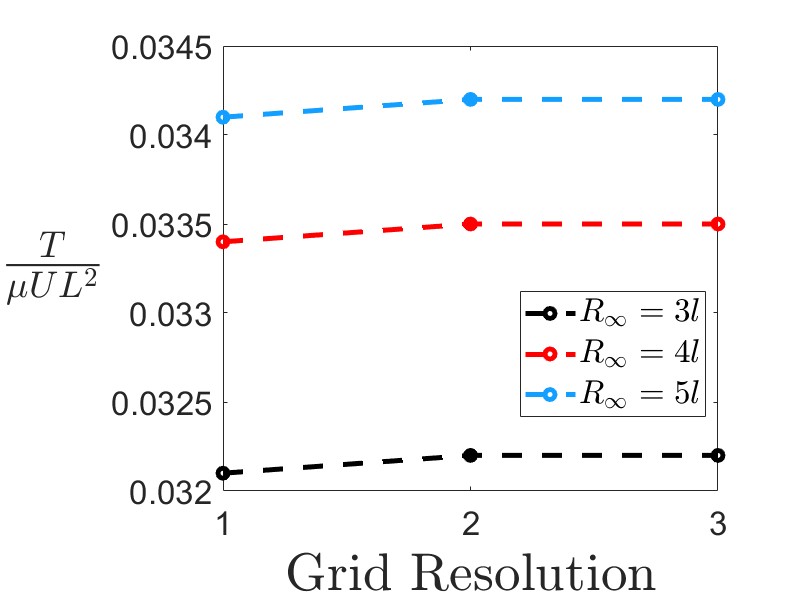}
  \caption{}
  \label{fig:sub3}
\end{subfigure}
\begin{subfigure}{0.31\textwidth}
  \includegraphics[width=\textwidth]{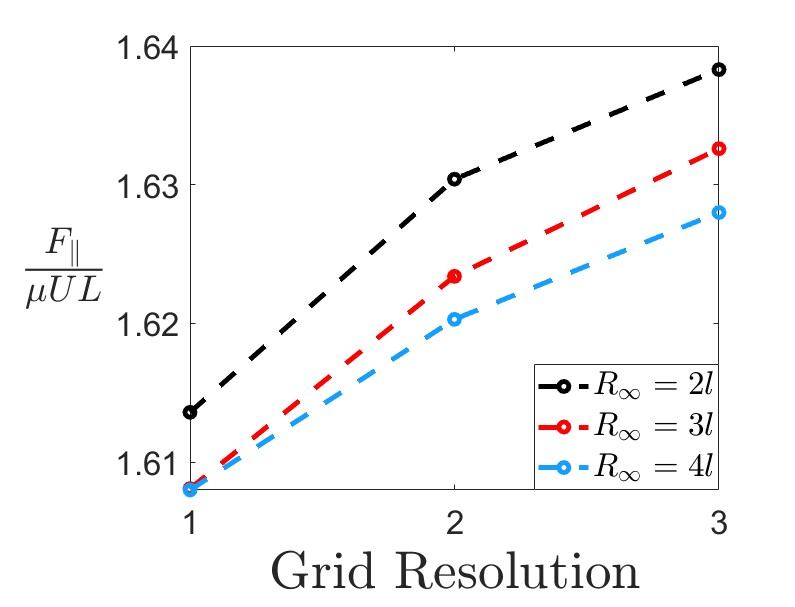}
  \caption{}
  \label{fig:sub4}
\end{subfigure}
\begin{subfigure}{0.31\textwidth}
  \includegraphics[width=\textwidth]{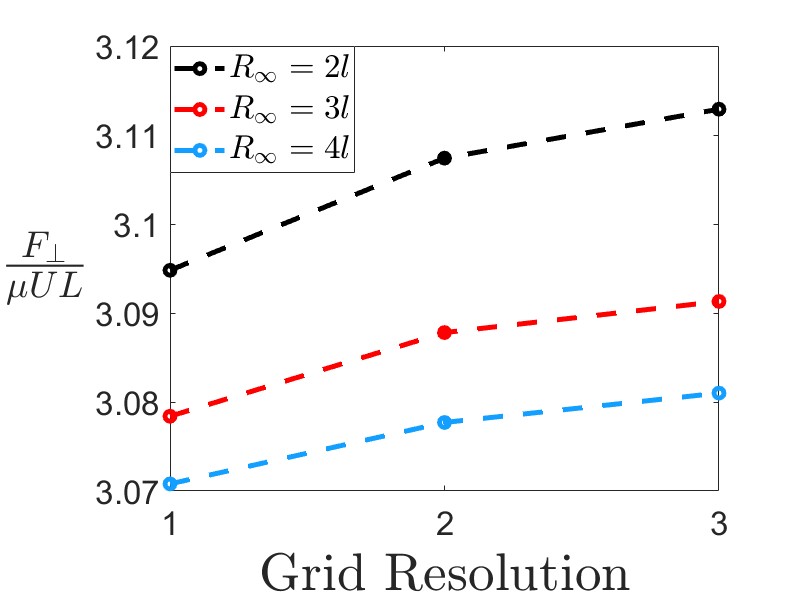}
  \caption{}
  \label{fig:sub5}
\end{subfigure}
\begin{subfigure}{0.31\textwidth}
  \includegraphics[width=\textwidth]{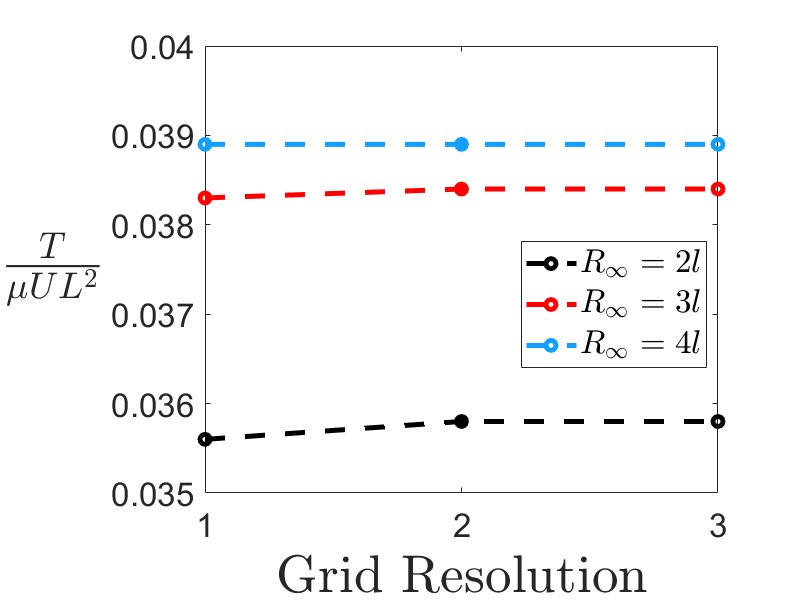}
  \caption{}
  \label{fig:sub6}
\end{subfigure}
\caption{Variation of the forces parallel ($F_{\parallel}$) and perpendicular ($F_{\perp}$) to the particle axis and the inertial torque ($T$) as obtained from the finite difference code of \cite{sharma2023finite} with the grid resolution and the outer boundary radius ($R_{\infty}$) of the computational grid. Plots (a), (b) and (c) are for $\n{Re}_D = 0.1$ and plots (d), (e) and (f) are for $\n{Re}_D = 1$. $l$ is the half-length of the spheroidal particle. } 
\fig{FD_code_validity} 
\end{figure} 
\begin{table}
  \begin{center}
\def~{\hphantom{0}}
  \begin{tabular}{lcccc}
      $\n{Re}_D$  & $R_{\infty}$   &   Grid resolution 1 & Grid resolution 2 & Grid resolution 3 \\[3pt]
       0.1   & $3l$ & 251, 151, 51 & 301, 201, 51 & 351, 231, 51 \\
       0.1   & $4l$ & 301, 201, 51 & 351, 231, 51 & 401, 261, 51 \\
       0.1  & $5l$ & 351, 231, 51 & 401, 261, 51 & 451, 301, 51 \\
       1   & $2l$ & 151, 91, 51 & 201, 121, 51 & 251, 151, 51 \\
       1 & $3l$ & 201, 121, 51 & 251, 151, 51 & 301, 201, 51 \\
       1 & $4l$ & 251, 151, 51 & 301, 201, 51 & 351, 231, 51 \\ 
  \end{tabular}
  \caption{Values of $R_{\infty}$ and the number of radial, polar and azimuthal grid points  ($N_1, N_2,$ and $N_3$) corresponding to each grid resolution type in figure \reffig{FD_code_validity}.}
  \label{tab:FD_code_grid}
  \end{center}
\end{table} 
The corresponding number of grid points along the three prolate spheroidal coordinate directions for each label is provided in table \ref{tab:FD_code_grid}. We see that as the grid resolution is refined from type 1 to type 3 and $R_\infty$ increases (from $3l$ to $5l$ for $\n{Re}_D = 0.1$ and from $2l$ to $4l$ for $\n{Re}_D = 1$), the forces and torque exhibit clear convergence toward their asymptotic values. For grid resolution type 3 and the largest $R_{\infty}$ considered, the deviation from the results at higher resolutions and larger outer boundary radii remains consistently within $2–3\%$ for both $\n{Re}_D = 0.1$ and $\n{Re}_D = 1$. This level of accuracy demonstrates that the code achieves satisfactory convergence with grid resolution type 3 and $R_{\infty} = 5l$ and $4l$ (for $\n{Re}_D = 0.1$ and $1$ respectively), capturing the essential flow features across both near-field and far-field scales, even with the computational challenges posed by a particle aspect ratio of 100. Consequently, we select these values for grid spacing and outer boundary radius to balance the trade-off between accurate force and torque predictions, ensuring numerical convergence, and managing computational cost. 

\end{document}